\documentclass[twocolumn,floatfix,showkeys]{emulateapj}
\usepackage{graphicx}
\usepackage{float}
\usepackage{rotating}
\usepackage{epsfig,floatflt}
\usepackage[latin1]{inputenc}
\usepackage{amsfonts}
\usepackage{amssymb}
\usepackage{amsthm}
\newcommand{\diff}{\,\mathrm{d}}
\usepackage{epsfig}
\theoremstyle{remark}

\theoremstyle{definition}

\newcommand{\Oo}{{\cal O}}

\begin{document}
\title{On the Magnitude of Dark Energy Voids and Overdensities}
\author{David F. Mota$^1$,  Douglas J. Shaw$^2$ and Joseph Silk$^3$}
\affil{$^1$ Institute for Theoretical Physics, University of Heidelberg, 69120 Heidelberg,Germany} 
\affil{$^2$ DAMTP, Centre for Mathematical Sciences,
University of Cambridge, Cambridge CB3 0WA, UK}
\affil{$^3$  Astrophysics, University of Oxford, Oxford OX1 3RH,UK} 

\begin{abstract}
We investigate the clustering of dark energy within matter overdensities and voids. 
In particular, we derive an analytical expression for the dark energy
density perturbations, which is valid both in the linear, quasi-linear and fully non-linear regime of structure formation. We also investigate the possibility of detecting such dark energy clustering
through the ISW effect.  In the case of uncoupled quintessence models, if the mass of the field
is of order the Hubble scale today or smaller, dark energy fluctuations are always small compared to the matter density
contrast. Even when the matter perturbations enter the non-linear regime, the dark energy perturbations remain linear. 
We find that virialised clusters and voids correspond to local overdensities in dark energy, with
$\delta_{\phi}/(1+w) \sim \Oo(10^{-5})$ for
voids, $\delta_{\phi}/(1+w) \sim \Oo(10^{-4})$ for super-voids and $\delta_{\phi}/(1+w) \sim  \Oo(10^{-5})$ for a typical virialised cluster.
If voids with radii of $100-300\,{\rm Mpc}$ exist within the visible
Universe then $\delta_{\phi}$ may be as large as $10^{-3}(1+w)$.
Linear overdensities of matter and super-clusters generally
correspond to local voids in dark energy; for a typical super-cluster:
$\delta_{\phi}/(1+w) \sim \Oo(-10^{-5})$.  
The approach taken in this work could be straightforwardly extended to study the clustering of more general dark energy models.
%We conclude that if dark
%energy is described by an uncoupled quintessence model, then today's
%dark energy is almost homogeneous on sub-horizon scales with the largest
%perturbations  at most of order $10^{-3}$ and associated with extreme large voids of matter.
\end{abstract}

%\pacs{98.80.-k,98.80.Jk}


\maketitle

\section{Introduction} \label{sec:intro}
It has been almost a decade since observations of Type Ia supernovae
(SNe Ia) were first found to support the notion that our universe is
currently undergoing a phase of accelerated expansion
\citep{sn1a,sn1aa}.  Since that time, evidence in favour of this
accelerated expansion has strengthened significantly as the result of
further SNe Ia observations \citep{sn1anew,wbounds,sn1anew1,sn1anew2}, improved
measurements of the cosmic microwave background (CMB) \citep{wmap,wmap3} and
surveys of large scale structure (LSS) \citep{lss,lss1}.  The precise
cause of this late-time acceleration, however, remains unknown.

If general relativity is accurate on astrophysical scales, then the assumptions of large-scale
homogeneity and isotropy require that the agent responsible for the
universe's acceleration, dubbed `dark energy', behave cosmologically
as a fluid with negative pressure.  The standard model of particle
physics predicts only one such fluid: the vacuum energy, or
cosmological constant, for which the pressure, $p$, is always equal to
minus the energy density, $\rho$.  However, if the vacuum energy
density is indeed non-zero then it is generally expected to be of the order of $M_{\rm Pl}^4$, where $M_{\rm Pl}$ is
the Planck-mass. This is some $120$ orders of magnitude larger than
the observed dark energy density.  A number of proposals have
therefore been made in the literature for models in which dark energy
is dynamical and associated with some new form of energy
\citep{quint,quint1}. In these models, the present small size of the
effective cosmological constant is by-product of the age of
Universe. The acceleration of the Universe might alternatively be
explained by modifying General Relativity rather than postulating a
new form of energy \citep[see][]{dgp,morad,mod,mod1,tomi1,tomi2}.  It also been
suggested that the Universe is not accelerating at all, but that a
large local inhomogeneity prevents the SNe Ia data from being
correctly interpreted in terms of a homogeneous and isotropic
cosmological model \citep{nodark,nodark1}.  This said, dynamical dark
energy (hereafter DDE) is by far the most popular
candidate to explain the current astronomical data.  In the simplest DDE models, known as quintessence, dark energy is associated with the energy density of a scalar field with a canonical kinetic structure.

If dark energy does indeed exist then astronomical observations
presently provide us with only hints as to its nature.  We know that
today it represents about $70\%$ of the total energy density of the
Universe, and that its equation of state (EoS) parameter,
$w\equiv p/\rho$, is fairly close to $-1$: $w = -1\pm0.1$ for $z<1$,
\citep{wbounds}. Furthermore, if matter dominates the expansion of the
Universe for $z>1.8$ then $w(z > 1) = -0.8^{+0.6}_{-0.1}$
\citep{wbounds}. Hence, dark energy has negative pressure at higher redshifts
with a $98\%$ confidence.  These bounds on $w$ are entirely consistent with a pure
cosmological constant. Whilst detecting either $w \neq -1$ or $\diff w
/ \diff z \neq 0$ would rule out a cosmological constant, in many DDE
models significant deviations from $w = -1$ only occur at early times
and as such would be difficult to detect.  The late time, background cosmology predicted by many DDE models is therefore very
similar to that of a universe with a true cosmological constant.  

DDE models generally cease to mimic a cosmological constant in inhomogeneous backgrounds or when one considers cosmological perturbation theory. In particular, a number of authors have
studied the effect of DDE on the formation of large scale structure \citep{lsss,lsss1,lsss2,georg}.
In the vast majority of these works the energy density of dark energy
is taken to be homogeneous i.e. it is assumed that DDE does \emph{not}
cluster.  The extent to which this assumption of homogeneity is valid
has been the subject of some interest and much debate in the
literature. In an inhomogeneous background, the DDE energy density and
EoS parameter should exhibit some spatial variations. The key issue of
how large these variations should be is however far from settled,
particularly when the matter perturbation goes non-linear
\citep{maor,carsten,linder,doran,nelson,liberato}. In this paper we attempt to settle this
issue for uncoupled quintessence models by deriving an analytical
expression for the dark energy contrast, $\delta_{\phi}$, in by
presence of a matter perturbation. Importantly, we do \emph{not} constrain the matter
perturbation to be small (i.e. linear).

Clustering of DDE over scales smaller than about $100\,\mathrm{Mpc}$
has been the subject of a number of recent articles
\citep{roberto,bala,Mai,perc,peng,nunes,bala1}.  Most attention
has been focussed on models in which the DDE couples to baryonic and/or
dark matter since DDE clustering is expected to be strongest in such
theories \citep{amendola,brook,manera,valeria}.  In other works, a more
phenomenological approach has been taken and the observables
associated with DDE clustering have been parametrized. In particular,
it has been shown that inhomogeneities in dark energy could produce
detectable signatures on the CMB \citep{weller,koivisto}.

Recently, \citet{dutta} considered the growth of spatial
DDE perturbations in the \emph{absence} of any matter coupling.  They
considered only those circumstances where both the density contrast of
matter, $\delta_{m}$, and that of the DDE, $\delta_{\phi}$, were small
enough to be treated as a linearized perturbations about a homogeneous
and isotropic cosmological background.  They studied the simplest
class of quintessence models where dark energy is associated with the
slow-roll of a scalar field $\phi$ down a potential $V(\phi)$; $\phi$
is minimally coupled to gravity. The authors linearized the full field
equations for $\phi$, the matter and the metric and solved the
resulting system numerically.  Intriguingly they found that, at late
times, a local overdensity of matter corresponded to a local
under-density, or void, of dark energy.  Conversely, a void in the
matter was seen to produce a local DDE overdensity. Although the DDE
density contrast, $\delta_{\phi}$, is initially very small compared to
the matter density contrast, $\delta_{m}$, they found that, when
$\delta_m \sim \Oo(1)$, $\vert \delta_{\phi}\vert \sim
\Oo(10^{-2})$. $\delta_{\phi}$ was also observed to be growing more
quickly than $\delta_{m}$ at late times.  Their results suggest that
DDE clustering may induce a positive correction to the value of
$(1+w)$ of more than $10\%$ at the centre of an inhomogeneity with
properties similar to that of the local supercluster (hereafter LSC).
If accurate, the results of \citet{dutta} imply that any
deviations from $w=-1$ would be significantly amplified by the
presence of a local overdensity of matter.  Furthermore, they suggest
that DDE clustering might be relatively strong when the matter
perturbation goes non-linear.

\citet{dutta} studied DDE clustering when the matter
perturbation is growing in the linear regime, which is only accurate
when $\delta_m \ll 1$. They considered the evolution $\delta_{\phi}$
for a cluster of matter with initial density profile (at $z=35$) of
$\delta_{m} = A \exp(-r^2/\sigma^2)$ where $A = 0.1$ and $\sigma =
0.01 H_{i}^{-1}$ with $H_{i}$ being the initial value of the Hubble
parameter.  However, in the absence of any DDE, the linear
approximation would generally cease to be accurate when $z \approx
2.8$; additionally the perturbation would be expected to turnaround
when $z \approx 1.0$ and virialise when $z \approx 0.3$.  It is
therefore far from clear whether or not the sharp late-time growth in
$\delta_{\phi}$ found by \citet{dutta} is indeed a physical
effect, or just a result of using linearized field equations outside
of realm in which they are valid. 
Although 
one might well expect there exist some mapping 
between the linear and non-linear regimes as it 
happens in an Einstein-de Sitter Universe. 

In this paper we investigate a similar problem to that considered by \citet{dutta}, i.e. the clustering of uncoupled quintessence on sub-horizon scales.  There are, however, two important differences between our approach to this problem and the one taken by \citet{dutta}:   
\begin{enumerate}
\item Firstly, we do \emph{not} use numerical simulations but instead use the method of matched asymptotic expansions (MAEs) to develop a analytical approximation to $\delta_{\phi}$.
\item Secondly, we do \emph{not} require the density contrast of the matter perturbation, $\delta_{m}$, to be small. Indeed, our analysis and results remain valid even after the virialisation of the matter overdensity.  
\end{enumerate}
Our approximation for $\delta_{\phi}$ is accurate provided that $\delta_m$ is only non-linear ($\gtrsim \Oo(1)$) on sub-horizon scales.  This provision is consistent with observations. For simplicity we take the matter perturbation to be spherical symmetric. We also require that gravity is suitably weak whenever the
inhomogeneity is non-linear i.e. $GM/R \ll 1$. We restate these requirements in a rigorous fashion later, however they are essentially equivalent to the statement that gravity is approximately Newtonian over scales smaller than $H^{-1}$. Since DDE clustering in the linear regime has been dealt with in great detail by \citet{dutta}, our main focus in this paper is on what occurs when the matter perturbation goes non-linear.  

This paper is organized as follows: in Section \ref{sec:model} we
describe our model for an inhomogeneous  spacetime and the
DDE.  We state the equations that must be satisfied by the metric
quantities and the dark energy scalar field $\phi$.  In Section
\ref{sec:small} we introduce the method of matched asymptotic
expansions (MAEs).  This method relies the existence of locally small
parameters, and we state what these are for our model and interpret
them physically. We also note what constraints the smallness of these
parameters places upon our analysis.  In Section \ref{sec:evo} we
apply the method of MAEs to the evolution of dynamical dark energy
perturbations. We derive a simple equation for the DDE density
contrast, $\delta_{\phi}$, in terms of the peculiar velocity of matter
particles, $\delta v$, and the matter density contrast, $\delta_{m}$.
Importantly this equation is equally as valid for $\delta_{m} \gtrsim
\Oo(1)$ (`the non-linear regime') as it is when $\delta_{m} \ll
\Oo(1)$ (`the linear regime').  In Section \ref{sec:DDEclust} we use
our results to study the evolution of
$\delta_{\phi}$ in the linear ($\delta_{m} \ll 1$), quasi-linear
($\delta_{m} \sim \Oo(1)-\Oo(10)$) and fully non-linear ($\delta_{m}
\gg 1$) regimes.  We compare our analytical results with those found
numerically in the linear regime by \citet{dutta}.  In
Section \ref{sec:app} we consider the spatial profile of the DDE
density contrast for realistic astrophysical inhomogeneities such as
voids, supervoids, clusters and superclusters. We conclude in Section
\ref{sec:con} with a discussion of our results and their observational
implications. We also note how our analysis might be extended to
include even more general DDE models.

Throughout the paper we use units where $c=1$.

\section{The Model}\label{sec:model}
\subsection{Geometrical Set-Up} \label{sec:geo}
Our aim in this paper is to derive an expression for the DDE density contrast, $\delta_{\phi} = \delta
\varepsilon^{(\phi)}/\varepsilon^{(\phi)}_c$, inside an over-density, or
under-density, of matter. Henceforth we use $\varepsilon^{(i)}$ to represent the energy density of a component $i$.  We are particularly interested in those
cases where the matter perturbation is non-linear
i.e. $\delta\varepsilon^{(m)}/\varepsilon^{(m)}_{c}\gtrsim \Oo(1)$.
Although we aim to remain suitably general in our treatment of the
density perturbation, we do make the following simplifying assumptions:
\begin{itemize}
\item We assume spherical symmetry.  We briefly discuss the extent to which the relaxation of this assumption would affect our results in Section \ref{sec:con} below, and conclude that the qualitative nature of our findings would be unaffected.
\item We define a `physical radial coordinate', $R$, by the requirement that a spherical surface with physical radius $R$ has surface area $4\pi R^2$.  We assume that all curvature invariants are
regular at $R=0$ i.e.  we do not consider those cases in which there is a central black hole.  We argue below, however, that our results are still accurate even when there is a central black-hole, provided they are only applied at radii that are large compared to the Schwarzschild radius of the black hole.
\item Finally we assume that for radii smaller than some $R_0 \ll H^{-1}$
gravity is suitably \emph{weak} in the inhomogeneous region; $H$ is the
Hubble parameter of the background spacetime.  We define
what we mean by \emph{weak} rigorously in Section
\ref{sec:small}. For radii $R > R_0$, we require that the matter density contrast is $\lesssim 1$.  This assumption holds for most realistic models of collapsing overdensities provided that the radius the overdense region is less than about $0.1/H$.
\end{itemize} 

\subsection{Einstein's Equations}\label{sec:ein}
We take the matter content of the Universe to be a mix of irrotational dust and dynamical dark energy, which is described by a scalar field $\phi$.   
For simplicity we restrict ourselves to considering only spherically symmetric spacetimes for which the most general line element in comoving coordinates is:
$$
\diff s^2  = \diff t^2 - U(t,r)\diff r^2 - R(t,r)^2\left(\diff \theta^2 + \sin^2 \theta\,\diff \varphi^2\right).
$$ 
We make the definitions $U(t,r) \equiv R_{,r}^2(t,r)/Q(t,r)$ and $k(t,r) \equiv 1-Q(r,t)$. This coordinate choice is unique up to $t \rightarrow t + t_{0}$, and $r \rightarrow r^{\prime}(r)$.  
With these definitions the 2-spheres $\left\lbrace t, r \right\rbrace = \mathrm{const}$ have surface area $4\pi R^2(t,r)$, and in this sense $R(t,r)$ 
represents the `physical radial coordinate'. The energy-momentum tensor of pressureless dust is given by
$$
T_{ab}^{(m)} = \mathrm{diag}\left(\varepsilon^{(m)},0,0,0\right),
$$
and the energy-momentum tensor of the scalar field, $\phi$, is
$$
T_{ab}^{(\phi)} = \partial_{a}\phi\partial_{b} \phi - g_{ab}\left(\frac{1}{2}\partial_{c}\phi\partial^{c}\phi - V(\phi)\right).
$$
The Einstein equations for this metric read:
\begin{equation} G_{ab} = R_{ab} - \frac{1}{2}Rg_{ab} = \kappa
\left(T_{ab}^{(m)} + T_{ab}^{(\phi)}\right), \label{eeqn}
\end{equation} where $\kappa = 8 \pi$.  The $tt$-component of
Eq. (\ref{eeqn}) gives:
\begin{eqnarray} 
\left(R_{,t}^2 R + k R\right)_{,r} = &&R_{,r}R^2\kappa
\varepsilon^{(m)} + R_{,r}R^2 \kappa \left(\frac{1}{2}\dot{\phi}^2 \right. \\ &&\left.+
\frac{Q}{2R_{,r}^2}\phi_{,r}^2 + V(\phi)\right) +
RR_{,r}R_{,t}\frac{\dot{Q}}{Q}, \nonumber\label{tteqn}
\end{eqnarray}
and the $tr$-component of the Einstein equations is:
\begin{equation}
\frac{R_{,r}}{R}\frac{\dot{Q}}{Q} = -\kappa \dot{\phi} \phi_{,r}. \label{treqn}
\end{equation}
The $rr$-component of Eq. (\ref{eeqn}) reads:
\begin{eqnarray}
\left(R_{,t}^2 R + k R\right)_{,t} &+& Q_{,t}R = \\ \nonumber &-&\kappa \left(\frac{1}{2}\dot{\phi}^2 + \frac{Q}{2R_{,r}^2}\phi_{,r}^2 - V(\phi)\right)R_{,t}R^2.\label{rreqn}
\end{eqnarray}
>From $T^{(m)\,a}{}_{b;a} = 0$, it follows that:
\begin{equation}
\kappa\varepsilon^{(m)} = \frac{F(r)\sqrt{Q(r,t)}}{R_{,r}R^{2}}, \label{mateqn}
\end{equation}
where $F(r)$ is an arbitrary constant of integration.  If we define $r$ so that $R(r,t_{i}) = r$ for some $t=t_{i}$, then $F(r) = 2M_{i,r}(r)/\sqrt{Q(r,t_i)}$ where $M_{i}(r)$ is the mass inside a shell of radius $r$ at $t=t_i$. 

>From $T^{(\phi)\,a}{}_{b;a} = 0$ one obtains:
\begin{equation}
-\square \phi = V_{,\phi}(\phi). \label{phieqn}
\end{equation}
Eq. (\ref{phieqn}) is subject to the boundary conditions: $\phi_{,r} = 0$ at $R=0$, and $\phi_{c}(t) = \lim_{r \rightarrow \infty} \phi(r,t)$, 
where $\phi_{c}(t)$ is the solution of Eq. (\ref{phieqn}) in the cosmological background. 
We make the following definitions:
\begin{eqnarray*}
2M(r,t) &=& \int^{r}_{r_0(t)}  F(r)\sqrt{Q(r,t)}\diff r, \\
\varepsilon_{c}^{(\phi)} &=& \frac{1}{2}\dot{\phi}_{c}^2 + V(\phi_c), \\
\delta\varepsilon^{(\phi)} &=& \frac{1}{2}\dot{\phi}^2 + \frac{Q}{2R_{,r}^2}\phi_{,r}^2 + V(\phi) - \varepsilon_{c}(t),\\
\delta \tilde{\varepsilon}^{(\phi)} &=& \delta\varepsilon^{(\phi)} - R_{,t}\dot{\phi}\frac{\phi_{,r}}{R_{,r}},
\end{eqnarray*}
where $r_0(t)$ is defined by $R(r_0(t),t) = 0$. We also define:
\begin{eqnarray*}
P_{c}^{(\phi)}(t) &=& \frac{1}{2}\dot{\phi}_{c}(t) - V(\phi_c), \\
\delta P^{(\phi)}(r,t) &=&  \frac{1}{2}\dot{\phi}^2 + \frac{Q}{2R_{,r}^2}\phi_{,r}^2 - V(\phi) - P_{c}^{(\phi)}(t).
\end{eqnarray*}
Integrating Eqs. (\ref{tteqn}) and (\ref{rreqn}) and using Eqs. (\ref{treqn}) and (\ref{phieqn}) we find that
\begin{eqnarray}
R_{,t}^2 &=& -k(r,t) + \frac{2M(r,t)}{R} + \frac{1}{3}R^2\kappa\varepsilon_{c}^{(\phi)}(t) \nonumber\\&+& \frac{\kappa}{R}\int_{r_0(t)}^{r} R_{,r}(z,t) R^2(z,t)\,
 \delta \tilde{\varepsilon}^{(\phi)}(z,t)\,\diff z, \label{Rteqn}
\end{eqnarray}
and
\begin{eqnarray}
&&R_{,tt}= -\frac{M(r,t)}{R^2} - \frac{1}{6}R\left(\kappa \varepsilon_c^{(\phi)} + 3\kappa P_c^{(\phi)}\right) -\\&& 
\left[\frac{\kappa}{2R^2}\int_{r_0(t)}^{r} R_{,r}(z,t) R^2(z,t)\, \delta \tilde{\varepsilon}^{(\phi)}(z,t)\,\diff z 
+ \frac{1}{2}R \kappa \delta P^{(\phi)}\right]\nonumber \label{Rtteqn}
\end{eqnarray}
As $R \rightarrow \infty$ we must recover the FRW background cosmology which implies:
\begin{eqnarray*}
\lim_{r \rightarrow \infty} M(r,t) &\sim & \frac{1}{2}\kappa \varepsilon_c^{(m)}(t_0) a^3(t_0)r^3, \\
\lim_{r \rightarrow \infty} k(r,t) &\sim & k_0r^2,
\end{eqnarray*}
where $t_0$ is an arbitrary time, $a(t)$ is the scale factor of the FRW background, and $k_0$ is the curvature of the background.  In line with current observations, and because it greatly simplifies the calculations, we take the cosmological background to be flat and set $k_0 = 0$.  We do not attempt to solve the Einstein or DDE equations exactly, but instead we develop asymptotic approximations to the true solutions which are accurate so long as a number of parameters, which we define and interpret below, remain small. 

\section{Small Parameters and the Matched Asymptotic Expansions.}\label{sec:small}

One can think of the small parameters approach and the matched asymptotic expansions 
as an expansion in the Newtonian potential. Although it is perhaps fairer to say that is closest 
to the context of General Relativity. The differences are not obvious at leading order but at next 
to leading order one will see that what appears 
in the equation for the acceleration is not necessarily what one would expect to appear in the 
Non-relativistic regime.

As stated above in Section \ref{sec:geo}, we require that gravity is `suitably weak' inside the density perturbation.  By suitably weak we mean that
$$
\vert \delta_1  \vert \ll 1,
$$
where
$$
\delta_1 = \delta v \equiv R_{,t} - HR.
$$
Physically $R_{,t}$ is the velocity of a particle of dust in
$\left\{t,R\right\}$ coordinates, where $R$ is the physical radial
coordinate, and $HR$ is the velocity that such a particle would have
in the cosmological background.  The small parameter $\delta_{1} =
\delta v$ is therefore the peculiar velocity of a matter particle with
respect to the cosmological background. Recall that we use units
where $c = 1$, hence the first expression above is in fact the
dimensionless quantity $\delta v / c \ll 1$. Our assumption requires
that whenever $HR \gtrsim \Oo(1)$, we have $\vert \delta_1/ HR \vert
\ll 1$.  This condition is equivalent to requiring that the mean
matter density contrast, $\bar{\delta}(R,t)$, inside the sphere with
radius $R \sim \Oo(1/H)$ or greater, is small enough that it may be
treated as a linear perturbation about the cosmological background.
The mean matter perturbation is however allowed to be non-linear,
$\bar{\delta} \gtrsim \Oo(1)$, when $HR \ll 1$ provided $\vert R_{,t}
\vert \ll 1$. We also define
$$
\delta_2 = \frac{1}{3} R^2 \kappa (\varepsilon^{(m)}-\varepsilon^{(m)}_c) \equiv \frac{1}{3}R^2 \varepsilon^{(m)}_c \delta_{m}.
$$
We require $\delta_{2} \ll 1$. This again require that $\delta_{m}
\gtrsim 1$ only when $HR \ll 1$ i.e. the matter perturbation is only
non-linear on sub-horizon scales.  We note that $\delta_2 \sim
\Oo((R_{,t}^2+k)-H^2R^2)$, and so generally $\vert \delta_1 \vert \ll
1$ provided $\delta_2 \ll 1$.

 In what follows we assume that $\delta_1$ and $\delta_2$ are small
\emph{everywhere}.  These assumptions can be checked once one
specifies initial conditions for the matter overdensity, but they
generally hold very well whenever the scale of the inhomogeneous
region is $< 0.1/H$.

 We now define two over-lapping regions which we shall refer to as the interior and the exterior. 

\begin{figure}
\begin{center}
\includegraphics[width=80mm]{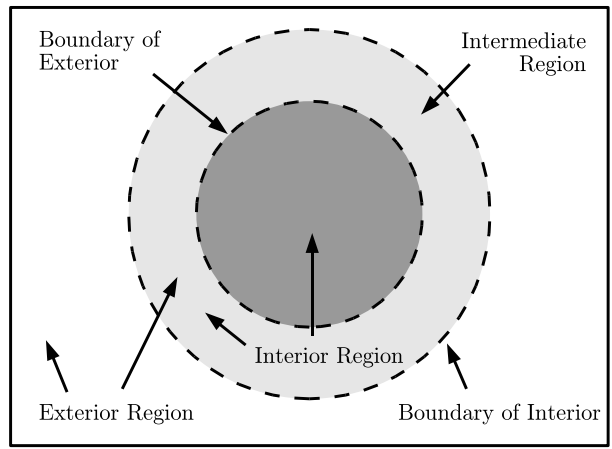}
  %% to include a figure, or
  %% to leave a blank space
 \caption{Sketch showing the interior, exterior and intermediate regions.  The interior is defined to be the region where the inner approximation is valid, and the exterior is the region where the outer approximation is applicable. In order for the matching procedure to work, we must require that the exterior and interior regions overlap in some intermediate region.  We must also require the exterior and interior regions are contiguous, so that one may move smoothly from the exterior, through the intermediate region, to the interior. }
 \label{figillu}
  \end{center} 
\end{figure}

\subsection{The Interior Region}
The \emph{interior region} is defined by
$$
\delta_3 \equiv HR \ll 1.
$$
We define the \emph{inner limit} of a quantity $F(r,t)$ to be $L_{int} (F) = \lim_{\delta_3 \rightarrow 0} F(r,t)$.  The inner limit can be imagined as the limit in which the cosmological background density of matter is taken to zero, and the cosmological horizon is taken to infinity.  In the interior we construct asymptotic approximations to quantities in the limit $HR \rightarrow 0$.  We also expand in $\delta_1$ and $\delta_2$.  We refer to this as the \emph{inner approximation}.
\subsection{The Exterior Region}
In the \emph{exterior region}, spacetime is required to be homogeneous and isotropic at leading order.  In other words, the matter perturbation must be linear in the exterior:
$$
\delta_4 \equiv  \vert \delta_1/\delta_3 \vert \ll 1, \delta_5 = \vert \delta_{m} \vert \ll 1.
$$  From the form of $R_{,t}$, it is clear that $\delta_5 \rightarrow 0$ implies $\delta_4 \rightarrow 0$. We define the \emph{exterior limit} of a quantity $F(r,t)$ to be $L_{ext} (F) = \lim_{\delta_5 \rightarrow 0} F(r,t)$. As in the interior limit we also expand in $\delta_1$ and $\delta_2$.   We construct asymptotic approximations to quantities in the exterior region in this exterior limit and refer to them as the \emph{outer approximations}.  

\subsection{Matching and the Intermediate Region}
We are primarily concerned with the behaviour of the DDE in the interior region. In Section \ref{sec:evo}, we find the inner approximation to $\phi$ by solving Eq. (\ref{phieqn}) order by order in the interior limit.  We cannot, however, apply both of the boundary conditions on $\phi$ directly to the inner approximation.  This is because the condition $\lim_{R \rightarrow \infty} \phi = \phi_c(t)$ must be applied at $R=\infty$, a point that is very clearly in the exterior and \emph{not} in the interior region.   As a result, the inner approximation will contain ambiguous
constants of integration. Fortunately, this ambiguity can be lifted by matching the inner and outer approximations to $\phi$ if there exists some \emph{intermediate region} where both approximations are simultaneously valid.  This matching of the inner approximation to the outer one is referred to as the method of matched asymptotic expansions (MAEs). It relies on the fact that, in any given region, the asymptotic expansion of a quantity is unique \citep[for a proof see][]{hinch}.  Thus if the inner and outer approximations of $\phi$ are both valid in the intermediate region, they must be equal in that region.  For the method of MAE to be applicable we must, of course, require that an intermediate region exists.  A necessary condition for an intermediate region to exist is that for some range of $R$:
$$
HR \ll 1, \quad \vert 3\delta_2 / \kappa \varepsilon_c^{(m)}(t) R^2 \vert = \vert 3\delta_m (t)/ \kappa \vert= \ll 1.
$$
This becomes a necessary and sufficient condition if the only boundary of the interior region is in the exterior one, and vice versa (see Figure \ref{figillu} for an illustration).   If the scale of the inhomogeneity, $R_{0}$, is taken to be the largest value of $R$ for which $\vert 3\delta_2 / \kappa \varepsilon_c^{(m)}(t) R^2 \vert > 0.3$, and then then an intermediate region generally exists provided that $R_{0} \lesssim 0.1 H$.  

In resume, in this section we have considered spherically symmetric
backgrounds and far from black hole horizons. We have basically
performed an expansion in the Newtonian potential at leading order, at
least as far as the evolution of scalar field perturbations are
concerned.  Notice, however, that this method can be straightforwardly
extended to allow for deviations from spherical symmetry \citep{shawbarrow,shawbarrowb,shawbarrowc}.

To be clear, we should point out that the expansion in $\delta_1$ and
$\delta_2$, which is a lot like expanding in the Newtonian potential,
is not really the key to our method.  That is just a
simplification.  The key is that we take: $\delta_3 = HR \ll 1$ in the
interior region and $\delta_m \ll 1$ in the exterior region. That is,
everything is linear in those two regions.  We then assume that there
is an intermediate region where both of these conditions hold. Which
implies that in the intermediate region: $\vert \delta_1 \vert \ll 1$
and $\vert \delta_2 \vert \ll 1$.  So the assumption that $\delta_1,
\delta_2 \ll 1$ everywhere ensures that we have an intermediate
region.    We could in fact
relax this but the calculations would become more difficult: often
still doable, but in almost all physically interesting cases totally
unnecessary. For instance the relaxation to $|\delta_1 | < 1$ and
$|\delta_2| \ll 1$, is very straightforward and allows one to go all
the way up to a black hole horizon \citep{shawbarrow,shawbarrowb,shawbarrowc}.

\section{Evolution of Dynamical Dark Energy Perturbations}\label{sec:evo}
In this section we find an asymptotic approximation to the DDE density contrast, $\delta_{\phi}$.  The discussion is fairly technical, and readers more 
interested in the results than the machinery used to derive them may prefer to focus on the statement, 
discussion and application of our results in Section \ref{sec:evo:dis} and following. 

The DDE is described by the field $\phi$ which satisfies:
$$
-\square \phi = V_{,\phi}(\phi).
$$
We solve this equation by constructing an asymptotic approximation to $\phi$ in the small parameter $\delta_{1}$.  We write:
$$
\phi \sim \phi_{c}(t) + \delta \phi(t,r)(1 + \mathcal{O}(\delta_{1})),
$$
where $\delta \phi \sim \Oo(\delta_{1})$.  Before solving for $\phi$, we make the dependence of the metric on $\delta_{1}$ explicit by transforming to a new radial coordinate, $\rho = R(t,r)/a(t)$,  where $a(t)$ is the scale factor of the Friedmann-Robertson-Walker (FRW) cosmological background.  In $\left\lbrace t, \rho \right\rbrace$ coordinates the metric is:
\begin{eqnarray}
\diff s^2 &=& \frac{\left(1-k(r,t)-\delta_1^2(r,t)\right)\diff t^2}{1-k(r,t)} + 2\frac{\delta_{1}(r,t)a(t) \diff t \diff \rho}{1-k(r,t)} -\nonumber\\ &&\frac{a^2(t)\diff\rho^2}{1-k(r,t)} - a^2 \rho^2 \left\lbrace \diff
 \theta^2 + \sin^2 \theta \diff \varphi\right\rbrace.
\end{eqnarray}
Consider $-\square \phi_{c}(t)$ in this metric:
$$
-\square \phi_{c}(t) = V_{,\phi}(\phi_c) + \frac{1}{2}\frac{\dot{Q}}{Q} \dot{\phi}_c(t)  -
 \frac{\dot{\phi}_c(t)}{a(t) \rho^2} \frac{\partial}{\partial \rho}\left(\rho^2\delta_1\right).
$$
where $\dot{Q}$ is evaluated at constant $r$ as opposed to constant $\rho$ or $R$ and is thus given by Eq. (\ref{treqn}). It follows that $\delta \phi$ satisfies:
\begin{eqnarray}
-&\square& \delta\phi \left(1+\mathcal{O}(\delta_{1})\right) = \left(V_{,\phi}(\phi)-V_{,\phi}(\phi_{c})\right)\\\nonumber
 &+& \frac{1}{2}\kappa \dot{\phi}_{c}^2(t) \rho\delta\phi_{,\rho} + \frac{\dot{\phi}_{c}(t)}{a(t) \rho^2} \left(\rho^2 \delta_{1}\right)_{,\rho}.\label{phieqn1}
\end{eqnarray}
We now solve for $\delta \phi$ in both the interior and exterior regions.
\subsection{Inner Approximation}
We note that $\frac{1}{2}\kappa \dot{\phi}_c^2 R^2 \lesssim \mathcal{O}(\delta_{3}^2)$, and that $\vert k
 \vert \sim \mathcal{O}(\delta_{3},\delta_{1}^2,\delta_{2})$, so that
\begin{eqnarray}
-\frac{\rho^2}{a}\partial_{t}\left(a^3(t)\delta\phi_{,t}\right) + \left(\rho^2 \delta \phi_{,\rho}\right)_{,\rho} 
\sim \left(V_{,\phi}(\phi)-V_{,\phi}(\phi_c)\right)a^2\rho^2 \nonumber  \\ \nonumber+ \dot{\phi}_{c}(t)a\left(\rho^2 \delta_{1}\right)_{,\rho} 
+\mathcal{O}(\delta_{1}\delta \phi, \delta_2\delta \phi, \delta_3^2 \delta \phi). \label{inteqn}
\end{eqnarray}
In many cases  $\delta \phi$ is small enough so that $$(V_{,\phi}(\phi)-V_{,\phi}(\phi_c))a^2\rho^2 \sim V_{,\phi \phi}(\phi_c)R^2 \delta \phi,$$ and $V_{,\phi \phi}\sim \mathcal{O}(H^2)$ so that $\phi_{c}(t)$ evolves over cosmological time-scales. In these cases it is clear that:
$$
(V_{,\phi}(\phi)-V_{,\phi}(\phi_c))a^2\rho^2 \sim \mathcal{O}(\delta_{3}^2 \delta \phi),
$$  
and so this term can be dropped at leading order.  Although it will often be the case that $V_{,\phi}(\phi)-V_{,\phi}(\phi_c) \approx V_{,\phi \phi}(\phi_c)\delta \phi$ we do not have to require such a strong assumption in order to justify ignoring the effect of the potential to leading order in the interior approximation. Instead, we make the far weaker assumption that:
\begin{equation}
\vert V_{,\phi}(\phi_c + \delta \phi)-V_{,\phi}(\phi_c)\vert R^2 / \delta \phi \sim \mathcal{O}(\delta_{1}, \delta_{2}, \delta_{3}). \label{Vassump}
\end{equation}
A physically-viable DDE theory for which this requirement does not hold would be difficult to construct. For instance, if this condition did not hold then even if, at one instant, $V_{,\phi}(\phi_c) \sim \mathcal{O}(H^2)$ or smaller, as must be required for $\phi_c$ to evolve over cosmological time scales, shortly afterwards small changes in $\phi_{c}$ would cause $V_{,\phi}(\phi_c)$ to grow to be many orders of magnitude greater than $H^2$. This would result in $\phi_{c}$ evolving over time-scales that are much shorter than the Hubble time.  The only DDE theories that we can imagine that could accommodate this and still be compatible with observations would involve $\phi$ being held almost completely fixed at a minimum of $V(\phi_c)$, in which case the DDE would be almost entirely indistinguishable from a cosmological constant.  

To leading order in the interior, we would actually be justified in setting $a(t)={\rm const}$ as $\dot{a} \rho = HR = \delta_{1} \ll 1$. However, since it might not be completely clear that these terms may be ignored, and because we can solve the $\delta \phi$ equation without having to ignore them, we continue to include them. 

We define $\diff \eta = \diff t / a(t)$ and $u = a\delta \phi$, and note that $\rho^2 a_{,\eta\eta}/a = \dot{H}R^2 + H^2 R^2 \sim \mathcal{O}(\delta_{3})$, so that:
\begin{eqnarray}
&&\left(-u_{,\eta \eta} + \frac{1}{\rho^2}\left(\rho^2 u_{,\rho}\right)_{,\rho} \right)\left(1+ \mathcal{O}(\delta_{1}, 
\delta_{2}, \delta_{3})\right)\sim\nonumber \\  &&\dot{\phi}_{c}(t) \frac{1}{\rho^2}\left(a(t)\rho^2 \delta_1\right)_{,\rho}.
\end{eqnarray}
which has as a  solution
\begin{eqnarray}
&\delta &\phi(t, \rho) \sim   -\frac{1}{2}\int_{0}^{\infty} \,\rho^{\prime\, 2}\,\diff \rho^{\prime}\, \int_{-1}^{1} 
\diff s\,\frac{f(\eta(t)-X, \rho^{\prime}) }{a(t) X(\rho,\rho^{\prime},s)} + \nonumber \\ && \frac{{\cal F}(\eta-\rho)}{a(t)\rho} 
+ \frac{{\cal G}(\eta+\rho)}{a(t)\rho} + \mathcal{O}(\delta_{1}\delta \phi, \delta_{2}\delta \phi, \delta_{3}\delta \phi),  
\label{intsol}
\end{eqnarray}
where $f(\eta,\rho) = \frac{\dot{\phi}_c(t) a(t)}{\rho^2} \left(\rho^2 \delta_{1}(t,\rho)\right)_{,\rho}$
and $X(\rho,\rho^{\prime},s) = \sqrt{ \rho^2 + \rho^{\prime\,2} - 2\rho \rho^{\prime} s}$; ${\cal F}$ and ${\cal G}$ represent waves in the scalar field and we discuss them further momentarily.  To go further, we make the following reasonable assumptions about the behaviour of $\delta_{1}$.  We assume that there exists some $R_{-2}$, in the interior region, such that for $R > R_{-2}$ in the interior region, $\delta_{1}$ decreases faster than $1/R^2$ as $R \rightarrow \infty$; $f$ then decreases faster than $1/R^{3}$ for $R > R_{-2}$. This assumption ensures that dominant contributions to the integral in Eq. (\ref{intsol}) comes from values of $\rho^{\prime}$ that are well inside the interior region.   Note that generally $f(\eta, \rho)$ varies over conformal times-scales of order $1/aH$ and $\delta_1/a\dot{\delta}_1$. Provided that $\delta_1$ is not momentarily zero, these scales are much larger than $X \lesssim \sqrt{2}R_{-2}/a$ so we can Taylor expand $f(\eta-X,\rho^{\prime})$ as
$$
f(\eta-X,\rho^{\prime}) = f(\eta, \rho^{\prime}) + f_{,\eta}(\eta,\rho^{\prime}) X + \mathcal{O}(f_{,\eta \eta}X^2),
$$
where the neglected terms are $\mathcal{O}(\delta_{1}^2, \delta_{3}^2, \delta_1\delta_{3},\delta_{2})$ times $f(\eta, \rho^{\prime})$.  After some algebra and integration we then arrive at:
\begin{eqnarray}
&\delta \phi(t,R)\left(1+\mathcal{O}\left(\delta_1, \delta_2, \delta_{3}\right)\right) \sim \dot{\phi}_{c}(t)
 \int^{R}_{\infty} \delta_{1}(t, R^{\prime})\,\diff R^{\prime} \nonumber& \\ &+ \frac{{\cal F}(\eta- \rho) + {\cal G}(\eta + \rho)}{R}.&
\end{eqnarray}
where the previous assumption relating to the Taylor expansion of $f$ can be seen to be equivalent to the statement that $\ddot{\delta}_{1} (t,R)R^2 \ll \delta_{1}$.   This assumption will generally break down if there is some initial instant, $t=t_i$ say, when $\delta_1 = 0$.  At $t \gg t_i$ however, $\ddot{\delta}_1 R^2 \ll \delta_1$ is equivalent to $\delta_{1}^2,\, \delta_{3}^2,\, \delta_{2} \ll 1$, which is certainly the case in the interior region.   

Generally, the functions ${\cal F}$ and ${\cal G}$ are related to the initial conditions on $\delta \phi$.  Because we have required $R_{-2} \ll H^{-1}$ our analysis will break down at early times since generally $HR_{-2} \rightarrow \infty$ as $t \rightarrow 0$. As a result, we cannot generally determine ${\cal F}$ and ${\cal G}$ by simply specifying  some initial conditions on$\delta \phi$, at $t=t_i$ say, and applying them to the interior approximation since the interior approximation by not be valid when $t=t_i$. The requirement that $\delta \phi(R=0)$ can however be applied to the interior approximation to give ${\cal F}(\eta) = - {\cal G}(\eta)$.  To the order at which we work, ${\cal F}(\eta)$ itself should be fixed by matching to the outer approximation.  
\subsection{Exterior Solution and Matching}
In order for $\phi_{c} = \lim_{R \rightarrow \infty} \phi(t,R)$ we must have $\delta \phi(t,R) \rightarrow 0$ as $R \rightarrow \infty$.   In the exterior region, spacetime is FRW to leading order in $\delta_1$ and $\delta_2$, and the equation describing for the leading order behaviour of $u = a(t)\delta \phi$ reads:
\begin{eqnarray}
-u_{,\eta \eta} &+& \frac{1}{\rho^2}\left(\rho^2 u_{,\rho}\right)= \left(V_{,\phi\phi}(\phi_c) - \dot{H} - H^2\right) a^2 u +\nonumber \\ &&
 \frac{1}{2}\kappa \dot{\phi}_{c}^2(t) a^2 \rho u_{,\rho} + \frac{\dot{\phi}_{c}(t)a^2}{\rho^2} \left(\rho^2 \delta_{1}\right)_{,\rho}.\label{phieqnext}
\end{eqnarray}
where $\diff\eta = \diff t / a(t)$.   Solving this equation is far from straightforward.  Fortunately, however, our main interest in not in how $\delta \phi$ behaves in the exterior region but in interior region where the perturbation in the matter density can be non-linear, and so it is not necessary to solve Eq. (\ref{phieqnext}) in the exterior region so long as we know how its solutions behave in some intermediate region where $\delta_{3} = HR \ll 1$.  We require that $\delta \phi \rightarrow 0$ as the matter perturbation is removed i.e. $\delta_1\rightarrow 0$. We have also required that $\delta_1$ drop off faster than $1/R$ for $R > R_{-2}$ where $R_{-2}$ is in the interior region i.e. $HR_{-2} \ll 1$. In this intermediate region, the solutions of Eq. (\ref{phieqnext}) for which $\delta \phi \rightarrow 0$ as $R \rightarrow \infty$ have the following form:
\begin{eqnarray}
\delta \phi(t,R) &\sim& \dot{\phi}_{c}(t) \int_{C(\eta)}^{R} \delta_1(t,R^{\prime})\diff R^{\prime} + \frac{K(\eta)}{R} +\nonumber \\ &&
 \mathcal{O}(\delta_1 \delta \phi, \delta_3 \delta \phi, \delta_2 \delta \phi),
\end{eqnarray}
where $C(\eta) \sim \mathcal{O}(1/H)$ or greater. We have assumed that $t$ is larger compared to any initial instant when one one or more of $\delta_1 = 0$, $\delta \phi=0$ or $\delta \dot{\phi} =0$ hold. With these assumptions, matching to the interior region gives:
$$
K(\eta) = {\cal F}(\eta) + {\cal G}(\eta) = 0,
$$
and
$$
\frac{2{\cal F}_{,\eta}(\eta)}{a(\eta)} = - \dot{\phi}_{c}(t) \int_{C}^{\infty} \delta_1(t,R^{\prime})\diff R^{\prime}. 
$$
We have required drop off of $\delta_1$ means that, to leading order, we can set $C=\infty$.  This implies that the $2{\cal F}_{,\eta}/a(\eta)$ is sub-leading order in the interior approximation and so may be neglected.  

Thus to leading order in the interior region we find:
%\begin{equation}
$$
\delta \phi(t,R) \sim \dot{\phi}_{c}(t) \int_{\infty}^{R} \delta v(R^{\prime},t)\diff R^{\prime}\left(1+ \mathcal{O}(\delta_1, \delta_2, \delta_3)\right). \label{deltaphieqn}
$$
%\end{equation}
Where $\delta v= R_{,t}-HR$ is the peculiar radial velocity of the matter particles relative to the expansion of the background Universe.  This approximation is valid in the interior region at late times compared to any initial instance when $\delta \phi = 0$ and/or $\delta \dot{\phi} = 0$ which is equivalent to $\vert \delta \ddot{v} \vert R^2 \ll \vert \delta \dot{v}\vert R  \ll \vert \delta v \vert $.  Note that, to this order, the surfaces of constant $\phi$ are surfaces of constant:
$$
t_{\phi} = t - \int^{\infty}_{R} \delta v \diff R^{\prime}.
$$
We use this asymptotic approximation for $\delta \phi$ to evaluate the DDE density contrast on surfaces of constant $t$ in the matter rest frame. In the rest frame of the matter particles:
$$
\varepsilon^{(\phi)} = \frac{1}{2}\dot{\phi}^2 + \frac{Q}{2}\phi_{,R}^2 + V(\phi),
$$
and the local inhomogeneity in $\varepsilon^{(\phi)}$ is therefore:
\begin{eqnarray}
\delta \varepsilon^{(\phi)} &\sim& \dot{\phi}_{c}^2(t)\left(-3H\int_{\infty}^{R}\delta v(R^{\prime},t)\diff R^{\prime} + (\delta v(R,t))^2 \right.\nonumber \\ &+& \left.\int_{\infty}^{R}\left(R_{,tt} - (\dot{H}+H^2)R^{\prime}\right) \diff R^{\prime}\right).
\end{eqnarray}
The corrections terms are $\mathcal{O}(\delta_1, \delta_2, \delta_3)$ times smaller than the leading order term.  Since $\dot{\phi}_{c}^2 = (1+w)\varepsilon^{(\phi)}_{c}$ we have
\begin{eqnarray*}
\delta_{\phi} \sim &&(1+w) \left(-3H\int_{\infty}^{R}\delta v(R^{\prime},t)\diff R^{\prime} + (\delta v(R,t))^2\right. \\ &&\left.+ \int_{\infty}^{R}\left(R_{,tt} 
- (\dot{H}+H^2)R^{\prime}\right) \diff R^{\prime}\right),
\end{eqnarray*}
$R_{,tt}$ is the acceleration of the matter particles in $(t,R)$ coordinates. Eq. (\ref{Rtteqn}) gives:
$$
R_{,tt}R - \dot{H}R^2 - H^2 R^2 \sim -\frac{\delta M}{R} + \mathcal{O}(\delta_3^3\delta_1,\,\delta_1^2\delta_3^2,\,\delta_2\delta_3^2).
$$
where $\delta M(R,t) = M(R,t) - \frac{1}{2}\Omega_m H^2 R^3 = \frac{1}{2}\Omega_m H^2 R^3 \bar{\delta}$ is the mass contrast at time $t$ inside the sphere with physical radius $R$.  

\subsection{Discussion}\label{sec:evo:dis}
In this section we found that the DDE density contrast can be expressed, to leading order, in terms of the radial peculiar velocity, $\delta v$ , the mean density contrast, $\bar{\delta}$ and the unperturbed DDE equation of state parameter $w$:
\begin{eqnarray}
\delta_{\phi} \sim &&(1+w)\left(3 H\int_{R}^{\infty} \delta v(R^{\prime},t) \diff R^{\prime} + (\delta v)^2\right.\nonumber\\ &&\left. + \frac{1}{2}\Omega_m H^2 \int_{R}^{\infty} \bar{\delta}(R^{\prime},t) R^{\prime} \diff R^{\prime}\right). \label{deltaphimain}
\end{eqnarray}
This expression for $\delta_{\phi}$ is our main result and it is valid for both linear and non-linear sub-horizon matter perturbations.  In order to evaluate the above expression, one must simply specify the peculiar velocity, $\delta v$, of the matter particles and the matter density contrast $\bar{\delta}$.  This exterior is valid in the interior region at late times compared to any initial instance when $\delta \phi = 0$ and/or $\delta \dot{\phi} = 0$ i.e. $\vert \delta \ddot{v} \vert R^2 \ll \vert \delta \dot{v}\vert R  \ll \vert \delta v \vert $.

Note that $H$ and $w$ will depend on the mass of the field $m$, so in
that sense, equation (\ref{deltaphimain}) is not completely model
independent. However, all the model dependent terms come from the
homogeneous and isotropic background cosmology.  So the expression for
$\delta_{\phi}$ is model independent provided $m R_c \ll 1$. Which is
true if $m \sim O(H)$ and $HR_c \ll 1$ as assumed.

The key point is that large deviations in $\delta_{\phi}$ from the
prediction using a LCDM background could only occur if $\delta v$ and
$\delta_m$ were predicted to change by an order of magnitude or more.
As far as we know, however, structure formation is compatible with
LCDM over the scales of clusters and so it seems unlikely that such
large deviations occur.

As we shall see below the DDE perturbation is generally small compared to the matter perturbation, 
so to leading order one may neglect any DDE perturbations when evaluating $\bar{\delta}$ and $\delta v$. 

It is interesting to note that the sign of $\delta_{\phi}$ depends on
the relative magnitudes of $\bar{\delta}$ and $\delta v$: From
equation (\ref{deltaphimain}) one deduces that the formation of local
overdensities or local voids of dark energy arise from two competing
effects: \begin{itemize} \item The Drag effect: associated with
deviations from $\delta v/HR = 0$.  If one has an overdense region,
then it expands more slowly. In agreement with \cite{dutta}, we find
this drag effect to produce a local underdensity of dark energy.
\item The Pull effect: associated with deviations from $\delta_m = 0$.
An overdensity of matter pulls the dark energy (and everything else)
towards it. So matter and fields tend to clump around it.  As a
result, the pull effect means that an overdensity of matter creates an
overdensity of DE.  \end{itemize} So there is one effect that pulls
the dark energy in and another that pushes it out.  As we will see, in
the linear regime the drag effect grows more quickly than the pull
effect, so one gets a dark energy void.  The onset of non-linear
structure formation, however, causes $|\delta v/HR|$ to reach a
maximum and then to tend to $1$ at late times, whereas $\delta_m$ just
keeps on growing. So the pull effect dominates at late times, so one
ends up with an overdensity of dark energy.

We have assumed above that the mass of the DDE scalar field is small compared to the inverse length scale of the cluster.  Provided this is 
$\delta_{\phi}$ in uncoupled DDE theories is \emph{independent} of the details of theory describing the dark energy. 
%Generally, the larger the mass of $\phi$ the more DDE clustering is suppressed.

 In the next section we use Eq. (\ref{deltaphimain}) to evaluate $\delta_{\phi}$ in both the linear and non-linear regimes. We find that $\vert \delta_{\phi}(R=0) \vert \sim \Oo(\Omega_m H^2 R_{c}^2 \bar{\delta}_{clust})$, where $R_{c}$ is the radial scale of the cluster, and $\bar{\delta}_{clust}$ is the mean matter density contrast in  $R \leq R_{c}$.
\section{Clustering of Dynamical Dark Energy}
\label{sec:DDEclust}
\subsection{Dynamical Dark Energy  Clustering in the Linear Regime}
Although one of our main aim in this work was to study DDE clustering
in the \emph{non-linear} regime, our analysis is also perfectly valid
in the linear regime, i.e. when $\delta_{m} \ll 1$, provided that the
matter inhomogeneity is sub-horizon at the instant when we wish to
evaluate $\delta_{\phi}$.  We found above that $\delta_{\phi}$ is
given by Eq. (\ref{deltaphimain}). In the linear regime $\vert \delta
v \vert \ll HR$, and so the term proportional $\delta v^2$ in
Eq. (\ref{deltaphimain}) is small compared to the other two terms and
should be neglected.  $\delta_{m}(r,t)$ is the matter density
contrast, and we define $\delta_{k}(t)$ to be its Fourier
transform. In the linear regime (in comoving-coordinates):
$$
\dot{\delta}_{m} = -\mathbf{\nabla}\cdot \delta \mathbf{v},
$$
and so:
$$
\delta v_{k} = -\frac{i H f(a) a \delta_{k}}{k},
$$
where $f(a) = \diff \ln \delta_{k}/\diff \ln a$ and $\delta v_{k}$ is the Fourier transform of $\delta v$. We recognize that $3\Omega_m H^2 \bar{\delta} R /2= \delta M / R^2 = \partial \delta \Phi/\partial R$ where $\delta \Phi$ is the leading order perturbation to the Newtonian potential.  $\delta \Phi$ is given by:
$$
\nabla^2 \delta \Phi = 4\pi G \delta_{m}.
$$
The Fourier transform of $\bar{\delta}(R,t)R$ is therefore: $i a \delta_{k}/k$, and the Fourier transform of the DDE density contrast, $\delta_{\phi}$, in the linear regime is given by:
\begin{equation}
\delta_{\phi\,k}^{({\rm lin})} \sim -\frac{3}{2}(1+w) g(a)\frac{a^2 \Omega_m H^2 \delta_{k}}{k^2}. \label{linmtm}
\end{equation}
where
\begin{equation}
g(a) = \frac{2f(a)}{\Omega_m} - 1. 
\end{equation}

\begin{figure}
\begin{center}
\includegraphics[width=80mm]{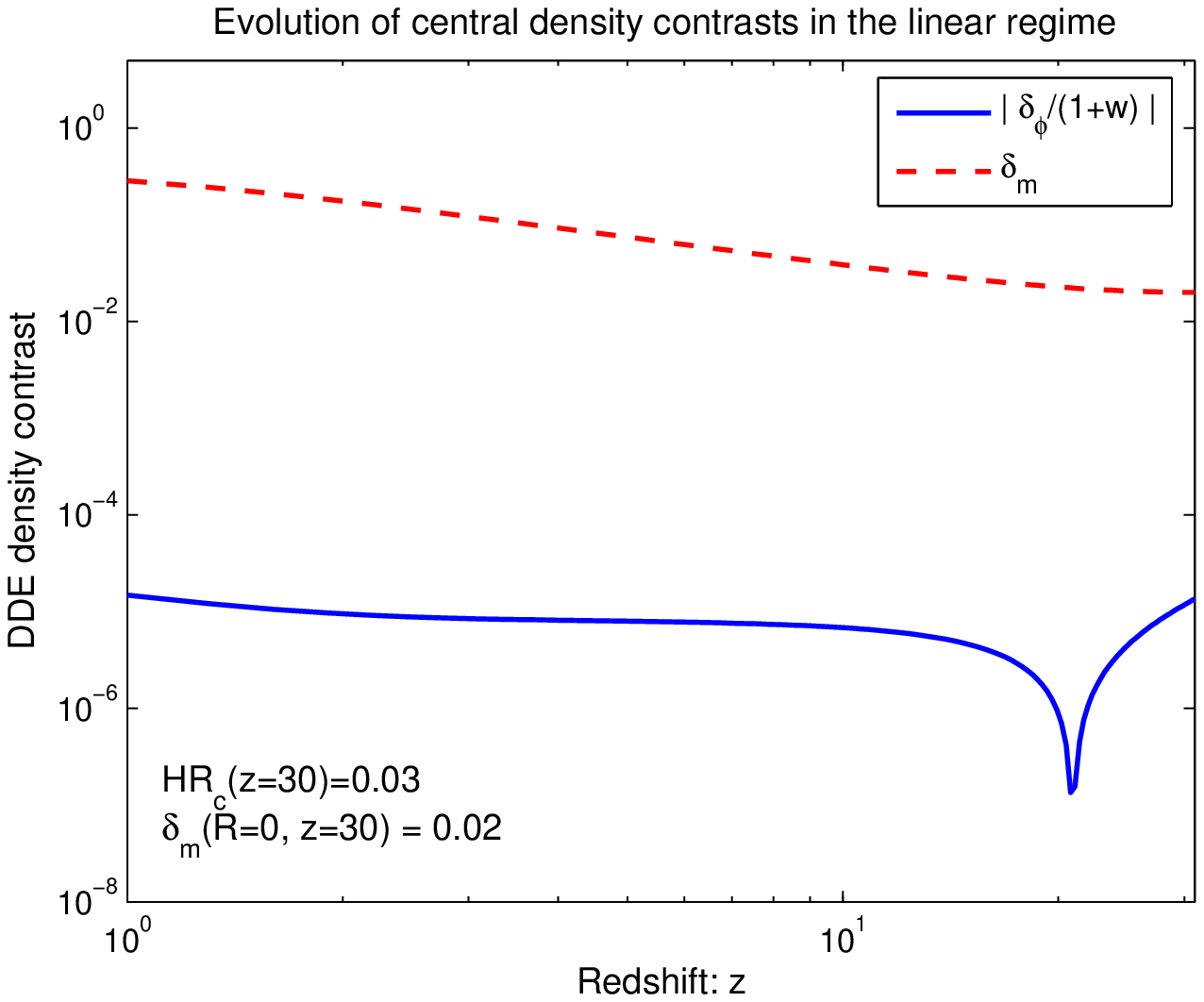}
\includegraphics[width=80mm]{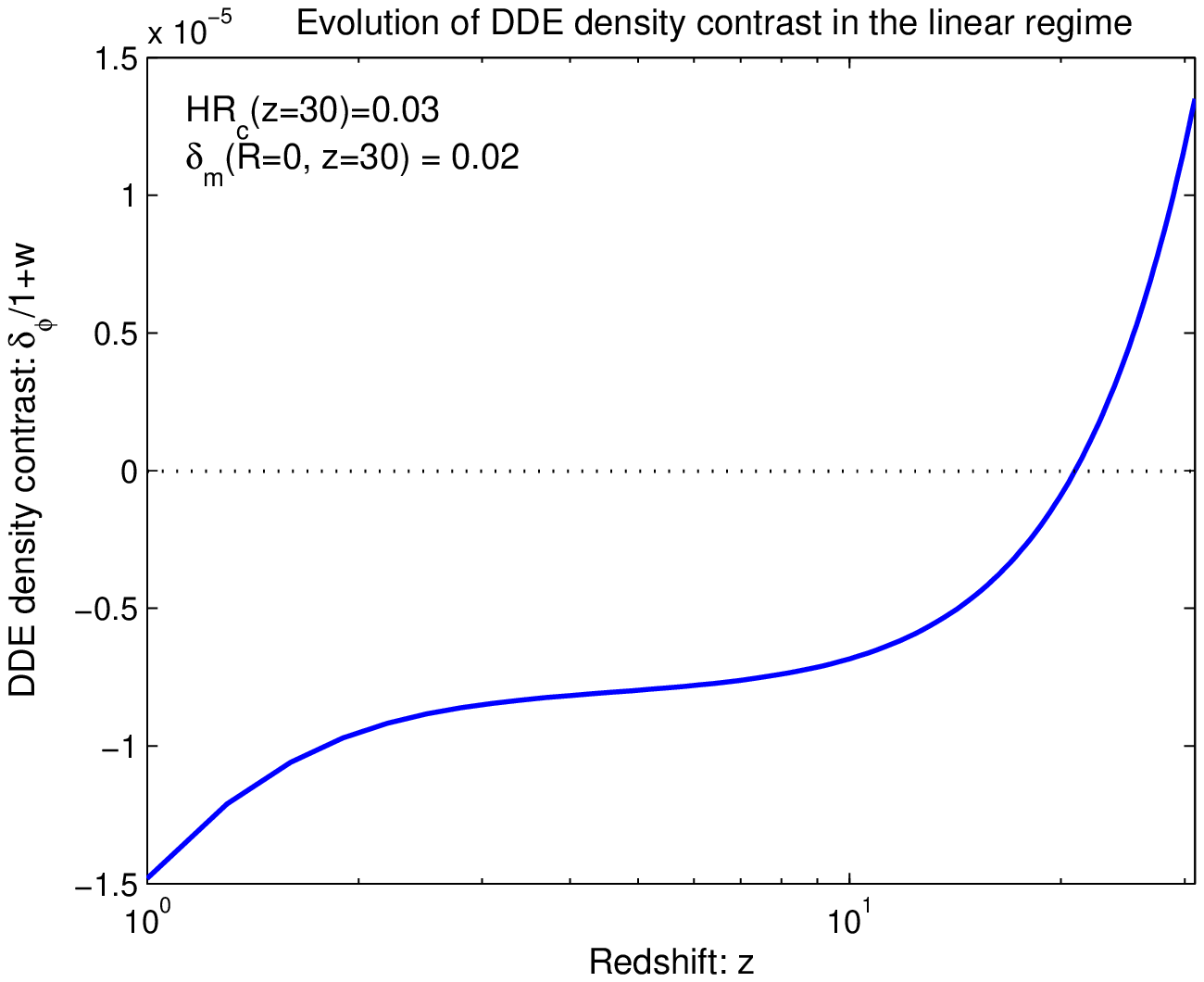}
  %% to include a figure, or
  %% to leave a blank space
  \end{center}
 \caption{Evolution of the DDE and matter density contrasts in the linear regime for an overdensity that begins to collapse at $z=z_{i}=30$. We have taken $\Omega_{m}(z=0) = 0.27$.  Initially the matter density profile is taken to be Gaussian with $\delta_{m}(z_{i},r=0) = 0.02$ and $H_{i} r_{c} = 0.03$. This corresponds $\delta_{m}(r=0) = 0.3$ and $R_{c} = a r_{c} \approx 31 h^{-1}\mathrm{Mpc}$ today.  $\vert \delta_{\phi} \vert$ is always small compared to $\delta_{m}$. At early times, the DDE density contrast is positive however it must be noted that our approximation is only valid in this case for $z \lesssim 29$. At late times it is negative.  We can see that $\vert \delta_{\phi}/(1+w) \vert$ is growing at late times in the linear regime; indeed it is in fact growing faster than $\delta_m$ for $z < 0.23$.}
 \label{figlin}
 \end{figure}

Motivated by observations, we assume $(1+w)$ to be small at late times, and $(1-\Omega_m)$ to be small at early times.  
To a first approximation then we approximate $f(a)$ by its $\Lambda$CDM value.  At late times a very good approximation to 
$f(a)$ in the $\Lambda$CDM model is $\Omega^{0.6}_m$ \citep{Peebles1980, Lahav1991}. Note that corrections to this fitting formulae occur for quintessence-like models \citep{wangstei}. However, these are generally too small to greatly affect our results. Hence we take the $\Lambda$CDM  expression for simplicity. Even when it is not acceptable to approximate $f(a)$ by its $\Lambda$CDM, we do not expect this approximation to greatly alter the qualitative nature of our results or the order of magnitude of $\delta_{\phi}$.

Transforming back into real space we have:
\begin{equation}
\delta_{\phi}^{({\rm lin})} \sim -\frac{(1+w)g(a)\Omega_{m}a^2 H^2}{2} \int_{r}^{\infty} r^{\prime} \bar{\delta}(r^{\prime},t)\diff r^{\prime}, \label{linreal}
\end{equation}
Our expression for $\delta_{\phi\,k}^{({\rm lin})}$ is independent of the mass of $\phi$. This is because we have assumed that $m_{\phi} \sim \Oo(H)$ in the linear regime and that the matter perturbation is small compared with $H^{-1}$.  The mass of the scalar field therefore has  only a negligible effect on the scales over DDE clustering occurs.   It is clear that the sign of $\delta_{\phi\,k}$ is the same as the sign of $-g(a)\delta_k$. At late times $g(a) >0$, and so a linear local overdensity of matter ($0 < \delta_m \ll 1$) corresponds to a DDE void.  Similarly, there is a DDE overdensity at late times (in the linear regime) if there is a local void of matter. If an initial mean density perturbation $\delta_{k}(z_i)$ begins to collapse at $z=z_i$ when $a=a_i=1$, $\Omega_m = \Omega_{mi}$ and $H_i$ we have:
\begin{eqnarray}
\delta_{k} &\approx&  \frac{1}{3\Omega_{mi}^{0.4}+2} \\ \nonumber &&\left(3\Omega_{mi}^{0.4}\left(\frac{\Omega_m}{\Omega_{mi}}\right)^{0.2} a + 2\left(\frac{\Omega_{mi}}{\Omega_m}\right)^{0.5} a^{-3/2}\right)\delta_{k}(z_i),
\end{eqnarray}
and so
\begin{eqnarray*}
\delta_{\phi\,k}^{({\rm lin})} \sim &&\frac{9(1+w)\Omega_{mi} H^2_{i}}{6\Omega_{mi}^{0.4}+4} \left(\Omega_{mi}^{0.4}(2\Omega_{m}^{-0.4}-1)\left(\frac{\Omega_m}{\Omega_{mi}}\right)^{0.2}\right. \\ &&\left.- \frac{8}{3}\left(\frac{\Omega_{mi}}{\Omega_m}\right)^{0.5}\left(\frac{1+z}{1+z_i}\right)^{5/2}\right)\frac{\delta_{k}(z_i)}{k^2}.
\end{eqnarray*}
Our expression for $\delta_{\phi}$ was derived under the assumption that $\vert \delta \ddot{v} \vert R^2 \ll \vert \delta \dot{v}\vert R  \ll \vert \delta v \vert $.  As result it will break down as we approach $z=z_i$. More precisely, we find that our approximation is valid at $R=0$ for $a \gtrsim a_{i}(1+2H_{i}R_{-2}(z_i))$   i.e. $z \lesssim z_{\rm app} \equiv z_{i}(1-2H_{i}R_{-2}(z_{i}))$ where $H_{i}$ is the initial value of $H$ and $R_{-2}(z_i)$ is the initial value of $R_{-2}$. 
>From this expression we can see that a matter \emph{overdensity} corresponds initially to a DDE \emph{overdensity} ($\delta_{\phi}^{({\rm lin})} >0$), however this DDE overdensity becomes a DDE void at some $z=z_{crit}$, and $\delta_{\phi}^{({\rm lin})} < 0$ for $z < z_{\rm crit}$.  We see that $z_{\rm crit}$ is given by:
$$
\frac{1+z_{\rm crit}}{1+z_i} \approx \Omega_{m}^{-0.6}(z_{\rm crit})\left(\frac{6 - 3\Omega_{m}^{0.4}(z_{\rm crit})}{8}\right)^{2/5},
$$
where we have taken $\Omega_{mi} \approx 1$.  If $\Omega_m(z_{\rm crit}) =1$, which should be a good approximation for $1.8 \lesssim z \lesssim 3000$, we have $z_{\rm crit}=(3/8)^{2/5}(1+z_i)-1$. For this analysis to be valid we must have $z_{\rm crit} < z_{\rm app}$ which with $\Omega_{m}(z_{\rm crit}) \approx 1$ holds $H_{i}R_{-2}(z_{i}) \lesssim 0.24$. For inhomogeneities that are larger than this at $z=z_{i}$ we do \emph{not} expect $z_{\rm crit}$ to be an accurate approximation to the redshift when the DDE density contrast changes sign.

Note that $z_{\rm crit}$ does not depend on the size of the inhomogeneity, although, of course, we must have $z_{\rm crit} \lesssim z_{\rm app}$ for our analysis to be valid, which implies that at $z=z{\rm crit}$, $HR_{-2} \lesssim 0.2$. 

If $\Omega_m  = 1$ exactly then $\delta_{\phi}^{({\rm lin})}/(1+w) \rightarrow \mathrm{const}$ at late times. If $\Omega_m < 1$, then $\vert \delta_\phi / (1+w) \vert$ grows likes:
$$
\left\vert \frac{\delta_{\phi}}{1+w} \right\vert \propto 2\Omega_{m}^{-0.2} - \Omega^{0.2}_{m}, 
$$
at late times in the linear regime.  This implies that $\vert \delta_{\phi}/(1+w) \vert$ tends to a constant in the linear regime if $\Omega_m = 1$, and is growing for $\Omega_m < 1$.  At late times in the linear regime then: $\vert \delta_{\phi}/\delta_{m}\vert\propto \Omega^{-0.4}_{m}a^{-1}$.  If we take $\Omega_{m} = 0.27$ today then $\vert \delta_{\phi}^{({\rm lin})} /(1+w)\vert$ is predicted to grow faster than $\delta_{m}$ when $z = 0.23$.  \citet{dutta} also found that at late times in the linear regime $\vert \delta_{\phi}/(1+w)\vert $ grows faster than $\delta_{m}$.  This is behaviour were to continue into the non-linear regime it would, of course, imply that $\vert\delta_{\phi}/(1+w)\vert$ would ultimately grow to be very large and indeed dominate over $\delta_{m}$.  As we show in this paper, however, before this can happen the matter perturbation goes non-linear and this slows the growth of $\vert \delta/(1+w)  \vert$. 

We conclude our analysis of the linear regime by noting that Eq. (\ref{linreal}) implies that at late times:
\begin{eqnarray}
&&\delta_{\phi}^{(lin)}(r=0,t) \approx \\ \nonumber &-&\frac{3C_{\delta}(1+w)}{2}(2\Omega_m^{0.6}-\Omega_m)H^2 (ar_{-2})^2 \bar{\delta}(r_{-2},t),
\end{eqnarray}
where $r_{-2}$ is defined the smallest value of $r$ for which $\partial \ln\bar{\delta} / \partial \ln r \geq -2$, and $C_{\delta} \sim \Oo(1)$ depends on the precise form of $\bar{\delta}$. 

We have shown that in the linear regime  $\vert \delta_{\phi}\vert$ at the centre of the inhomogeneity is always smaller than $\vert \bar{\delta}(r_{-2},t)\vert$ by a factor of about $(1+w)(H a r_{-2})^2$; $Ha r_{-2}$ is roughly equal to the physical size of the inhomogeneity as a fraction of the horizon size.  
If the initial density perturbation is Gaussian i.e. 
$$\delta_{m}(z_i) = \delta_{0} \exp(-r^2/r_{c}^2)$$ then 
$$\delta_{i}(r) = \frac{3\delta_0 r_c^3 }{4r^3}\sqrt{\pi}\mathrm{erf}(\frac{r}{r_c}) - 2\frac{r}{r_c}e^{-(r/r_c)^2)}$$ and:
$$
\delta_{\phi}^{(lin)}(r=0,t) \approx -\frac{9(1+w)\delta_{0}H_{i}^2r_c^2}{20}(2\Omega_{m}^{-0.2}-\Omega_{m}^{0.2}),
$$
at late times. 

We plot the evolution of $\delta_{m}(r=0)$ and $\delta_{\phi}(r=0)$
for a Gaussian initial matter density perturbation in the linear
regime in Figure \ref{figlin}.  We have assumed that
$\delta_{m}(r=0,z_{i}) = 0.02$ at $z_{i}=30$, and $H_{i}r_c = 0.03$,
which corresponds to and $\delta_{m}(r=0,z=0) = 0.3$ and $H a r_{c} =
0.010$ today i.e. $R_{c}= ar_{c} \approx 31 h^{-1}\,\mathrm{Mpc}$. For this choice of inhomogeneity, our approximation for $\delta_{\phi}$ is only good for $1+z \lesssim 0.94(1+z_{i}) \approx 29$.
Just as \citet{dutta} saw in their numerical simulations, we find that close to $z=z_i$, $\delta_{\phi}/(1+w)>0$.  It then
becomes negative, and at late times it continues to grow more
negative, and it appears as if it might overtake $\delta_{m}$ at some
point in the future.  However, as we shall show below, this does not
occur and $\vert \delta_{\phi}\vert$ remains small at all times.  Note
that the apparent increase in $\delta_{\phi}$ at early times is due in part to fact that our approximation breaks down as one approaches $z=z_{i}$.  If one wishes to study the evolution of $\delta_{\phi}$ all the way back to $z=z_{i}$, the numerical approach taken by \citet{dutta} provides accurate results.  Our real focus in this work is, however, not on what occurs at very early times in the linear regime, but how the DDE density contrast evolves when the matter inhomogeneity goes non-linear. We consider this below.

\subsection{Dynamical Dark Energy Clustering in the Quasi-Linear Regime}
We begin our investigation of how $\delta_{\phi}$ evolves when the matter perturbation exits the linear regime by considering a matter perturbation that is in the weakly non-linear or quasi-linear regime i.e. $-1 <
\bar{\delta} \lesssim 10$. Eqs.(\ref{treqn}) and (\ref{rreqn}) give $M(r,t)/R \sim M(r)/R + \mathcal{O}(\delta_3^3\delta_1,\,\delta_1^2\delta_3^2,\,\delta_2\delta_3^2)$.  We suppose that at some initial time, $t_{i}$, $R_{,t} = HR$ and $R = r$. $M(r)$ is then the mass inside the shell with radius $r$ at $t=t_i$.  We write:
$$
M(r) = \frac{\Omega_{m}(t_i)}{2}H_{i}^2 r^3 (1+\delta_{i}(r)).
$$
$\delta_{i}(r)$ is interpreted as the initial mean matter density
contrast.  In the full non-linear regime, which we consider in the next subsection,
we are only able to evaluate $R(r,t)$ and hence, via
Eq. (\ref{deltaphimain}), $\delta_{\phi}$, analytically in a matter-dominated
background ($\Omega_{m}=1$).  In many cases of interest, however, $\delta_{m}
\sim \Oo(1)$, e.g. superclusters and voids. Although the linear approximation
fails for such objects, a good leading approximation to the true mean
density contrast of matter, $\bar{\delta}$, in the range, $-1 <
\bar{\delta} \lesssim 10$ is given by:
$$
1+\bar{\delta}(r,t) \approx \left(1-2\bar{\delta}_{\rm lin}(r,t)/3\right)^{-3/2},
$$
where $\bar{\delta}_{\rm lin}$ is the linear mean density contrast.  This approximation improves as $\Omega_{m} \rightarrow 0$ and  is exact for $\Omega_{m} = 0$.  When this approximation holds we say that we are in the quasi-linear regime. 

In this section we find $\delta_{\phi}(\bar{\delta})$ for inhomogeneities in the quasi-linear regime. We are again assuming that any deviations from the $\Lambda$CDM model for structure formation due the background DDE evolution are sub-leading order  i.e. $(1-\Omega_m)(1+w)$ is small. If this is not the case then our expression for $\delta_{\phi}$ in terms of $\bar{\delta}$ and $\delta v$ is still valid, but the evolution of $\bar{\delta}$ and $\delta v$ would change; even still the order of magnitude of $\delta_{\phi}$ would not be greatly effected.

\begin{figure}
\begin{center}
\includegraphics[width=80mm]{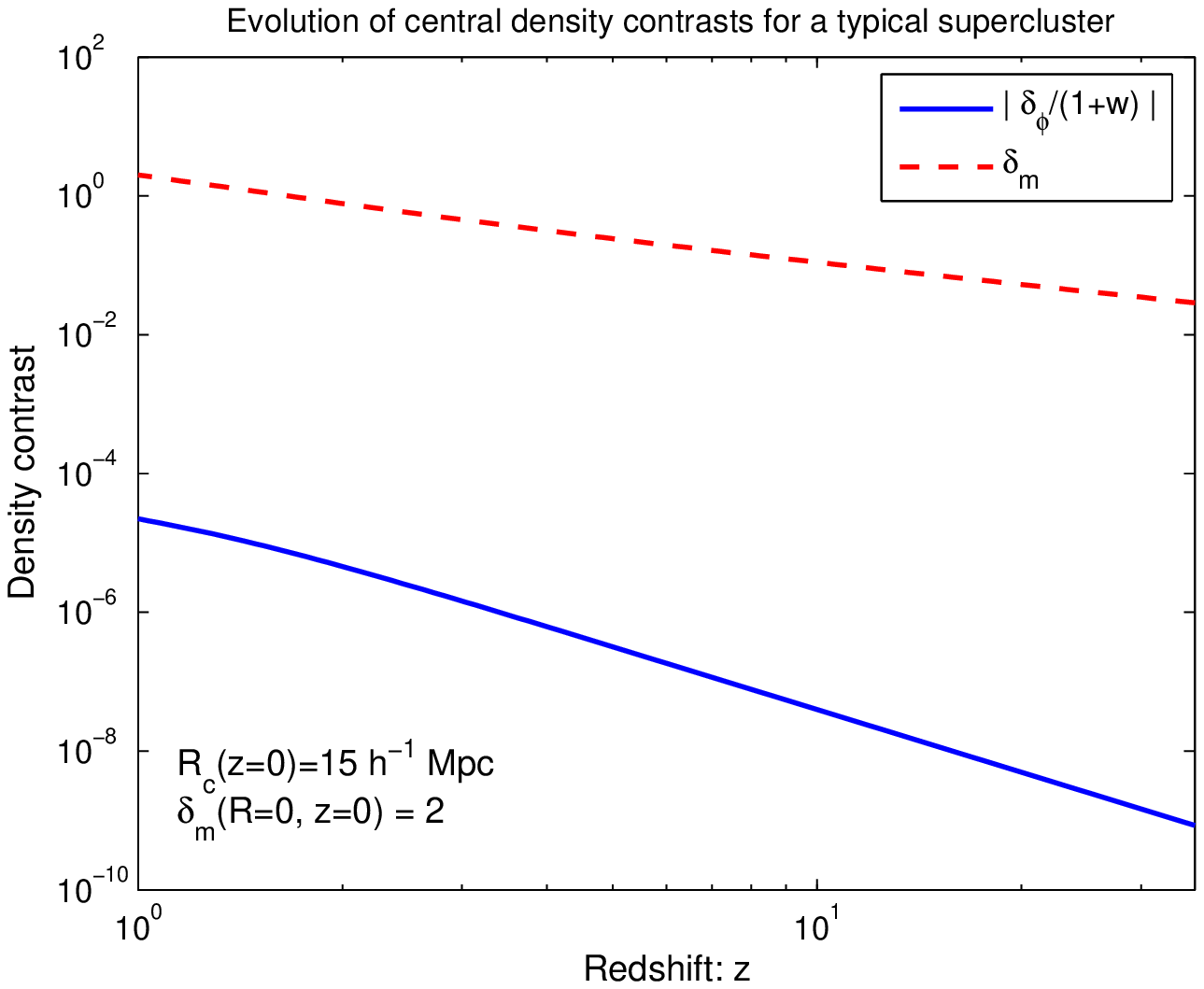}
\includegraphics[width=80mm]{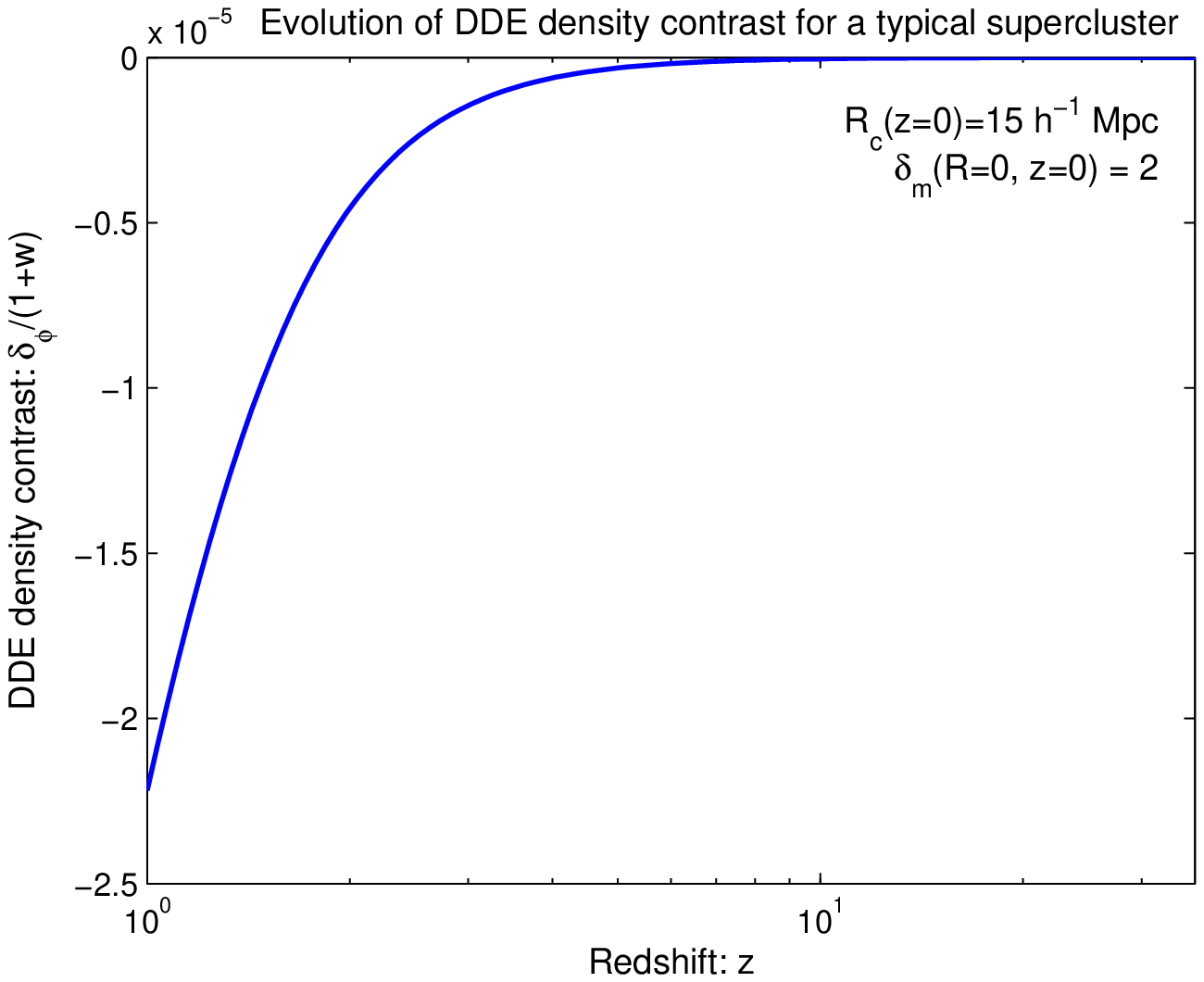}
  %% to include a figure, or
  %% to leave a blank space
  \end{center}
 \caption{Evolution of the DDE and matter density contrasts for a supercluster which today has a core matter overdensity of $2$ and a core radius of $15h^{-1}\,\mathrm{Mpc}$.  We have take $\Omega_{m} = 0.27$ today.  $\vert \delta_{\phi} \vert$ is always $\ll \delta_{m}$, and $\delta_{\phi}/(1+w) < 0$ which corresponds to a DDE void. As in the linear regime, $\vert \delta_{\phi}/(1+w)\vert$ grows monotonically with time, and at late times $\vert \delta_{\phi}/(1+w)\vert$ is growing more quickly than $\delta_{m}$.}
 \label{QlinSupercluster}
 \end{figure}

The peculiar velocity, $\delta v = R_{,t} - HR$, is related to $\bar{\delta}(r,t)$ thus:
$$
\delta v = -\frac{\bar{\delta}_{,t}R}{3(1+\bar{\delta})}  \approx -\frac{1}{2}HRf(a)((1+\bar{\delta})^{2/3}-1),
$$
where, as above, $f(a) \approx \Omega_{m}^{0.6}$ at late times.  In the quasi-linear regime the \emph{physical radius}, $R(r,t)$, of a shell with mass $M(r)$ is given by:
\begin{eqnarray}
R &=&
a\left(1-\frac{2\delta_{\rm lin}(r,t)}{3}\right)^{1/2}\left(\frac{2M(r)}{a_{i}^3\Omega_{m\,i}H_i^2}\right)^{1/3}\nonumber \\
&=& ar\left(1-\frac{2\delta_{\rm lin}(r,t)}{3}\right)^{1/2}\left(1+\delta_i(r)\right)^{1/3}.
\end{eqnarray}
Using Eq. (\ref{deltaphimain}), we find that the dark energy density contrast, $\delta_{\phi}$, in the quasi-linear regime is well approximated by $\delta_{\phi}^{({\rm ql})}$ where:
\begin{eqnarray} \label{deltaphiql}
&&\delta_{\phi}^{({\rm ql})} \equiv \\ \nonumber &&
-\frac{(1+w)\Omega_{m}H^2}{2}\int_{R}^{\infty} \diff R^{\prime}\,R^{\prime}
\left(\frac{3f(a)((1+\bar{\delta})^{2/3}-1)}{\Omega_m} -
\bar{\delta}\right) \nonumber\\ &&\nonumber+ \frac{(1+w)\Omega_{m}H^2R^2}{4} \left(\frac{f(a)^2}{\Omega_m}\right)((1+\bar{\delta})^{2/3}-1)^2.
\end{eqnarray}

\begin{figure}
\begin{center}
\includegraphics[width=80mm]{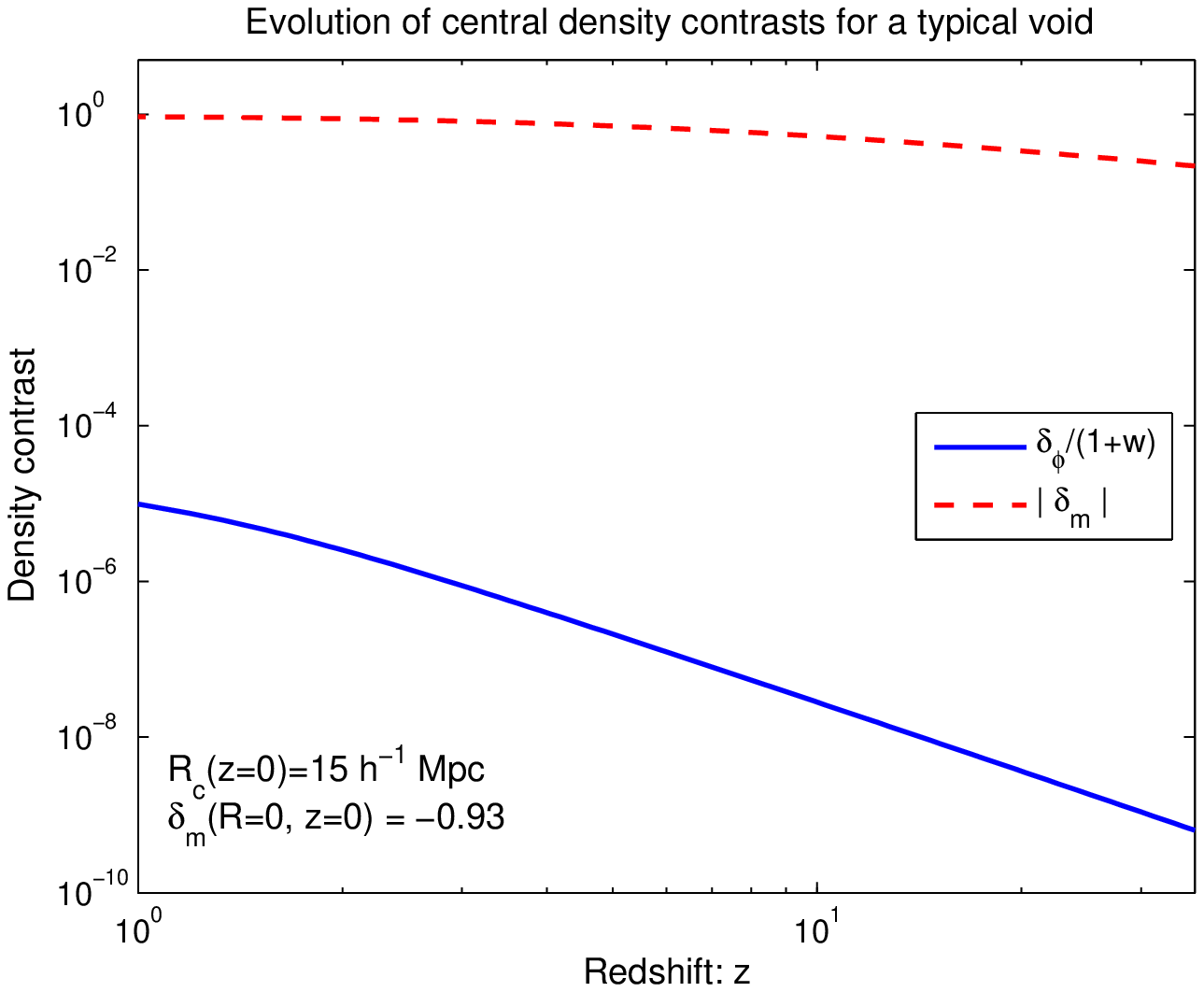}
\includegraphics[width=80mm]{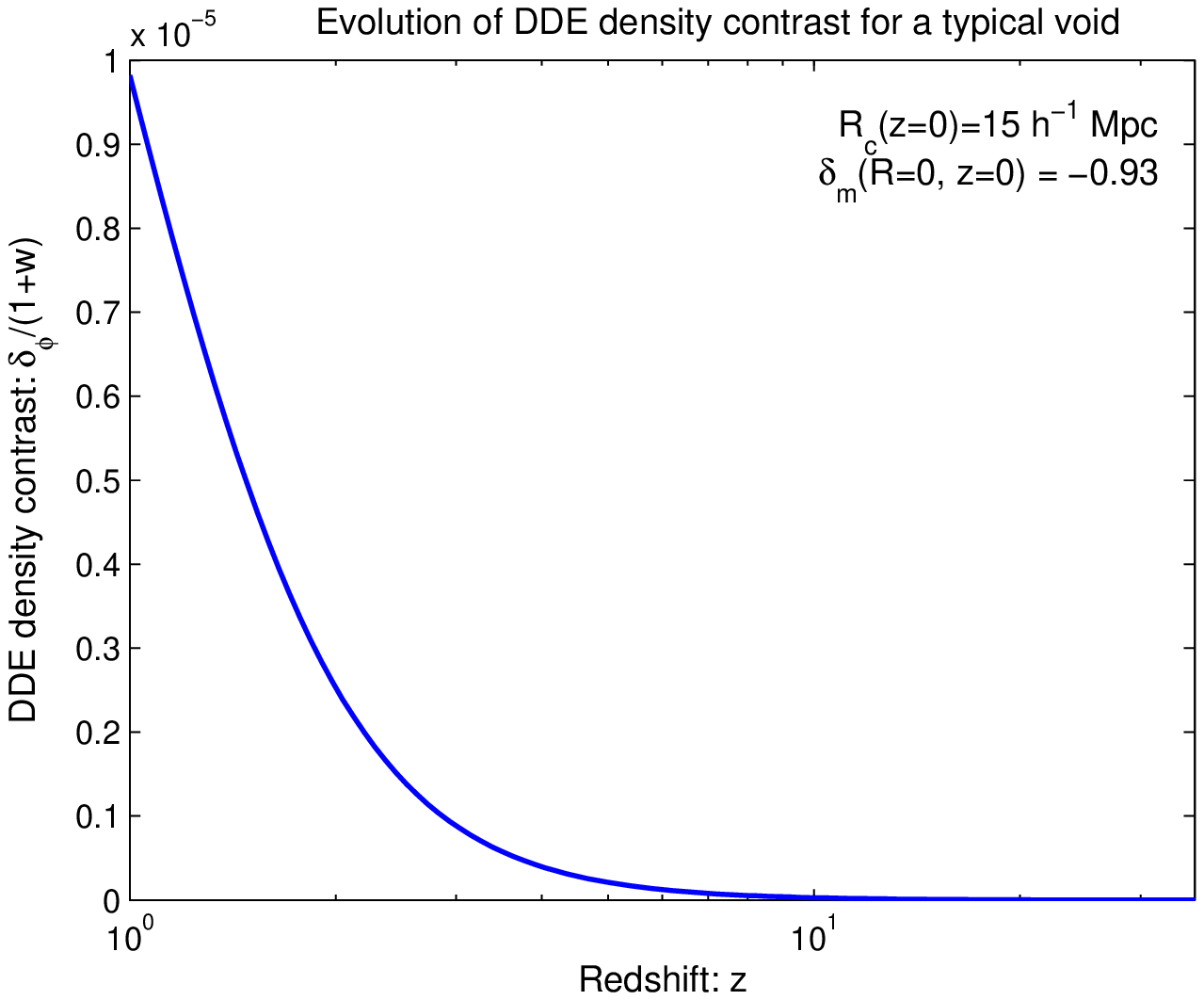}
  %% to include a figure, or
  %% to leave a blank space
  \end{center}
 \caption{Evolution of the DDE and matter density contrasts for a typical void which today has a core matter density contrast of $-0.93$ and a core radius of $15h^{-1}\,\mathrm{Mpc}$. We have taken $\Omega_{m}=0.27$ today.  $\vert \delta_{\phi} \vert$ is always $\ll \vert\delta_{m}\vert$, and $\delta_{\phi}/(1+w) > 0$ which corresponds to a DDE overdensity. As in the linear regime, $\vert \delta_{\phi}/(1+w)\vert$ grows monotonically with time and at late times $\vert \delta_{\phi}/(1+w)\vert$ is growing more quickly than $\vert \delta_{m} \vert$.}
 \label{QlinVoid}
 \end{figure}

For $0.27 \leq \Omega_{m} \leq 1$, the first term in
Eq. (\ref{deltaphiql}) is negative definite for all positive values of
$\bar{\delta}$ for which the quasi-linear approximation holds ($0 <
\bar{\delta} \lesssim 10$) and positive definite for $-1 < \bar{\delta} < 0$. The second term in
Eq. (\ref{deltaphiql}) is clearly positive definite but vanishes at
$R=0$.  The relative magnitude of the two terms depends on the choice
of initial density profile. However, at $R=0$ it is clear that an overdensity of matter,
$\bar{\delta}>0$, corresponds to a dark energy void, $\delta_{\phi}(R=0,t) <
0$.   Similarly, a void of matter corresponds to a local DDE overdensity at $R=0$. In Figures \ref{QlinSupercluster} and \ref{QlinVoid} respectively show the evolution of $\delta_{\phi}/(1+w)$ and $\bar{\delta}$ at $R=0$ for a typical supercluster and a typical void.  For both the supercluster and the void we assume an initial density contrast profile that is flat for $r < r_{c}$ and drops as $1/r^3$ for $r > r_{c}$. We take the initial instant to be in the far past i.e. $z_{i} \rightarrow \infty$. The supercluster is taken today to have a central density contrast of $\bar{\delta}(R=0) = 2$ and $r=r_{c}$ to corresponds to a physical radius of $R_{c} = 15 h^{-1}\,\mathrm{Mpc}$, which are appropriate for an object such as the local super-cluster.  For the void we take typical values of $\bar{\delta}(R=0) = -0.93$ today and $R_{c} = 15h^{-1}\,\mathrm{Mpc}$. In each case we have taken $\Omega_{m}(z=0)=0.27$.  It is clear that in both cases $\vert\delta_{\phi}/(1+w) \vert$ grows faster than $\vert \delta_{m} \vert$, however the DDE density contrast remains much smaller than $\vert\delta_{m}\vert$ at all times. Note that $\delta_{\phi}$ evolves monotonically for objects in both the linear and quasi-linear regimes.

\subsection{Dynamical Dark Energy Clustering in the Non-Linear Regime}
The quasi-linear approximation to $\delta_{\phi}$ is valid for
$-1 \lesssim \bar{\delta} \lesssim 10$, however it breaks down for large positive
values of the matter density contrast, $\bar{\delta}$.    To go further we  evaluate
$\delta_{\phi}$ in the fully non-linear regime.  To do this we need to
calculate both the evolution of the peculiar velocity, $\delta v$,
and that of the mean density contrast mass contrast, $\bar{\delta}$.    The non-linear
evolution of the matter overdensity is significantly more simple to
calculate analytically in a matter dominated Universe ($\Omega_{m} =
1$).    Fortunately, when the matter density contrast is large today
$\bar{\delta} \gtrsim 100$, the evolution of both $R_{,t}$
and $\varepsilon^{(m)}$ in the inhomogeneous region are well-approximated by the $\Omega_{m} = 1$ solution. This is because the epoch of dark energy domination begins sometime after the cluster has turned around. 

\begin{figure}
\begin{center}
\includegraphics[width=80mm]{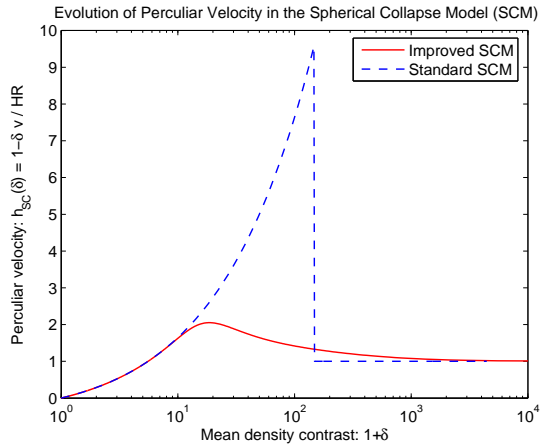}
  %% to include a figure, or
  %% to leave a blank space
  \end{center}
 \caption{Plot showing how the peculiar velocity, $h_{SC} = 1-\delta v / HR$, varies with the mean density contrast, $\bar{\delta}$, in both the standard spherical collapse model (dashed line) and the improved one (solid line) of \citet{improvesc}.  The improved model is a good fit to the results of N-body simulations for $\bar{\delta} \gtrsim 15$.  We note that $h_{SC}$ reaches a maximum value of about $2$ when $\bar{\delta} \approx 18$ in the improved model, whereas in the standard collapse model $h_{SC}$ continues to grow until $\bar{\delta} \approx 140$, where it reaches a maximum value of about $9.5$.  In the non-linear regime, the leading order expression for $\bar{\delta}_{\phi}$ contains a term that is proportional to $\delta v^2 \propto h_{SC}^2$.  Therefore if the standard spherical collapse model were used to evaluate $\delta_{\phi}$ it could result in it being over-estimated, for $20 \lesssim  \bar{\delta}\lesssim 140$, by roughly  1-2 orders of magnitude.}
 \label{figisc}
 \end{figure}

If $\Omega_{m}=1$, we are effectively dealing with a Tolman-Bondi \citep{tolbondi,tolbondi1} background at leading order. We therefore have analytical solutions for $R$ and $t$ in terms of $r$ and a parameter $\eta(r,t)$:
\begin{eqnarray}
R = \frac{1+\delta_{i}(r)}{2\delta_{i}(r)}r\left(1-\cos \eta\right), \label{Reqn}\\
t = t_{s}(r) + \frac{(1+\delta_{i}(r))}{2H_{i}\delta_{i}^{2/3}(r)}\left(\eta - \sin \eta\right), \label{teqn}
\end{eqnarray}
where
\begin{eqnarray}
t_{s}(r) &=& t_{i} - \frac{(1+\delta_{i}(r))}{2H_{i}\delta_{i}^{2/3}(r)}\left(\eta_{i}(r)-\sin\eta_{i}(r)\right) \\
\nonumber &\sim& t_{i}\left(\frac{\delta_{i}(r)}{5} + \mathcal{O}(\delta_{i}^2)\right)
\end{eqnarray}
and
$$
\eta_{i}(r) =\cos^{-1}\left(\frac{1-\delta_{i}(r)}{1+\delta_{i}(r)}\right) \sim 2\delta_{i}^{1/2}(r) - \frac{2}{3}\delta_{i}^{3/2}(r) + \mathcal{O}(\delta_{i}^{5/2}).
$$
We assume that at early times, $(z > 1.8)$, $\Omega_{m} \approx 1$. These solutions then provide an excellent approximation to the true early time evolution of the inhomogeneity.  These solutions also provide an good approximation at late times, provided we are dealing with an overdensity of matter that turns around prior to the onset of dark energy domination ($\bar{\delta} \gtrsim 100$ today). Assuming that we are dealing with such an overdensity, we find:
\begin{eqnarray}
\frac{\delta M}{R^2} &=& \frac{1}{2} \Omega_{m} H^2R
\left(\frac{9(a/a_i)^3(\eta-\sin\eta)^2}{2(t/t_i)^2(1-\delta_t)^2(1-\cos \eta)^3}
-1\right) \nonumber  \\ &=& \frac{1}{2}\Omega_m H^2 R \bar{\delta}.
\end{eqnarray}
When the background is matter-dominated, $(a/a_i)^3/(t/t_i)^2 = 1$ and
$\Omega_m = 1$. In the non-linear regime $\delta_{t} = t_{s}(r)/t \ll 1$, and terms proportional to $\delta_{t}$ in the above expression can be dropped. 

We have so far modelled the matter as being a spherically symmetric,
pressureless perfect fluid.  This approximation neglects the random
motion of the matter particles, and other effects due to the break
down of spherical symmetry.  These are small prior to turnaround but
thereafter act to slow down the collapse of the inhomogeneity.
Without any such corrections the matter perturbation would eventually
collapse to a point.  When these effects are included however the
inhomogeneity virialises and relaxes to a steady-state.  In a
matter-dominated background, each shell of constant $M$ ceases to
collapse and virialises when its radius, $R_{\rm vir}$, is about one half
of the radius at which it turned around, $R_{\rm ta}$.  N-body simulations
suggest that $R_{\rm ta} / R_{\rm vir} \approx 1.8$ \citep[see][]{hamilton}.

\begin{figure}
\begin{center}
\includegraphics[width=80mm]{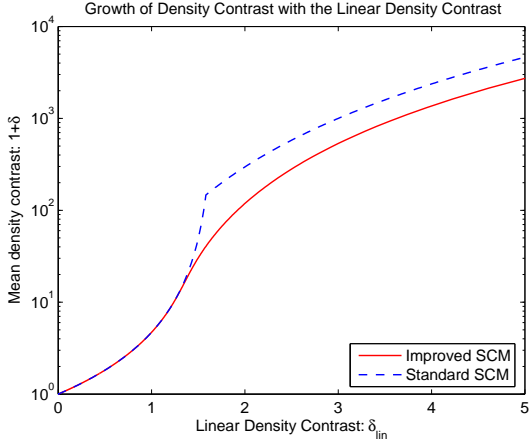}
  %% to include a figure, or
  %% to leave a blank space
  \end{center}
 \caption{Plot showing how the true mean density contrast, $\bar{\delta}$, depends on the linear one, $\bar{\delta}_{\rm lin} \propto a(t)$ for $\Omega_{m} = 1$.    We show both the standard spherical collapse model (SCM), the dashed line, and the improved SCM proposed by \citet{improvesc}, the solid line.  For $15 \lesssim \bar{\delta} \lesssim 140$, the $\bar{\delta}$ grows faster in the standard SCM than in the improved one, however the overall discrepancy between the models' predictions for $\bar{\delta}(\bar{\delta}_{\rm lin})$ is not nearly as great as it is for their predictions for $h_{\rm SC}(\bar{\delta}) = 1- \delta v /HR$: see FIG. \ref{figisc}.}
 \label{figisc2}
 \end{figure}

Since process of virialisation is not accounted for in our matter
model, we must account for it in some fashion if we are to make accurate
predictions.  This is generally done via the rather ad hoc process of
manually halting the collapse when each shell reaches its virial
radius, $R_{\rm vir}$; in this model one takes $R_{\rm vir} = R_{\rm ta}/2$; this
occur at $\eta= 3\pi/2$.  Whilst this provides a reasonably good
approximation for the magnitude of the density contrast,
$\bar{\delta}$, the peculiar velocity, $\delta v$, is discontinuous at
instant of virialisation ($\eta = 3\pi /2$). This ad hoc procedure
therefore gives widely inaccurate predictions for the peculiar
velocity, $\delta v = R_{,t}-HR$ (see Figure \ref{figisc}).  Since our
formula for $\delta_{\phi}$ depends on both $\delta v$ and
$\bar{\delta}$ we require a more realistic model for virialisation in
the context of spherical collapse.  Such an improved model was
proposed recently by \citet{improvesc}.  In this improved spherical
collapse model, it is assumed that $h_{\rm SC}\equiv \delta v / HR$ is a
function of $\bar{\delta}$ i.e. $h_{\rm SC}=h_{\rm SC}(\bar{\delta})$.  $R$ is
still parametrized according to Eq. (\ref{Reqn}), and it is found
that:
$$
t = t_{s}(r) + \frac{(1+\delta_{i}(r))}{2H_{i}\delta_{i}^{2/3}(r)}T(\tau), 
$$
where $\tau = \eta -\sin \eta$.   The function $T(\tau)$ is then fixed so that: Prior to turnaround, when deviations from spherical symmetry are small, $T(\tau) \sim \tau$. For larger values of $\bar{\delta}$ ($\gtrsim 15$), $ T(\tau)$ is chosen so that $h_{\rm SC}(\bar{\delta})$ agrees, under certain reasonable and common assumptions, with the behaviour that is seen in N-body simulations of \citet{hamilton}.   \citet{improvesc} provide the following fitting formula for $T(\eta)$:
\begin{equation}
T(\eta(\tau)) = \tau + \frac{3.468(\tau_f - \tau)^{-1/2}\exp\left(-\frac{15(\tau_f - \tau)}{\tau}\right)}{(1+0.8(\tau_f-\tau)^{1/2}-0.4(\tau_f - \tau))},\label{Tfit}
\end{equation}
where $\tau_{f} = 5.516$.  In this model $R_{\rm vir}/R_{\rm ta} \approx 0.59$ which compares more favourably with the value of $0.56$ suggested by N-body simulations \citep{hamilton} than the value of $0.50$ that is generally used. In this improved model, $\delta_{\phi}$ is still given by Eq. \ref{deltaphimain}, and for $\Omega_{m} = 1$ the peculiar velocity is given by:
$$
\frac{\delta v}{HR} = \frac{R_{,t}}{HR}-1 = \sqrt{\frac{1+\bar{\delta}}{2}} \frac{\sin \eta}{(1-\cos \eta)^{1/2}} \frac{d \tau}{d T}-1,
$$
and the density contrast is given by:
$$
1+\bar{\delta} = \frac{9 T(\eta)^2}{2(1-\cos \eta)^3}.
$$

Figure \ref{figisc} shows $h_{\rm SC} \equiv 1-\delta v /HR $ as a
function of $\bar{\delta}$ in both the improved spherical collapse
model, and the standard one (with virialisation occurring suddenly at
$R=R_{\rm vir}$).  Note that in the standard spherical collapse model
(SCM), $h_{\rm SC}$ continues to grow until $\bar{\delta} \approx 140$,
and then drops sharply, whereas in the improved model $h_{\rm SC}$ reaches
a maximum when $\bar{\delta} \approx 18$. Note also that the maximum value of $h_{\rm SC}$ predicted by standard SCM is almost five times as large as the maximum value gives by the improved model.  Since $\delta_{\phi}$ depends in part on $\delta v^2$ and hence $h_{\rm SC}^2$, using the standard SCM instead over the improved model could lead to $\delta_{\phi}$ being over estimated by as much as $2500\%$.

 Figure \ref{figisc2} shows $\bar{\delta}$, for $\Omega_m =1$, as a function of the mean linear
density contrast, $\bar{\delta}_{\rm lin} \propto a$, in both the improved
model and the standard one with virialisation put in by hand. We use
our modified spherical collapse model to find the evolution of
$\delta_{\phi}$. We find then that the DDE density contrast in the
non-linear regime is given by:
\begin{eqnarray}
\delta_{\phi}(R,t) \sim &&\frac{\Omega_{m} H^2(1+w)}{2}\int_{R}^{\infty}
\bar{\delta}(R^{\prime},t) G(\bar{\delta}; \Omega_{m}) R^{\prime} \diff R^{\prime}\nonumber \\ &&+ H^2(1+w) R^2 h(\bar{\delta})^2, \label{deltaphinon}
\end{eqnarray}
where
$$
G(\bar{\delta}; \Omega_{m}) = -\frac{6h(\bar{\delta})}{\bar{\delta}}+1.
$$
When $\bar{\delta} \ll 1$, $G(\bar{\delta}) \sim -1 - 2(\Omega_{m}^{-0.4}-1)$, and deep in
the non-linear regime, when $\bar{\delta} \gg 1$, $G(\bar{\delta}) \sim
1$. In the $\Omega_m = 1$ case we find that the following fitting
formula is accurate to within $2\%$:
\begin{eqnarray}
&&G(\bar{\delta};\Omega_m=1, \alpha = 1) \approx \tanh\left(-0.9218 +\right. \\ && \left.\nonumber \ln(\bar{\delta})\left(\frac{195 + 0.543 \bar{\delta}^6}{403 + \bar{\delta}^6}\right)-\frac{0.38\sin^3\left(\ln(\bar{\delta})-2.49\right)}{(\ln(\bar{\delta})-2.49)^3}\right). \label{Gfiteq}
\end{eqnarray}
We note that when $\Omega_m = 1$, $G(\bar{\delta}) < 0$ for
$\bar{\delta} \lesssim 10.6$ and is otherwise positive.  In a cluster with a core overdensity $\gtrsim 20$,
$\delta_{\phi}$ is therefore generally positive at $R=0$ i.e. there is DDE
overdensity at $R=0$, whereas for clusters with a core overdensity
$\lesssim 10 - 20$, there is generally a DDE void around $R=0$.   Even in the deep
non-linear regime, the central DDE overdensity is still surrounded by a void, with $\delta_{\phi}$ becoming negative when
$\bar{\delta}(R) \approx 10-20$. We see that $\vert \delta_{\phi}
\vert \sim \Oo(\Omega_m (1+w) H^2 R^2_{c}\bar{\delta} /2)$, where $R_{c}$
is the radius of the core of cluster (roughly defined by smallest radius after which $\bar{\delta}$ decreases faster than $1/R^2$), and $\bar{\delta}_c$
is the core density contrast i.e. the density contrast at $R=R_c$. 

\begin{figure}
\begin{center}
\includegraphics[width=80mm]{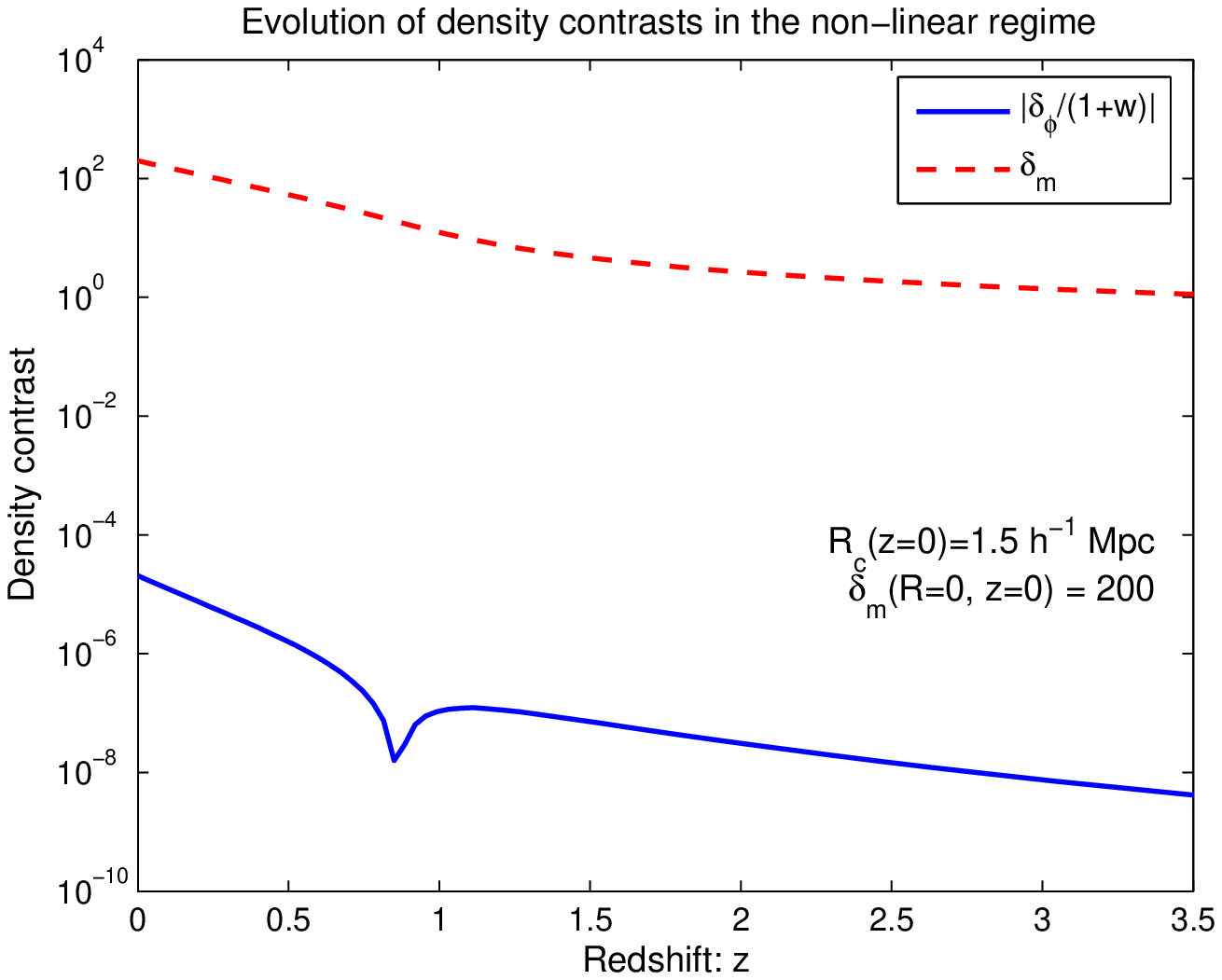}
\includegraphics[width=80mm]{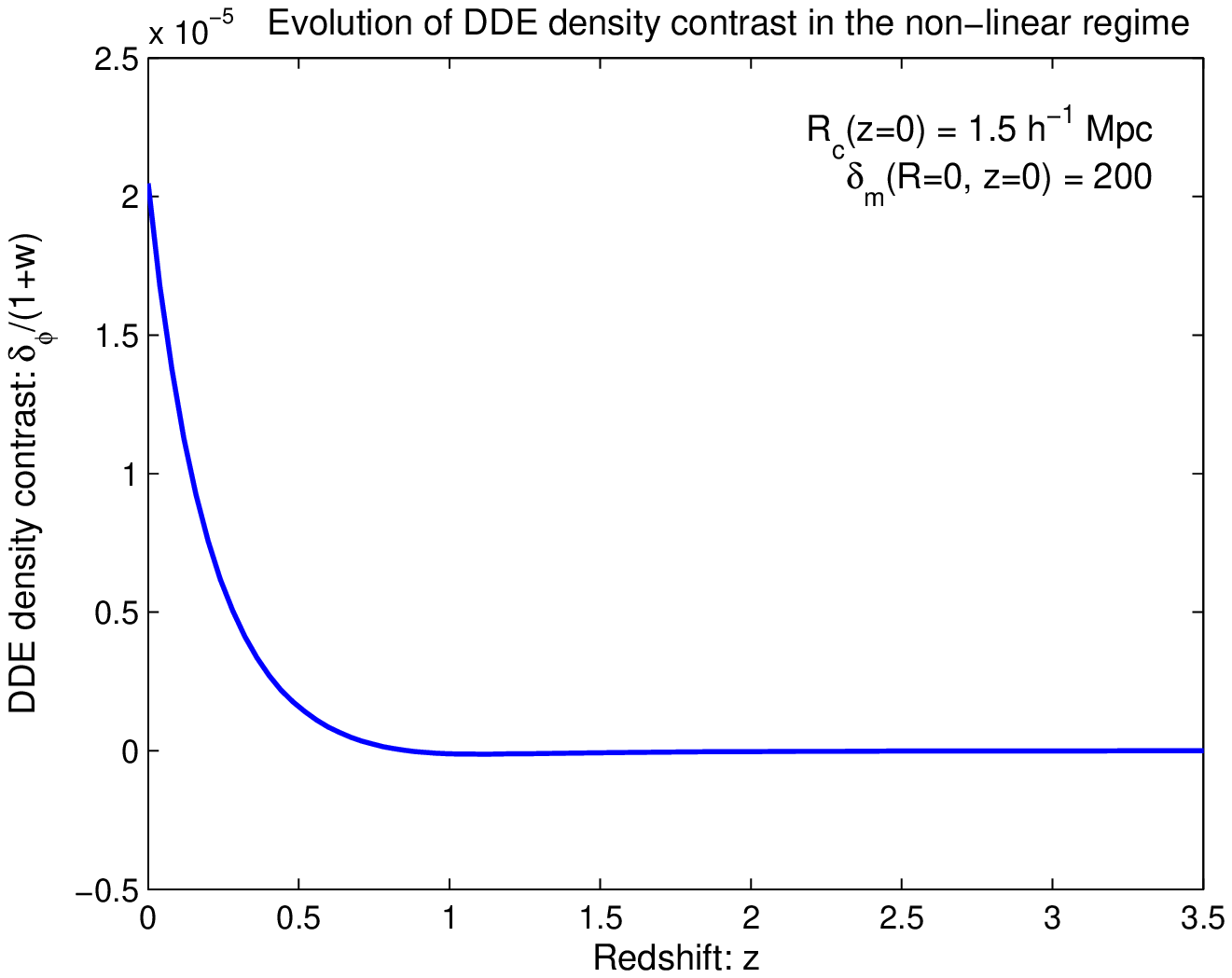}
  %% to include a figure, or
  %% to leave a blank space
  \end{center}
 \caption{Evolution of the DDE and matter density contrasts for a cluster which today has a core matter overdensity of $200$ and a core radius of $1.5 h^{-1}\,\mathrm{Mpc}$.  We have taken $\Omega_{m} = 0.27$ today.  $\vert \delta_{\phi} \vert$ is always $\ll \delta_{m}$. At late times $\delta_{\phi}/(1+w) > 0$, which corresponds to a central DDE overdensity, however at early times it is negative and there is a DDE void.  We see that $\delta_{\phi}$ changes sign when the mean core overdensity of matter is $\sim \Oo(10)$.}
 \label{nlevo1}
 \end{figure}

We use Eq. (\ref{deltaphinon}) to plot the evolution of $\delta_{\phi}(R=0)$ in the non-linear regime.  For $\bar{\delta} \gg 1$, the matter overdensity in a background with $0.2 \leq \Omega_{m} < 1$ evolves, to a good approximation, according to:
$$
1+\bar{\delta}(\eta)(\Omega_{m}) =  F(a)(1+\bar{\delta}_{0}(\eta)),
$$
where  $\bar{\delta}_{0} = 9T^2 / 2(1-\cos \eta)^3$ and $F(a) =\frac{4}{9t^2 \Omega_m H^2} \approx \Omega^{-0.4}$. We expect the largest contributions to $\delta_{\phi}(R=0)$ to come from regions where $\bar{\delta} \gg 1$ which implies that the we expect $G(\bar{\delta}; \Omega_{m}) \approx 1$ for all $\Omega_{m}$.  
Therefore, using Eq. (\ref{deltaphinon}), we approximate:
\begin{eqnarray}
\delta_{\phi}(R=0,t) &\approx& \frac{\Omega_{m}^{0.6} H^2 (1+w)}{2} \\ \nonumber &&
\int_{0}^{\infty} \bar{\delta}_{0}(\eta) G(\bar{\delta}_{0}(\eta),\Omega_{m} = 1) R^{\prime}(r,\eta) dR^{\prime}.
\end{eqnarray}

\begin{figure}
\begin{center}
\includegraphics[width=80mm]{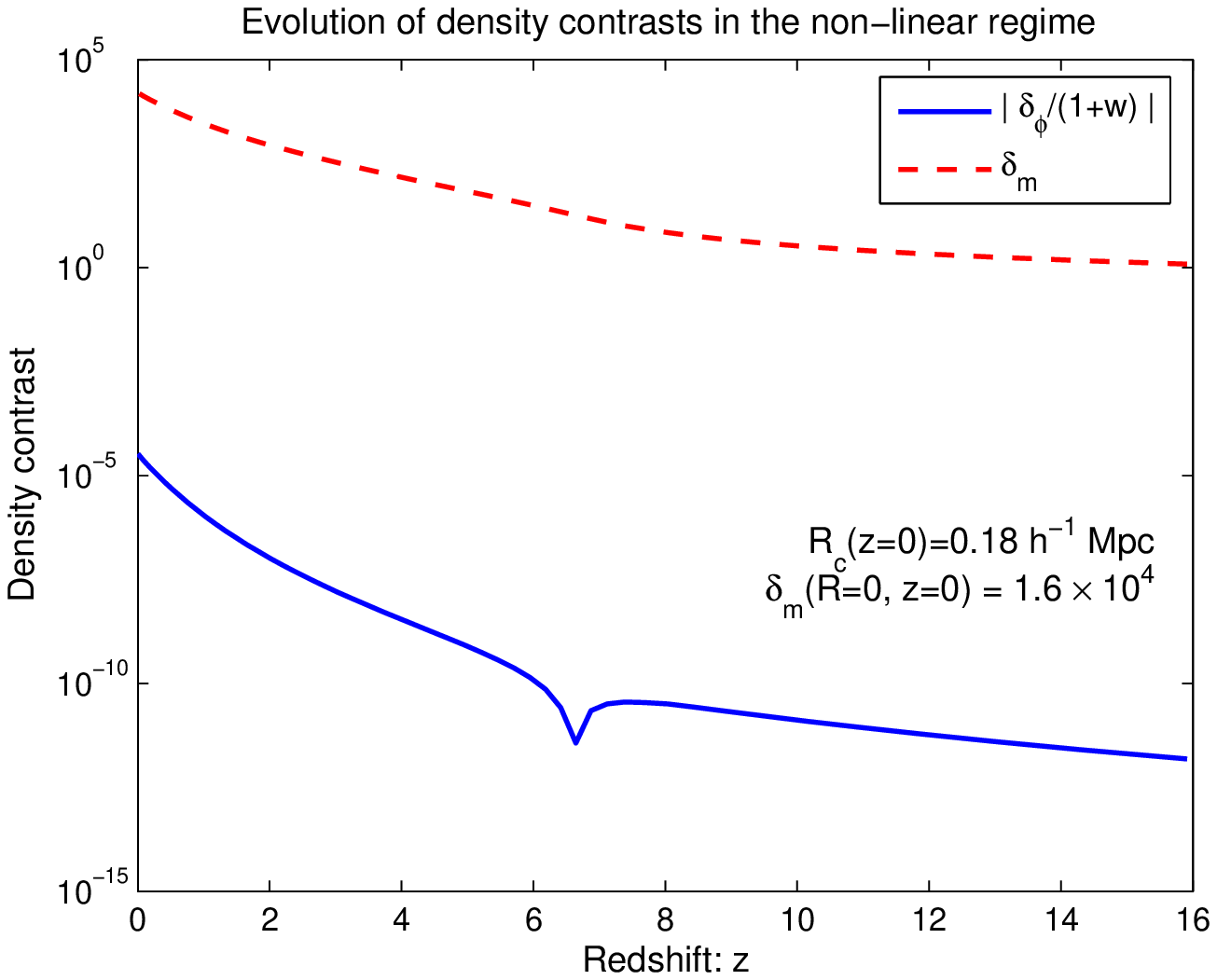}
\includegraphics[width=80mm]{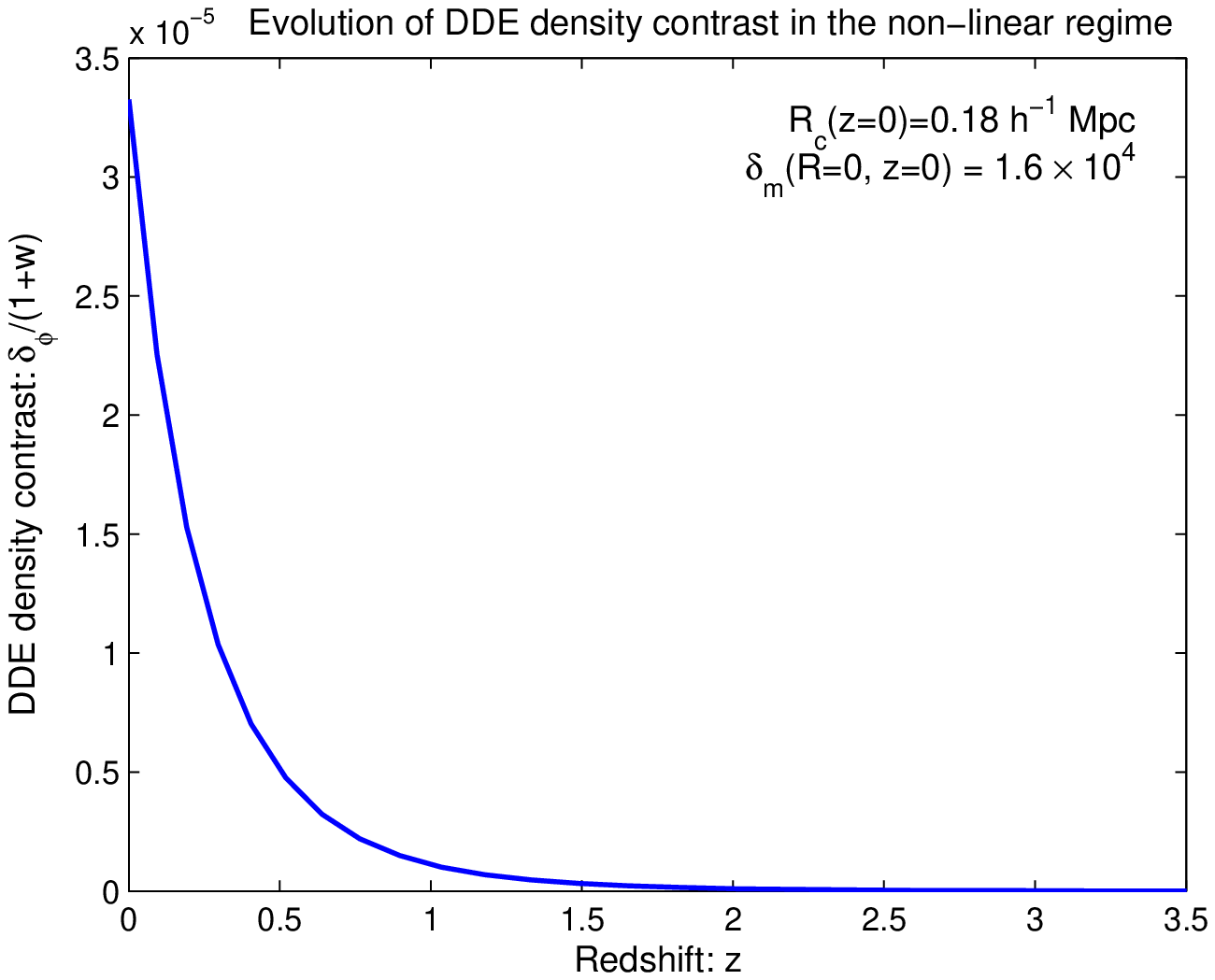}
  %% to include a figure, or
  %% to leave a blank space
  \end{center}
 \caption{Evolution of the DDE and matter density contrasts for a cluster which today has a core matter overdensity of $1.6 \times 10^{4}$ and a core radius of $0.18 h^{-1}\,\mathrm{Mpc}$.  We have taken $\Omega_{m} = 0.27$ today.  $\vert \delta_{\phi} \vert$ is always $\ll \delta_{m}$. At late times $\delta_{\phi}/(1+w) > 0$, which corresponds to a central DDE overdensity, however at early times it is negative and there is a DDE void.  We see that $\delta_{\phi}$ changes sign when the mean core overdensity of matter is $\sim \Oo(10)$.}
 \label{nlevo2}
 \end{figure}

For simplicity, we take the initial density profile to have a central core with  $\delta_{i}(r) = \delta_{i}$ for $r < r_c$ and for $r > r_c$, we take: $\delta_{i} \propto r^{-3}$. 

In Figure \ref{nlevo1} we plot $\bar{\delta}(R=0)$ and $\delta_{\phi}(R=0)$ for a cluster where today $r = r_{c}$ corresponds to a physical radius of $1.5 h^{-1} Mpc$, and the mean density contrast in $R < R_c$ today is $200$.  In Figure. \ref{nlevo2} we plot  $\bar{\delta}(R=0)$ and $\delta_{\phi}(R=0)$ for a smaller but denser cluster with $R_c = 0.18 h^{-1} \mathrm{Mpc}$ and $\bar{\delta}(R=0) = 1.6 \times 10^{4}$ today.  These values are fairly typical of clusters such as Coma which has been estimated to have a mean density contrast of $200$ inside a radius of $1.5 h^{-1}\,\mathrm{Mpc}$, and a core mean density contrast of about $1.6 \times 10^{4}$ inside a radius of $0.18 h^{-1} \, \mathrm{Mpc}$.  We can see that in both cases $\delta_{\phi}/(1+w) \sim \Oo(10^{-5})$ today and that, as expected, it is positive.  In both cases, $\delta_{\phi}/(1+w)$ was negative in the past, changing sign when the mean overdensity of the core was $\bar{\delta} \sim \Oo(10)$.  The core in these cases corresponds to the flat part of $\bar{\delta}$ i.e. $R < R_{c}$. For more general density profiles, the core radius would be taken to correspond to $R \lesssim R_{-2}$ where $R_{-2}$ is the smallest value of $R$ for which $\bar{\delta}$ faster as $R^{-2}$.  In all of the plots we have taken $\Omega_{m}(z=0) = 0.27$.  

We can see that in both cases $\delta_{\phi}/(1+w)$ continues to increase today.  This increase is due to the continued collapse of matter onto the collapsed core. At very late times, however, this accretion will cease, and $\delta_{\phi}(R=0)/(1+w) \rightarrow \mathrm{const}$ for all $\Omega_{m}$.  We can see this from Eq. (\ref{deltaphinon}).  At late times, and assuming minimal accretion: $\bar{\delta} = a^{3} \Delta(R)$, for some function $\Delta(R)$, and so:
\begin{equation}
\delta_{\phi}/(1+w) \approx  \frac{a^{3} \Omega_{m} H^2}{2} \int_{0}^{\infty} R^{\prime} \Delta(R^{\prime}) \, dR^{\prime}, 
\end{equation}
and $a^{3} \Omega_{m} H^2 = \mathrm{const}$ which implies $\delta_{\phi}/(1+w) \rightarrow \mathrm{const}$ at late times. It is clear that $\delta_{\phi}/(1+w)$ is smaller than $\bar{\delta}_{-2} = \bar{\delta}(R=R_{-2},t)$ by at a factor of about $\Omega_m H^2 R_{-2}^2$.  In follows that even at late times $\delta_{\phi}/(1+w) \ll 1$, as $\delta_{\phi}/(1+w) \sim \Oo(\Omega_{m} H^2 R_{-2}^2 \delta_{-2}) \sim \Oo(GM_{-2}/R_{-2})$, where $M_{-2} = M(R=R_{-2})$.   

\subsection{Discussion}
We have seen that in all regimes (linear, quasi-linear and fully non-linear regimes), the magnitude of DDE density contrast scales as: $\Omega_m H^2 R_{c}^2 \bar{\delta}_{c}$, where $R_{c}$ is the radial scale of the cluster, and $\bar{\delta}_{c}$ is the mean matter density contrast in $R<R_{c}$.  $R_{c}$ is roughly defined to be equal to $R_{-2}$, which is the smallest radius for which $\bar{\delta}_{c}$ first decreases faster than $1/R^2$. If an inhomogeneity is sub-horizon when it begins to collapse, we found that shorter after the collapse begins $\bar{\delta}$ and $\delta_{\phi}$ have the same sign i.e. a matter overdensity equates to a DDE overdensity, and the same for voids.    However, a certain time after collapse begins, $\delta_{\phi}$ changes sign.   This change of sign is related to the fact that at late times the decaying mode of the matter perturbation is negligible.  When this occurs, a mean over density of matter results in a DDE void, and vice versa.  Deep in the non-linear regime, when $\bar{\delta}_{c} \gtrsim 10-20$, however, $\delta_{\phi}(R=0)$ changes sign again. A large mean overdensity of matter then corresponds to an overdensity of dark energy at $R=0$.  

In both the linear and the quasi-linear regimes,
$\vert\delta_{\phi}/(1+w)\vert$ increases faster than
$\bar{\delta}_{c}$, although it is always much smaller than
$\bar{\delta}_{c}$.  This behaviour was also observed by \citet{dutta} in their numerical study of the linear regime, and it lead them to wonder whether $\vert \delta_{\phi}
/ (1+w) \vert$ might grow to be $\Oo(1)$ or larger in the non-linear
regime.  Using our results, we have shown that the
onset of non-linear evolution for the matter perturbation slows down
the growth of $\vert \delta_{\phi}/(1+w) \vert$ and that at very late
times $\delta_{\phi}/(1+w) \rightarrow \mathrm{const}$ and remains
$\ll 1$.  In the next section we consider the profile of
$\delta_{\phi}/(1+w)$ in and around typical astrophysical objects.

\section{Astrophysical Applications}
\label{sec:app}
In this section we use our results to evaluate the profile of the DDE density contrast around typical clusters, 
superclusters and voids of matter. We also investigate the possibility of detecting the dark energy clustering using the Integrated Sachs Wolfe (ISW) effect.
\subsection{Galaxy Clusters}

\begin{figure}
\begin{center}
\includegraphics[width=80mm]{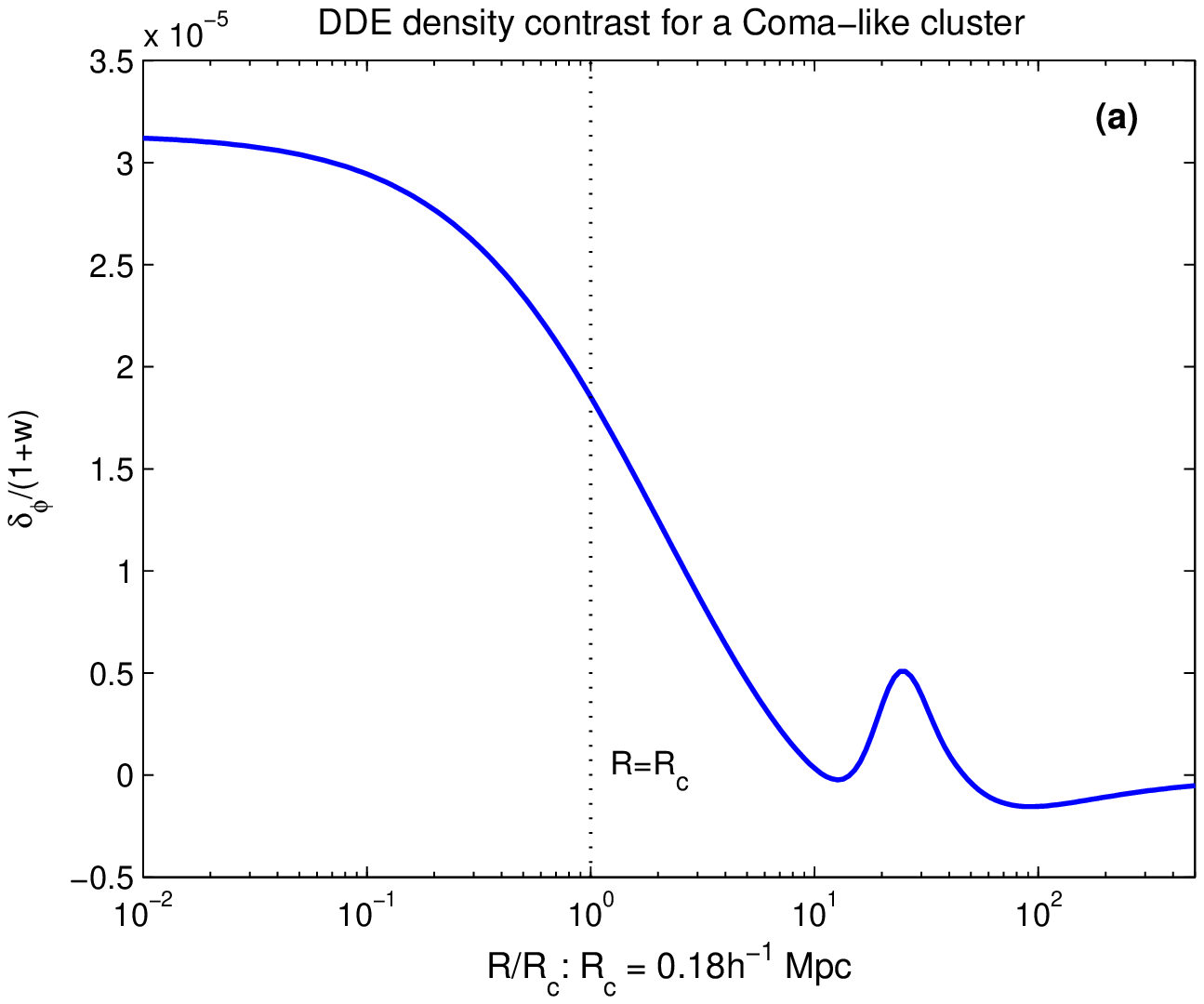}
\includegraphics[width=80mm]{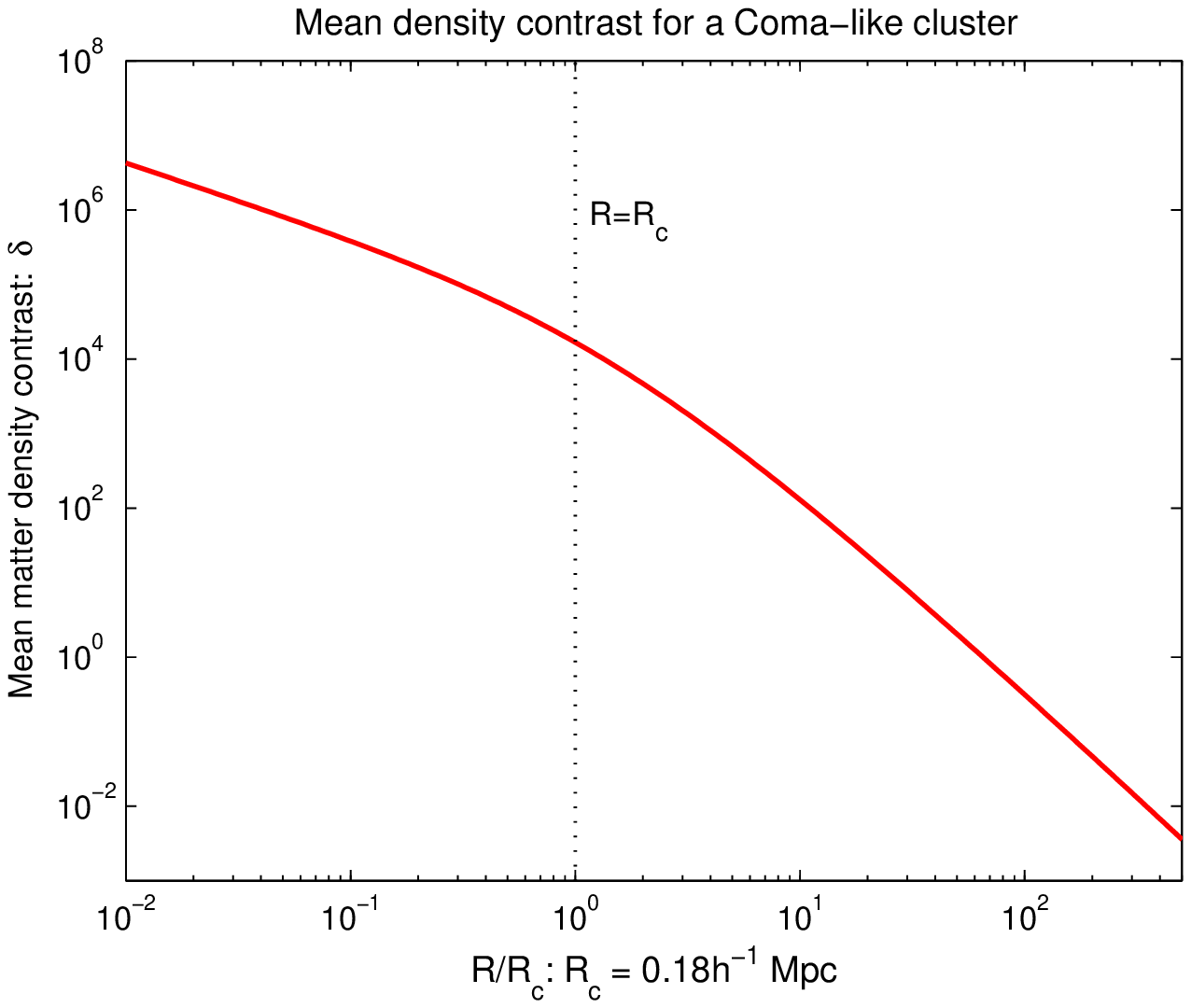}
  %% to include a figure, or
  %% to leave a blank space
  \end{center}
 \caption{The DDE (a) and matter density (b) profiles for a Coma-like cluster with NFW profile: $R_c = 0.18h^{-1} \mathrm{Mpc}$, $R_{200} = 1.5 h^{-1} \mathrm{Mpc}$. We have taken $\Omega_m = 0.27$. See text for discussion.}
 \label{comaplot}
 \end{figure}

We begin by considering the DDE density contrast in a virialised
galaxy cluster with similar properties to the Coma cluster.  \citet{nfw} (hereafter NFW) used high resolution N-body
simulations to derive a universal density profile for the dark-matter
halos of virialised structures such as galaxies and clusters.  The
resulting NFW profile is given by:
$$
\delta_{m} = \frac{\delta_{c}}{(R/R_{c})(1+(R/R_c))^2}, 
$$
where $R_{c}$ defines the scale at which $\delta_{m}$ changes from a $1/r$ drop-off to a $1/r^3$ one, and the characteristic density contrast $\delta_c$ is given by:
$$
\delta_{c} = \frac{200}{3} \frac{c^3}{\ln(1+c) - c/(1+c)},
$$
where $c = R_{200}/R_{c}$ defines the concentration of the halo, and
$R_{200}$ is roughly the virial radius and is defined to be the
largest value of $R$ for which the mean density contrast is $\geq
200$.  The mean density profile, $\bar{\delta}$, is then given by:
$$
\bar{\delta} = \frac{3\delta_{c}(\ln(1+R/R_c) -
R_c/(1+R/R_c))}{(R/R_c)^3}.
$$
Even though $\bar{\delta} \propto 1/R$ for $R < R_{c}$, the integrands
in Eq. (\ref{deltaphinon}) still behave well enough for small $R$ for
our analysis and our expression for $\delta_{\phi}$ to be valid and
applicable.  The DDE density contrast in a virialised cluster is given
by Eq. (\ref{deltaphinon}).  We note that since $\delta v^2 \propto h
\propto \Omega_m^{1/2}$, in the non-linear regime, and $\bar{\delta}$
depends on $R$ only though $R/R_{c}$ that, for fixed $\delta_c$,
$\delta_{\phi} \propto \Omega_m(1+w)H^2 R_c^2$.

\citet{geller} studied the mass profile of the Coma
galaxy cluster and found that over the entire range of $R <
10\,\mathrm{Mpc} \,h^{-1}$ it increased with $r$ at the rate predicted
by the NFW profile.  They found that three different observational
samples were well fitted by $R_{c} = 0.182 \pm 0.030$, $R_c = 0.167\pm
0.029$ and $0.192\pm 0.035 h^{-1}\,\mathrm{Mpc}$ with, in all cases,
$R_{200} = 1.5 h^{-1}\,\mathrm{Mpc}$. In what follows we take $R_c$
to be the mean of three values, $R_c \approx 0.18
h^{-1}\,\mathrm{Mpc}$, and $R_{200} = 1.5 h^{-1}\,\mathrm{Mpc}$, which
corresponds to $c \approx 8.3$.  We plot the profile of DDE density
contrast for a Coma-like cluster in Figure \ref{comaplot}; $R=R_c$
and $R=R_{200}$ are indicated on the plots.  We have taken $\Omega_m
=0.27$ in line with WMAP \citep{wmap,wmap3}. We note that $\delta_{\phi}$ is
positive around $R=0$ and also that it is very small
$\delta_{\phi}(R=0) \approx 3 \times 10^{-5}$. As $R$ increases,
$\delta_{\phi}$ becomes negative when $R \approx 1.5R_{200}$, and for
$R \gtrsim 4R_{200}$ it tends to $0$ from below as $R \rightarrow
\infty$ in line with our expectations.

\subsection{Superclusters}
Superclusters are the largest known gravitationally bound massive
structures.  The dark matter halos of superclusters typically have
radii of about $10-25 h^{-1}\,\mathrm{Mpc}$ and density contrasts of
about $1 \lesssim \bar{\delta} \lesssim 15$.  We consider two examples
of such objects: the local supercluster (LSC), of which our galaxy is
a part, and the Shapley supercluster (SSC) which lies about
$650\,\mathrm{Mlys}$ away.  The LSC has a mean overdensity of
$\bar{\delta}\approx 2-3$ over a scale $\sim 15h^{-1}\,\mathrm{Mpc}$
\citep{hoffman, tully}.  The SSC has been found to have an overdensity
of $\bar{\delta} \sim 10.3$ over a scale of $10.1h^{-1}\,\mathrm{Mpc}$
\citep{bardelli}.  \citet{bardelli} also found that if
the SSC had evolved linear then it would today have a linear
overdensity of $\delta_{SSC\,{\rm lin}} \approx 1.3$ and a linear scale of
$\approx 21.6h^{-1}\,\mathrm{Mpc}$.  For simplicity we model the
matter density contrast of the clusters thus:
\begin{itemize}
\item The clusters have a homogeneous core with physical radius
$R_{c}$.  In $R < R_{c}$, $\bar{\delta}=\bar{\delta}_{0}$ and
$\bar{\delta}_{{\rm lin}}=\delta_{{\rm lin}\,0}$.
\item For $R>R_c$, the initial density contrast drops off as $1/x^3$
where $R = R_{c} x (1+\bar{\delta}_{0})/(1+\bar{\delta}(R))^{1/3}$.
\end{itemize} With this choice of density contrast, the evolution of the
cluster is free from both shell-crossing and shell-focusing
singularities in the linear and quasi-linear regimes. For the LSC we
take $\bar{\delta}_{0} = 2$ and $R_c = 15h^{-1}\,\mathrm{Mpc}$ and for
the SSC we take $\bar{\delta}_0 = 10$ and $R_c =
10h^{-1}\,\mathrm{Mpc}$.  We additionally considered the S300 structure,
in the vicinity of the SSC, which was identified by \citet{bardelli}.  This structure has a mean overdensity of
$\bar{\delta}\approx 1.9$ on a scale of $24.8h^{-1}\,{\rm Mpc}$
\citep{bardelli}; we take $\bar{\delta}_{0} = 2$, $R_c =
25h^{-1}\,\mathrm{Mpc}$ for this structure.

\begin{figure}
\begin{center}
\includegraphics[width=80mm]{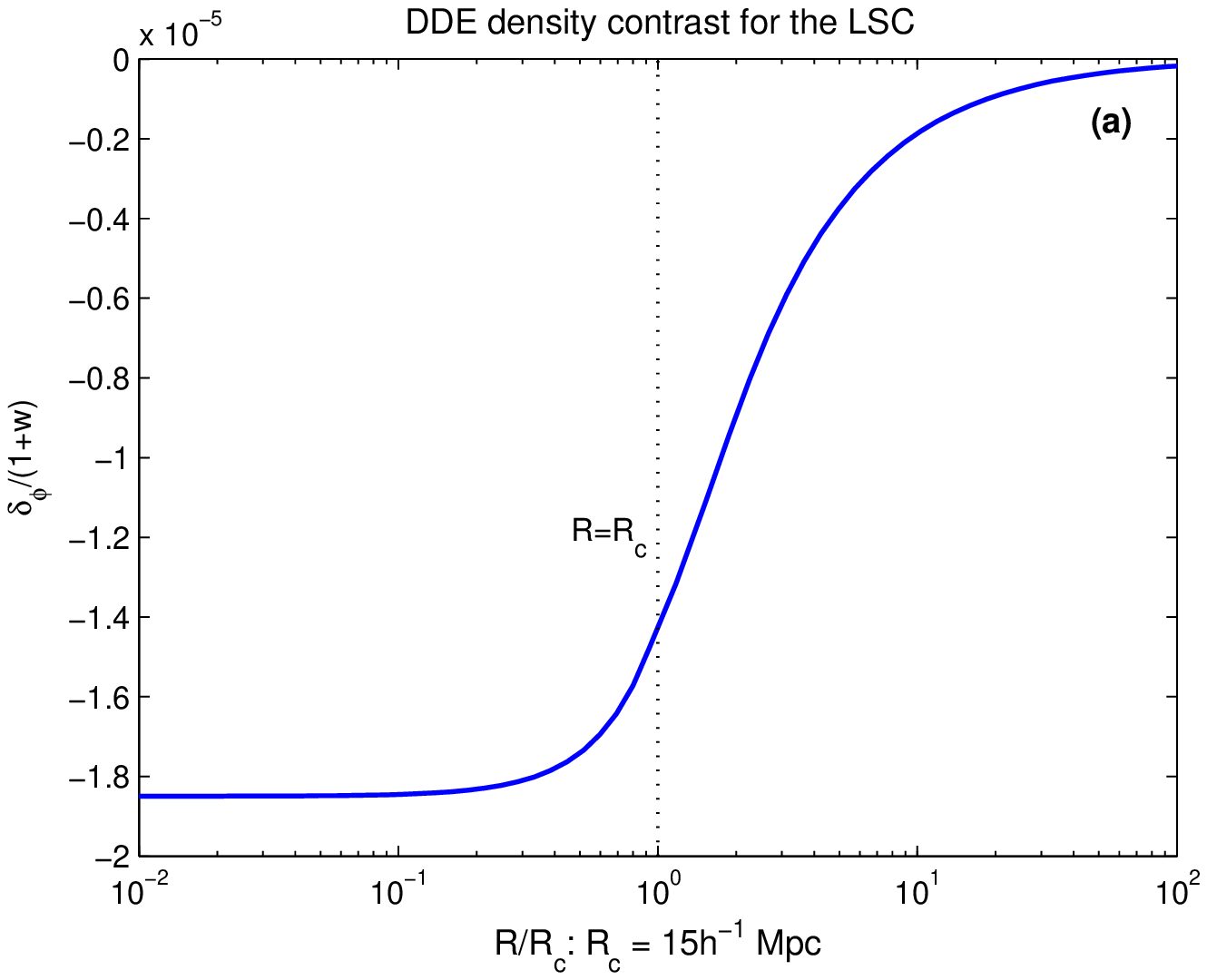}
\includegraphics[width=80mm]{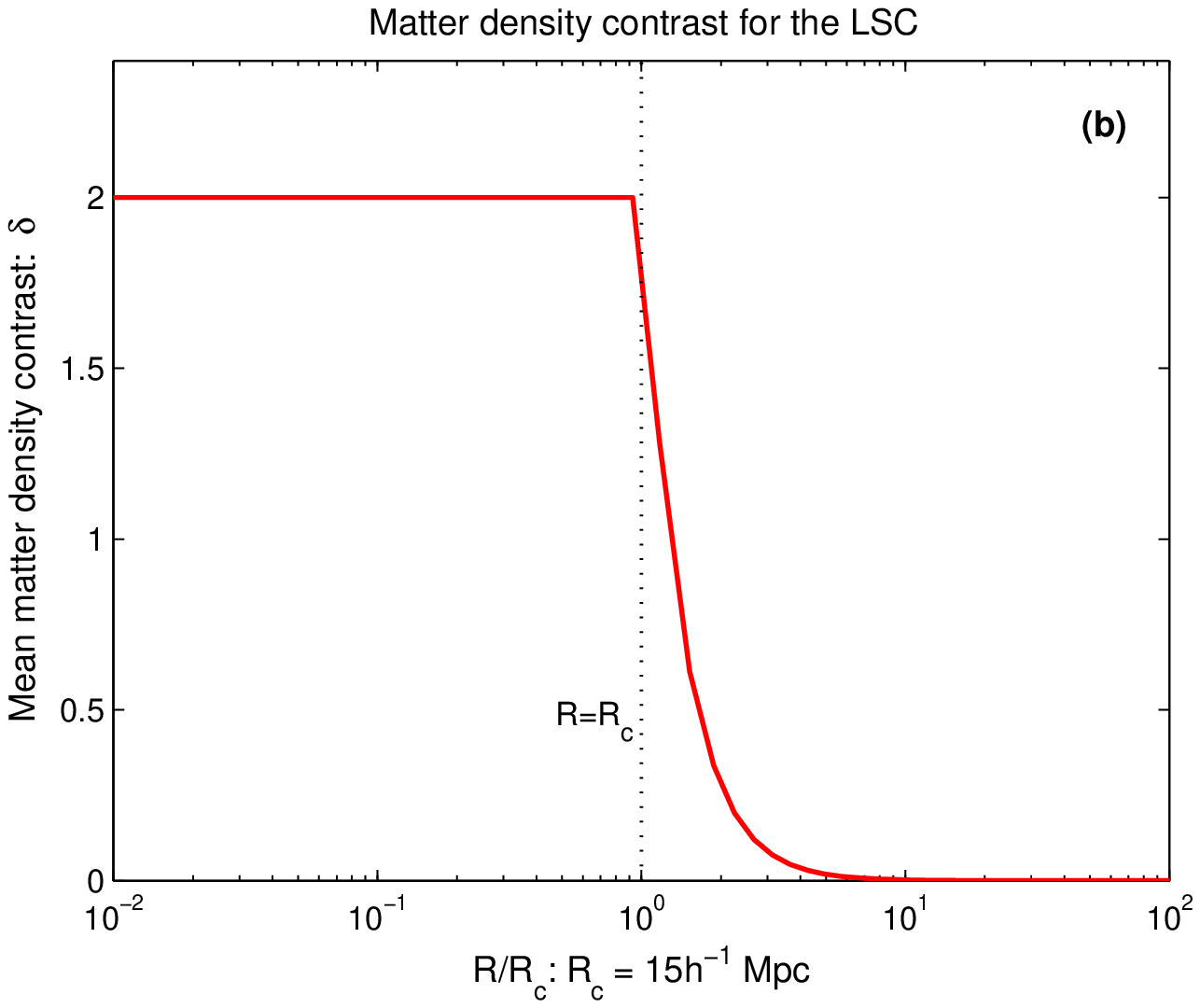}
  %% to include a figure, or
  %% to leave a blank space
  \end{center}
 \caption{The profiles of DDE and matter density contrasts for superclusters. We have taken $\Omega_m = 0.27$. 
 Figures (a) the DDE for the local supercluster (LSC) and Figure (b) shows our model for the LSC's matter density profile. 
We have taken the LSC to have a mean matter density contrast of $2$ inside a radius of $15h^{-1}\,\mathrm{Mpc}$.}
 \label{SCplot}
 \end{figure}

\begin{figure}
\begin{center}
\includegraphics[width=80mm]{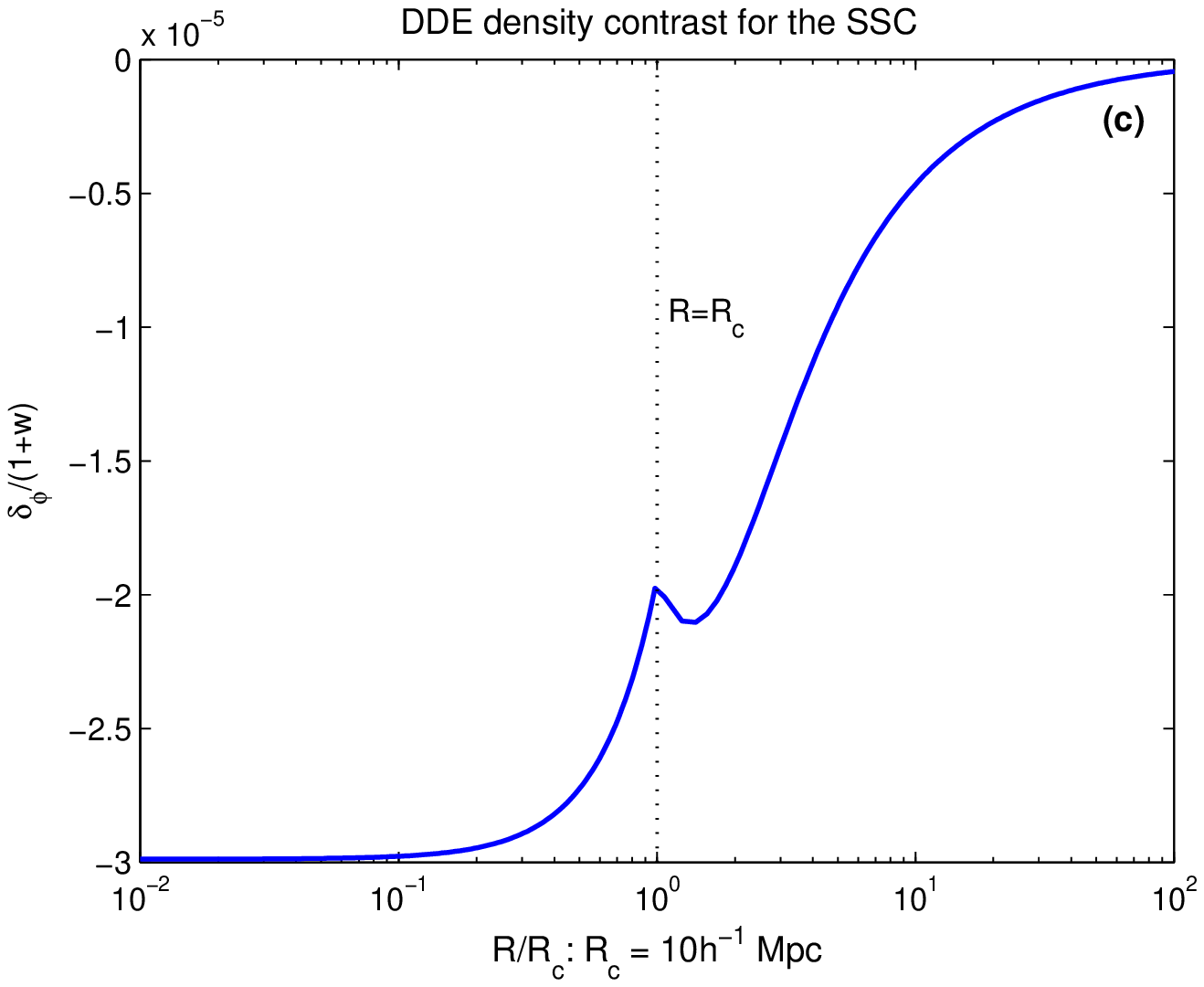}
\includegraphics[width=80mm]{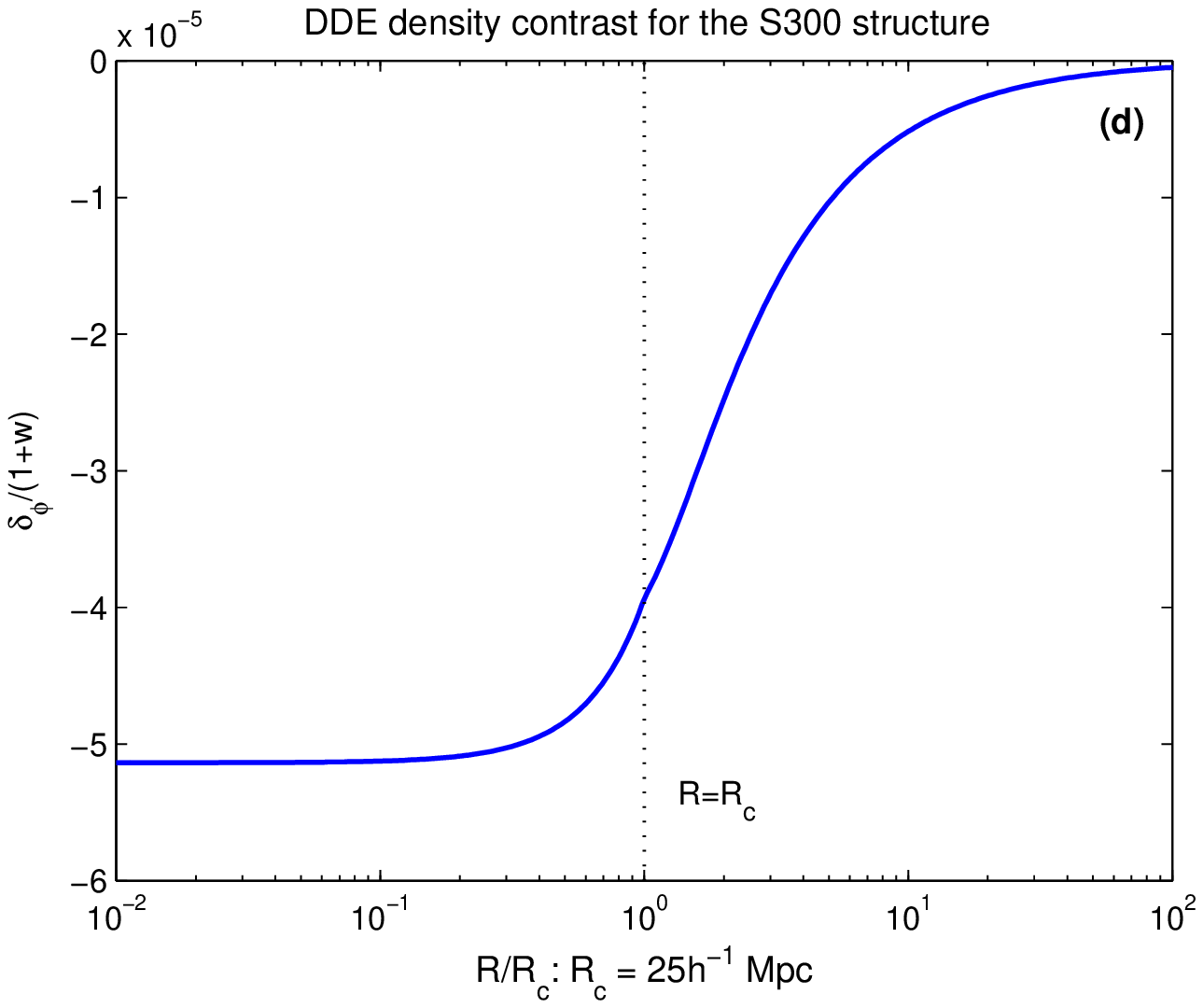}
  %% to include a figure, or
  %% to leave a blank space
  \end{center}
 \caption{The profiles of DDE and matter density contrasts for the Shapley supercluster (SSC). We have taken $\Omega_m = 0.27$. 
Figure (a) shows the DDE profile for the Shapley supercluster, which has been modelled as having a mean density contrast of $10$ 
inside a radius of $10h^{-1}\,\mathrm{Mpc}$. Figure (b) shows the DDE profile  the S300 structure (with mean density contrast of $2$ 
inside a radius of $25h^{-1}\,\mathrm{Mpc}$).}
 \label{SCplot2}
 \end{figure}

We plot the DDE density contrast profile for the LSC, SSC and S300
structures in Figure \ref{SCplot} and  Figure \ref{SCplot2}.  The shape of $\delta_{\phi}$ is
similar for all three objects, and the magnitude of the
DDE density contrast scales as $(1+w)\bar{\delta}_0 H^2 R_c^2$; it is
largest for the S300 structure and smallest for the LSC. In all three
cases, it is clear
that $\vert \delta_{\phi} / (1+w) \vert$ is small ($\lesssim 6.2
\times 10^{-5}$) and negative which corresponds to a local DDE void, 

\subsection{Voids of Matter}
In addition to local overdensities of matter, the Universe also
contains localized underdensities of matter or voids.  Voids typically
have the similar radii to superclusters, although they can also be much larger.
Since $\vert \delta_{\phi}/(1+w)\vert$ is proportional to the square
of the radius of the inhomogeneity, we expect voids to induce some of
the largest DDE inhomogeneities.  The classic historical example of a
void is also one of the largest and was discovered in 1981 by \citet{kirscher} in the constellations of Bo\"otes and Corona
Borealis.  Recent measurements of this Bo\"otes void have found it to
be roughly spherical with radius $62h^{-1}\,\mathrm{Mpc}$
\citep{kirscher2}.  21 galaxies have been observed in the Bo\"otes void
and its mean density contrast is estimated to be: $-0.80 <
\bar{\delta} < -0.66$ \citep{dey}.  The Bo\"{o}tes void is particularly
large and is sometimes termed a \emph{supervoid}.  Data from the
2dFGRS shows that the average radii of voids in NGP (North Galactic
Pole) and in SGP (South Galactic Pole) are $14.89\pm
2.67h^{-1}\,\mathrm{Mpc}$ and $15.61\pm 2.48h^{-1}\,\mathrm{Mpc}$
respectively \citep{hoyle, bolejko}.  The average mean density contrast
for these voids is $\bar{\delta} = -0.94 \pm 0.02$ in NGP and
$\bar{\delta} = -0.93 \pm 0.02$ in SGP \citep{hoyle}. 

Despite the large radii of these voids and supervoids, their scales
are still very much sub-horizon and, as such, our results are
applicable to them.  The evolution of voids is well approximated by
the quasi-linear regime and so we evaluate $\delta_{\phi}$ using
Eq. (\ref{deltaphiql}).  We model both voids and supervoids as having a
Gaussian initial mean density profile i.e. $\bar{\delta}_{\rm lin} \propto
\exp(-x^2/2)$ where $R = R_{c} x
(1+\bar{\delta}_{0})/(1+\bar{\delta}(R))^{1/3}$ defines $x$. With this
choice the evolution of the void is free from both shell-crossing and
focusing singularities provided $\bar{\delta}_{0} \gtrsim -0.95$.

\begin{figure}
\begin{center}
\includegraphics[width=80mm]{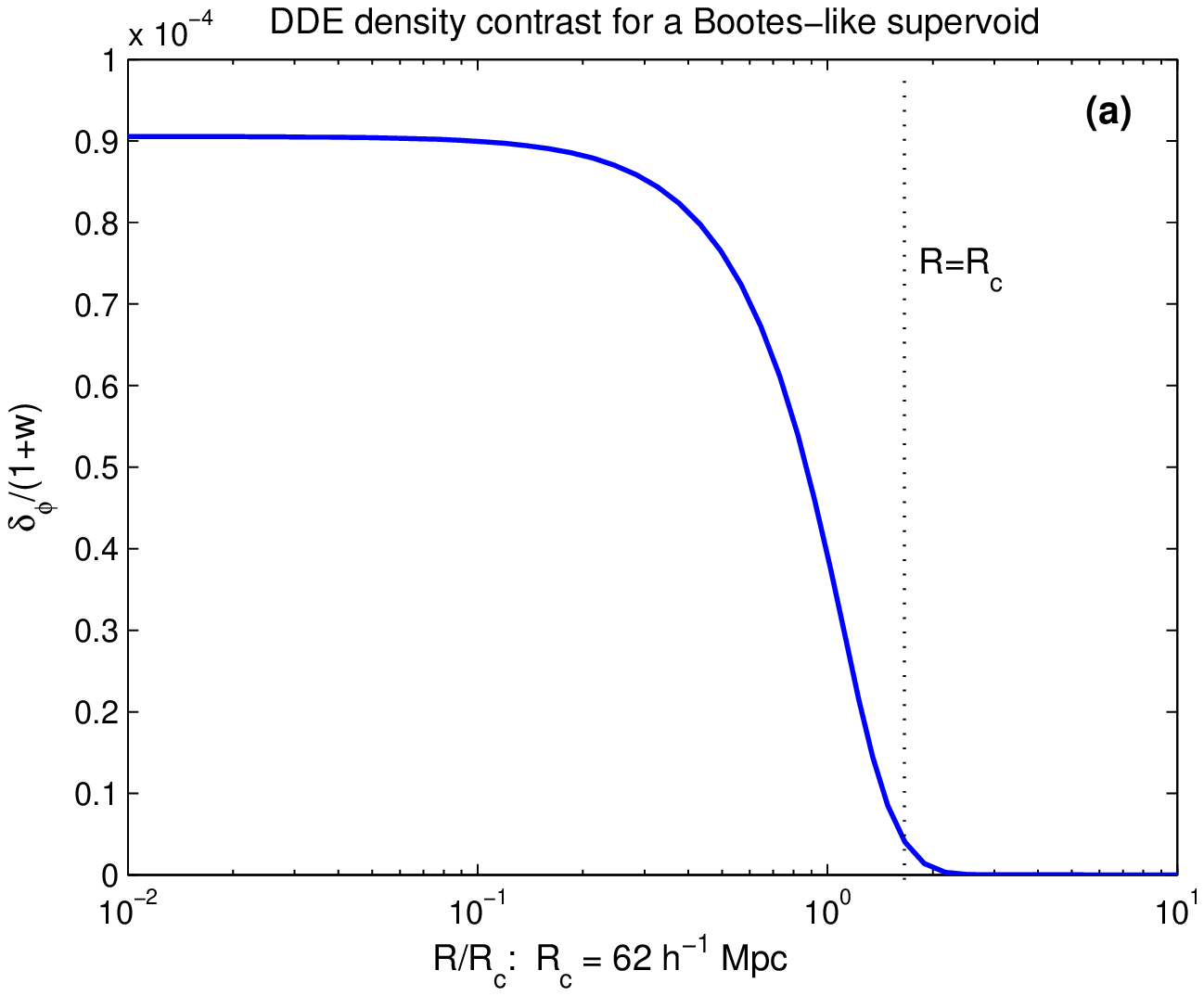}
\includegraphics[width=80mm]{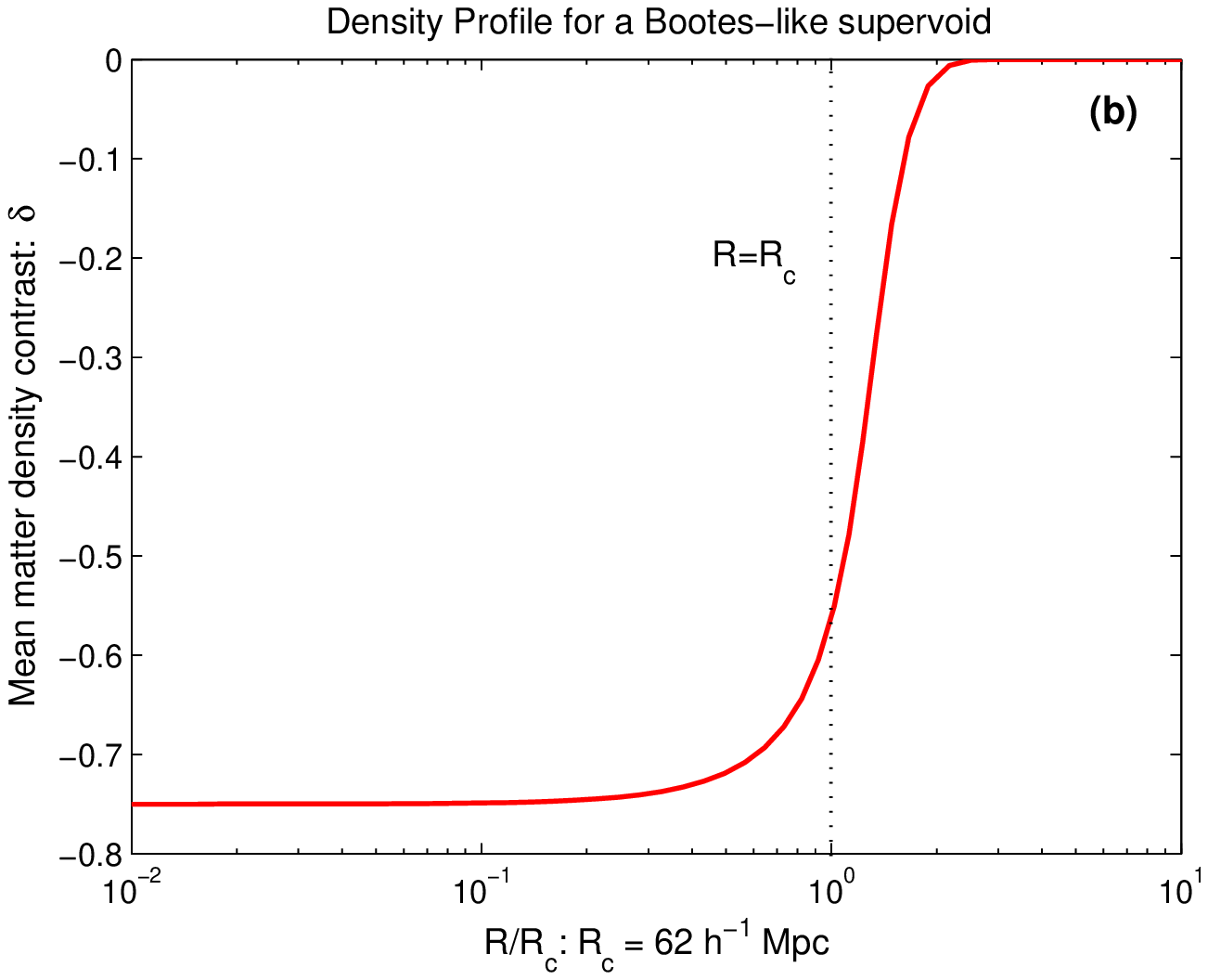}
  %% to include a figure, or
  %% to leave a blank space
  \end{center}
 \caption{The DDE (a) and matter (b) density contrast profiles for a supervoid. We taken $\Omega_m = 0.27$ and modelled the supervoid on the Bo\"{o}tes void: $\bar{\delta}(0) = -0.75$ and $R_{c} = 62h^{-1}\,\mathrm{Mpc}$. See text for further discussion.}
 \label{SVoidplot}
 \end{figure}

\begin{figure}
\begin{center}
\includegraphics[width=80mm]{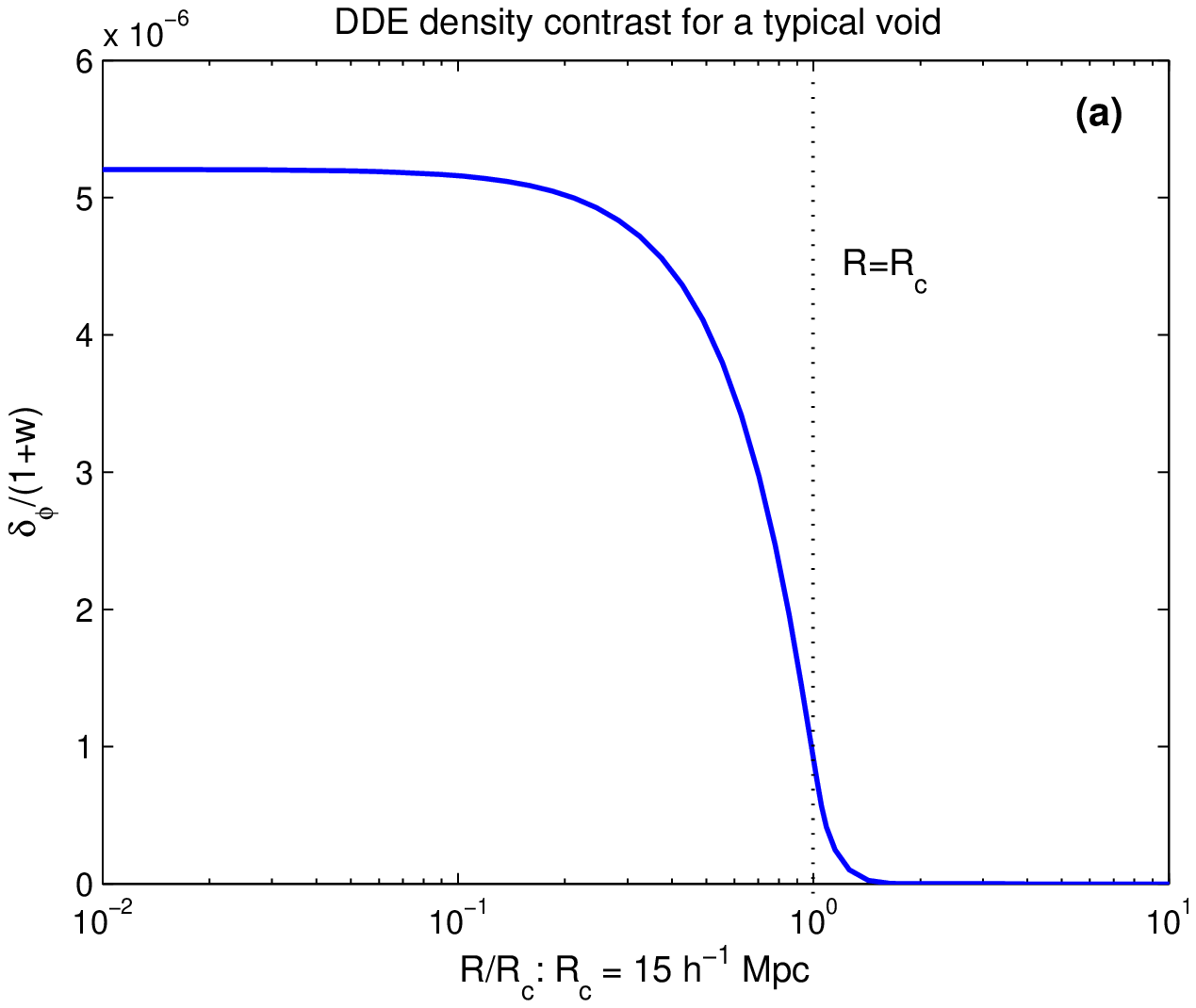}
\includegraphics[width=80mm]{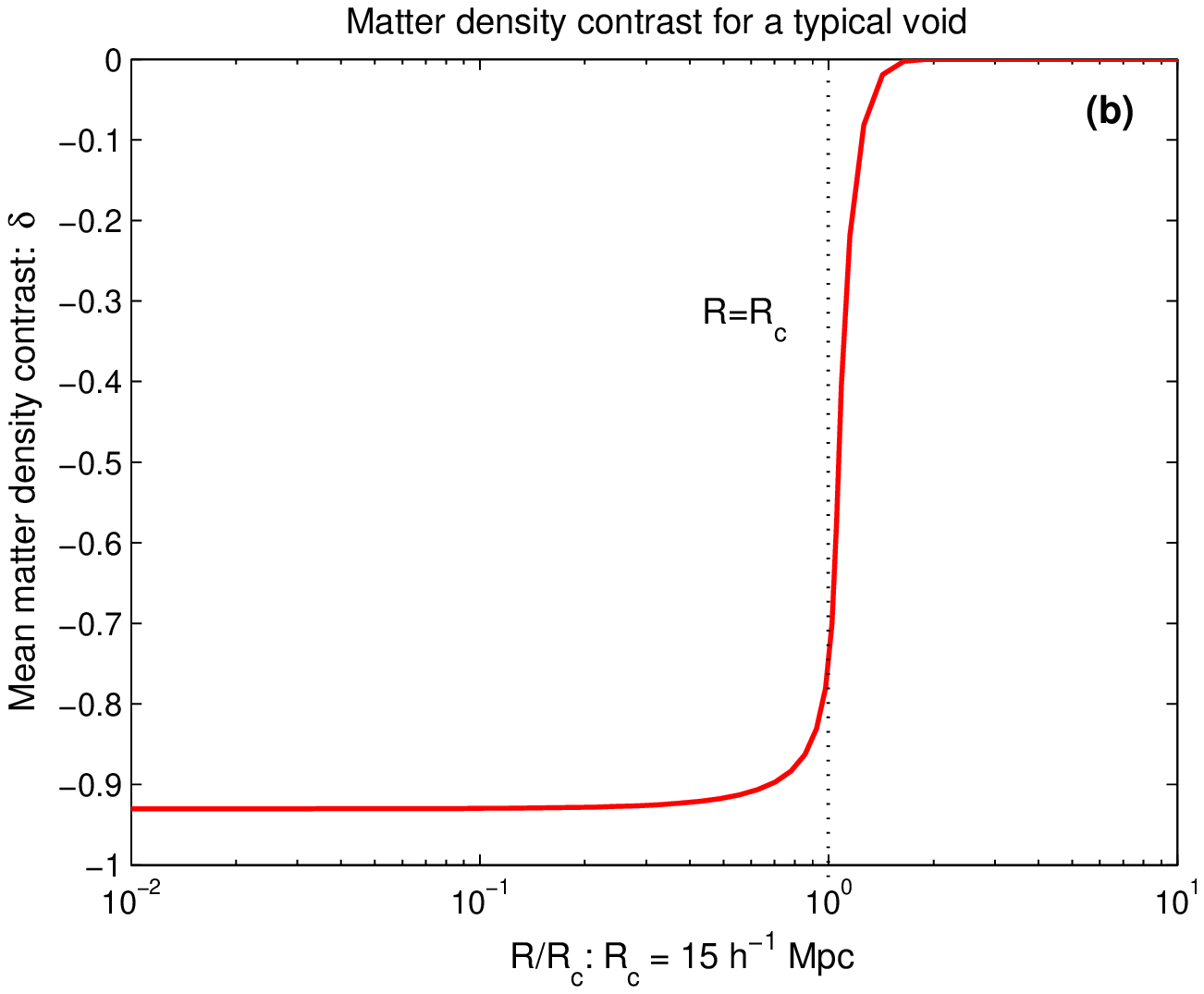}
  %% to include a figure, or
  %% to leave a blank space
  \end{center}
 \caption{The DDE (a) and matter (b) density contrast profiles for a typical void. We have taken $\Omega_m = 0.27$ and the void to have typical properties: $\bar{\delta}(0) = -0.93$ and $R_{c} = 15h^{-1}\,\mathrm{Mpc}$. See text for further discussion.}
 \label{Voidplot}
 \end{figure}

In Figure \ref{SVoidplot} we show $\delta_{\phi}$ and $\bar{\delta}$ for both a supervoid with similar properties to the Bo\"{o}tes void ($R_{c} = 62h^{-1} \mathrm{Mpc}$, $\bar{\delta}_0 = -0.75$), and Figure \ref{Voidplot} shows the same thing but for an average void ($R_{c} = 15h^{-1}\,\mathrm{Mpc}$, $\bar{\delta}_0 = -0.93$). In both cases we see, as expected, that $\delta_{\phi} > 0$ and that a void in the matter distribution corresponds to a DDE overdensity. We note that the magnitude of $\vert \delta_{\phi}/(1+w) \vert$ is an order of magnitude larger for a $60h^{-1}\,\mathrm{Mpc}$ radius supervoid than for an average  $15h^{-1}\,\mathrm{Mpc}$ void. At the centre of a supervoid we see that $\delta_{\phi} \approx  9 \times 10^{-5}(1+w)$, whereas at the centre of an average void $\delta_{\phi} \approx 5 \times 10^{-6}(1+w)$.   In all cases $\delta_{\phi}$ is both positive and very small.

\begin{figure}
\begin{center}
\includegraphics[width=80mm]{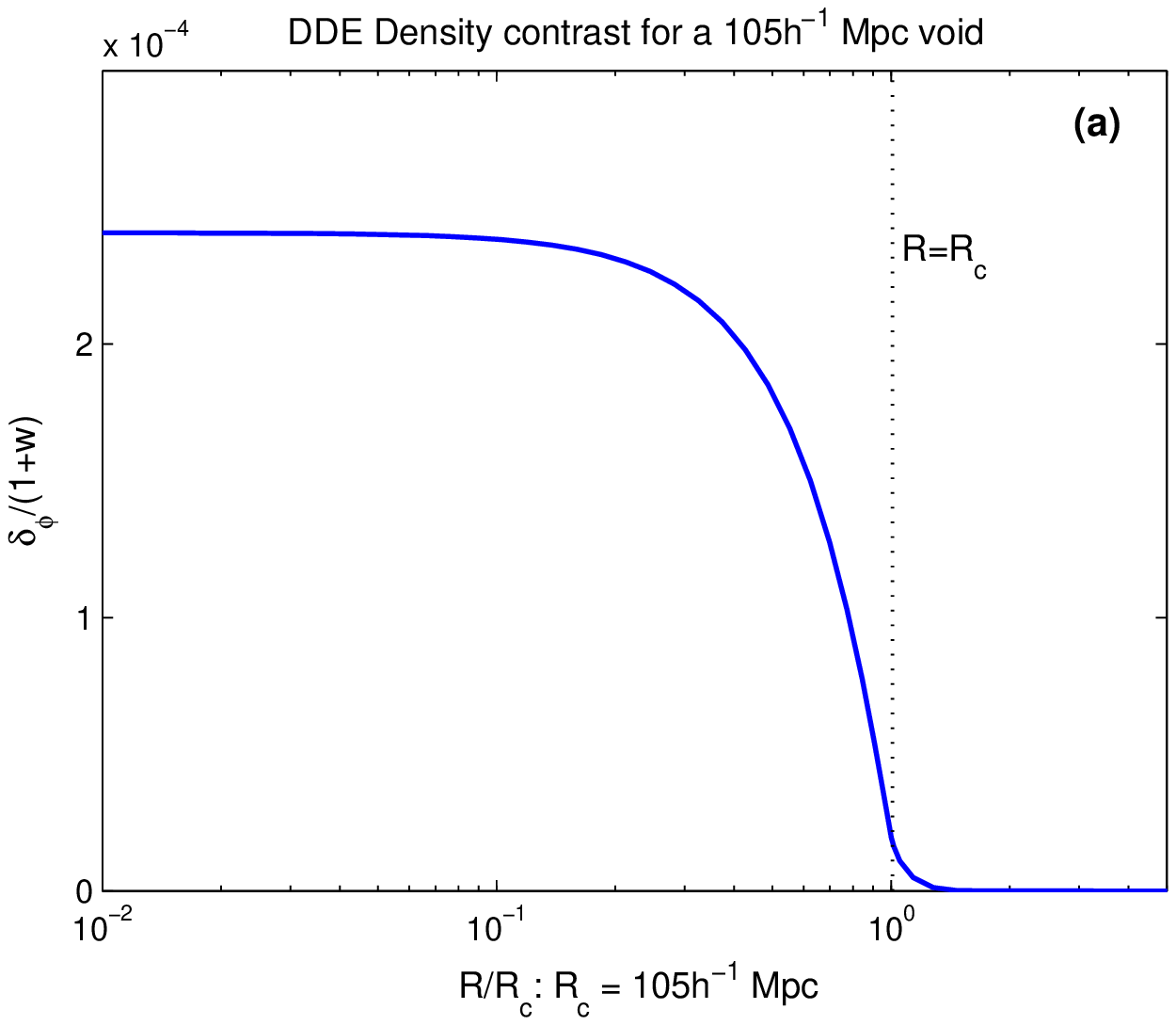}
\includegraphics[width=80mm]{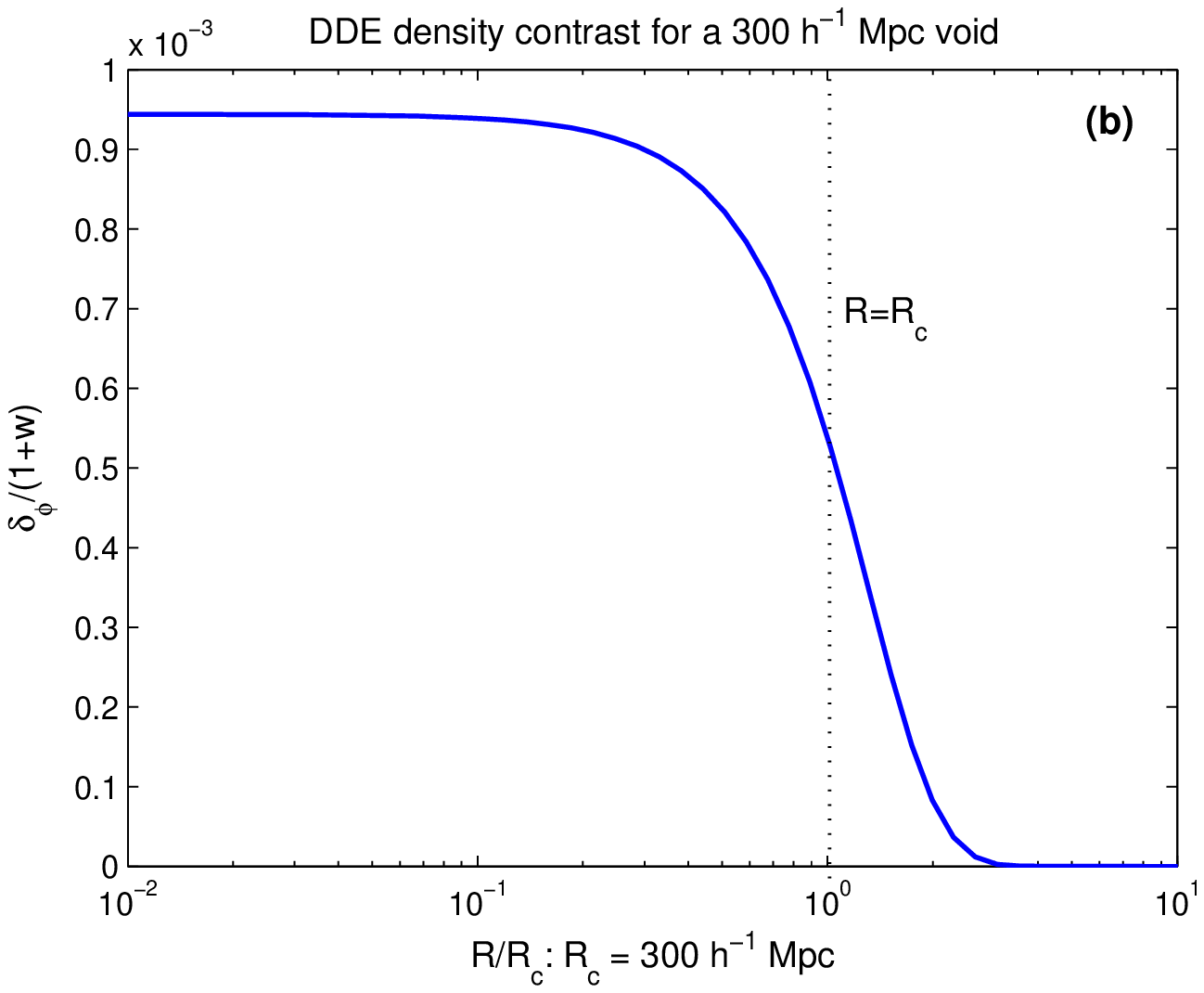}
  %% to include a figure, or
  %% to leave a blank space
  \end{center}
 \caption{The DDE density contrast for extremely large voids. In FIG. (a) we consider a $105 h^{-1}$~Mpc void with $\bar{\delta}(0)=-0.95$ and in FIG. (b) we have taken void to have radius $300 h^{-1}$~Mpc and central density contrast $\bar{\delta}(0) = -0.3$. The existence of voids with these properties as recently been postulated as a mean to explain the observed large-angle CMB anomalies.}
 \label{HBubbleFig}
 \end{figure}

Recently, there has been a great deal of discussion, both theoretical
and observational, about the possibility that extremely large voids
($R_{c} \approx 100-300 h^{-1}\,\mathrm{Mpc}$) exist within the
visible Universe \citep{WMAPcold,SilkVoid}. Additionally, a number of authors have
speculated that we may even be living inside such an object
\citep{HubbleBubble,HubbleBubble1,HubbleBubble2,HubbleBubble3}. For
example, \citet{WMAPcold} showed that both the WMAP cold spot
and an observed dip in the surface brightness and number counts of
extragalactic radio sources could be explained by the presence of an almost
completely empty $R_c \approx 105h^{-1}\,\mathrm{Mpc}$ at $z <1$. \citet{SilkVoid} showed that the ISW effect due to a void
with $\bar{\delta}_0 = -0.3$ and $R_{c} = 200 -300 h^{-1} \,
\mathrm{Mpc}$ would be observed as cold spots with a temperature
anisotropy $\Delta T/T\sim \mathcal{O}(10^{-5})$, which might explain
the observed large-angle CMB anomalies.  If extremely large voids such
as these exist they would be associated with the largest DDE
overdensities within the visible Universe. In Figure \ref{HBubbleFig}
we plot the $\delta_{\phi}$ profile for both the $R_{c} =
105h^{-1}\,\mathrm{Mpc}$, $\bar{\delta}_0 \approx -1$ object
postulated by \citet{WMAPcold} and for the $R_{c} =
300h^{-1}\,\mathrm{Mpc}$, $\bar{\delta}_0 \approx -0.3$ void
considered by \citet{SilkVoid}. In the former case we find $\delta_{\phi}(R=0)/(1+w) \approx 2.4 \times 10^{-4}$ and in the latter $\delta_{\phi}(R=0)/(1+w) \approx 9.5 \times 10^{-4}$.  These values of $\delta_{\phi}/(1+w)$ are about two orders of magnitude larger than those associated with a typical void. However, even around such extremely large voids we still find that the DDE perturbation is small: $\delta_{\phi}/(1+w) \lesssim 10^{-3}$.  

\subsection{ISW Effect due to DDE clustering}
\label{appA}
We have seen that the DDE density contrast,
$\delta_{\phi}$, is always small with magnitude $\lesssim 10^{-5}$ for
average sized structures in the Universe.  The Integrated Sachs Wolfe
(ISW) effect is highly sensitive to the presence of dark energy.  In a
purely matter dominated universe, the ISW effect vanishes entirely for
linear perturbations. We shall see that the DDE contribution to the
ISW from a cluster with radius $R_{c}$ is proportional to $(HR_c)^{5}$
and so we consider only large (but still sub-horizon) scales.  On such
scales the matter perturbation is small and we are in the linear
regime.  The ISW effect is most easily derived in the conformal
Newtonian gauge in which the metric takes the form:
$$
\diff s^2 = a^2 \left[(1+2\Psi)\diff \tau^2  - (1-2\Phi)\diff \mathbf{x}^2\right].
$$
Since we are dealing with shearless matter, to leading order, $\Psi = \Phi$ and, in momentum space:
\begin{eqnarray}
k^2 \Phi  &=&  -\frac{3 \Omega_m {\mathcal{H}}^2}{2} \delta_{m} - \frac{3 (1-\Omega_m) {\mathcal{H}}^2}{2} \nonumber \\ && \left(\delta_{\phi} +  \frac{3(1+w)\mathcal{H}^2 f(a) \delta_{m}}{k^2}\right)
\end{eqnarray}
where ${\mathcal{H}} = a H = \diff \ln a / \diff \tau$ and $\delta_{m}$ and $\delta_{\phi}$ are defined in the rest frame of the matter particles.   
Using Eq. (\ref{linmtm}) we see that:
\begin{eqnarray}
k^2 \Phi &=& -\frac{3 \Omega_m {\mathcal{H}}^2}{2} \delta_{m} - \frac{9(1+w) (1-\Omega_m)\Omega_m {\mathcal{H}}^4 \delta_{m}}{4k^2} \nonumber \\ &=& -\frac{3 \Omega_m {\mathcal{H}}^2}{2} \left(\delta_{m} - \frac{(1-\Omega_m)}{2f(a)-\Omega_{m}}\delta_{\phi}\right).
\end{eqnarray}
The temperature anisotropy due to the ISW effect is given by:
$$
\delta_{ISW}(\hat{\mathbf{n}}) = 2\int_{\tau_{LS}}^{\tau_0} \Phi^{\prime}((\tau_0-\tau)\hat{\mathbf{n}},\tau)\,\diff \tau,
$$
where $\tau_0$ is the value of $\tau$ today and $\tau_{LS}$ is its value on the surface of last scattering and $\Phi^{\prime} = d \Phi / d \eta$.  The contribution to $\Phi^{\prime}$ from the matter perturbation is:
$$
\Phi^{\prime}_{m} = \frac{3 \Omega_m {\mathcal{H}}^3}{2} \left(1-f(a)\right)\delta_{m},
$$
and $f(a) \approx \Omega_m^{0.6}$ at late times, whereas the contribution to $\Phi^{\prime}$ from the DDE clustering is:
\begin{eqnarray*}
\Phi^{\prime}_{\phi} &=& \frac{9(1-\Omega_m)\Omega_m {\mathcal{H}}^5 \delta_m}{4k^4}\left((1+w)(2+3w - f(a))-w_{p})\right) \\ &=& -\frac{3\Omega_{m}\mathcal{H}^3(1-f(a))}{2k^2} V(\Omega_m,w,w_{p}) \delta_{\phi},
\end{eqnarray*}
where $w_{p} = \diff w / \diff \ln a$ and
$$V(\Omega_m,w,w_{p}) = \frac{(1-\Omega_m)(2+3w-f(a)-w_{p}/(1+w))}{(2f(a)-\Omega_m)(1-f(a))}.$$
It follows that:
\begin{eqnarray*}
\frac{\Phi^{\prime}_{\phi}}{\Phi^{\prime}_{m}} &\approx& -\frac{3(1+w){\mathcal{H}}^2}{2 k^2} \left[\frac{1-\Omega_m}{1-\Omega_m^{0.6}}\right]  \nonumber \\ &&\left(1+\Omega_m^{0.6} + \frac{w_{p}}{1+w}-3(1+w)\right),
\end{eqnarray*}
assuming that $w_{p}/(1+w) \sim \Oo(1)$ or smaller in the recent past then:
$$
\left \vert \frac{\Phi^{\prime}_{\phi}}{\Phi^{\prime}_m} \right \vert \lesssim \frac{3(1+w){\mathcal{H}^2}}{k^2},
$$
We found above that the largest DDE perturbations would mostly likely be associated with large voids of matter, with $\delta_{\phi} \approx 9.5\times 10^{-4}(1+w)$ for a $300\,\mathrm{Mpc}$ void with central matter density contrast $\delta_{0}=-0.3$ \citep[such as that considered by][]{SilkVoid}.  In this case we would have:
$$
\frac{\Phi^{\prime}_{\phi}}{\Phi^{\prime}_m} \approx -7\times 10^{-3}(1+w)\left(1.5 - 3(1+w) + \frac{w_{p}}{1+w}\right),
$$
which is at most $\mathcal{O}(-10^{-2}(1+w))$.  In the absence of DDE clustering, the ISW temperature anisotropy of such a void was found to be $\vert \Delta T/ T \vert \sim \mathcal{O}(10^{-5})$. We therefore predict that the DDE clustering contribution to the temperature anisotropy would be at most $\mathcal{O}(10^{-7}(1+w))$.  For smaller objects (with radius $R_c$) the DDE ISW temperature anisotropy scales as $R_c^4$. 

In conclusion: the ISW due to sub-horizon DDE clustering in models where the dark energy is due to a minimally coupled scalar field with canonical kinetic term is always suppressed by a factor of about $(1+w) (HR_{c})^2$  relative to the ISW effect to the matter perturbation itself.  It therefore seems unlikely that the DDE contribution detected against the background of the ISW effect caused by the matter.

\subsection{Discussion} In this section we have applied our results to
evaluate the profile and evolution of $\delta_{\phi}/(1+w)$ in the
vicinity of typical galaxy clusters, superclusters and voids.  We
found that, today, the largest inhomogeneities in the DDE energy
density value should occur in the vicinity of supervoids such as the
one in Bo\"{o}tes \citep{kirscher}.  At the centre of a supervoid we
predict $\delta_{\phi}/(1+w) \approx 1.6 \times 10^{-4}$.  At the
centre of a galaxy cluster, supercluster or average sized void, we
found $\vert\delta_{\phi}/(1+w)\vert \approx 1-6 \times 10^{-5}$; with
$\delta_{\phi} > 0$ for both virialised clusters and voids, and
$\delta_{\phi} < 0$ for superclusters.  Generally $\vert \delta_{\phi}
/(1+w) \vert \lesssim 10^{-4}$ for supervoids, and $\lesssim 1 - 6
\times 10^{-5}$ for other structures.  Recently there has been a fair
amount of interest in the possibility that extremely large voids might
exists with radii $R_c = 100 - 300\,\mathrm{Mpc}$.  Even for objects
as large as this, however, we found that $\delta_{\phi}/(1+w) \lesssim
10^{-3}$.

Given the current bounds on the dark energy (EoS) parameter: $w(z<1) =
-1\pm 0.1$, our results imply that if dark energy is described by a
canonical scalar field that does not couple to (or only couples very weakly to) matter and is minimally
coupled to gravity, then its energy density today is virtually
homogeneous with largest fluctuations no larger than 1 part in
$10^{5}$ (or possibly one part in $10^{4}$ if $300\,\mathrm{Mpc}$
voids exist). Indeed, typical fluctuations in the DDE energy density will
be smaller than a few parts in $10^{6}$. We found that superclusters
are generally associated with small, local DDE voids whereas most
other structures (galaxies and voids of matter) are associated with
local overdensities of dark energy.

\section{Summary and Conclusions}
\label{sec:con} In this paper we have studied how dynamical dark
energy clusters over sub-horizon scales in theories where the dark
energy is described by a canonical scalar field with no coupling to
matter.  Sub-horizon dark energy clustering in such models has always
been expected to be small, but previous studies of it have assumed the
matter perturbation to be small and have, for the most part relied on
numerical simulations.  In this paper, however, we used the powerful
method of matched asymptotic expansions to derive an analytic
expression, Eq. (\ref{deltaphimain}), for the DDE density contrast in
terms of the peculiar velocity and density contrast of the matter.
Our equation for $\delta_{\phi}$ holds not only in the linear regime,
where the matter density contrast is small, $\delta_{m} \ll 1$, but
also in the far more interesting but far less studied quasi-linear and
non-linear regimes where $\delta_{m} \sim \Oo(1)$ or greater.

There were then essentially two parts to this paper, finding an
expression for $\delta_{\phi}$ as a function of $\delta v$, $\delta_m$
and $H$, and then evaluating this.  The evaluating part is only done
approximately for models that look sufficiently like $\Lambda CDM$. So
that to a first approximation we can use $\Lambda CDM$ to get $H$,
$\delta v$ and $\delta_m$.  Different dark energy models would give
different predictions but just from the form of $\delta_{\phi}$ one
can see that unless very large deviations from the $\Lambda CDM$
predictions of $\delta v$, $\delta_m$ and $H$ occur then
$\delta_{\phi}$ will be well approximated by taking a $\Lambda CDM$
background.

Our key assumption in deriving our formula for $\delta_{\phi}$ was that the
matter density contrast is only $\Oo(1)$ or greater on scales that are
small compared to the Hubble length, $H^{-1}$.  This assumption holds
very well for typical structures such as galaxies, clusters,
superclusters and voids.  We also made a number of simplifying
assumptions:
\begin{itemize}
\item We assumed spherical symmetry.  Although our interest is in dark
energy rather than varying-constants, the sub-horizon dynamics of
perturbations in the DDE scalar field are very similar to those of the
scalar fields in varying-constant models provided one sets the matter
coupling in such theories to zero.  \citep{shawbarrow,shawbarrowb,shawbarrowc} studied the evolution of a light scalar
field in an inhomogeneous background in the context of
varying-constant. Both spherically symmetric backgrounds and ones
without any symmetries were considered, and it was found that
deviations from spherically symmetry did not alter the leading order
behaviour of the scalar field perturbations when the matter coupling
vanishes. Therefore, whilst the assumption of spherical symmetry can
be relaxed, we do not expect it to greatly alter our results.
\item We assumed that we were not near a black hole horizon.  This
assumption allowed us to take $GM/R \ll 1$ on sub-horizon scales.
This assumption could also be relaxed, and once again both the form of
$\delta_{\phi}$ and its magnitude are not expected to be greatly
effected.  We again base this expectation on the results found for
light scalars in the context of varying-constants by \citet{shawbarrow1}.
\item We assumed that the mass of the scalar field, $m$, was very
small compared to $1/R_{c}$, where $R_{c}$ in the length scale of the
peak in the matter density contrast.  This assumption could be done
away with, however it would significantly complicate the analysis in
the non-linear regime.  Moreover since we have require $H R_{c} \ll 1$
and in most quintessence models $m \sim \Oo(H)$, it is for most models
generally true that $m R_{c} \ll 1$.  If $m R_{c} \sim \Oo(1)$ or
larger, then we expect that it will result in a smaller DDE density contrast,
i.e. smaller values of $\vert\delta_{\phi}\vert$.
\end{itemize} 

One of the main motivations for this work was the recent work of \citet{dutta}.  They studied the clustering of a subclass of
the DDE models which we have looked at in this work. They numerically
evolved the linearized field equations for both $\phi$ and the
metric. In the regime where the linearized equations are valid
(i.e. $\delta_{m} \ll 1$), our analytical results agree with theirs: shortly after the collapse begins,
$\delta_{\phi}/(1+w)$ is positive, but then as the dying mode
of an overdensity of matter becomes small, $\delta_{\phi}/(1+w)$
changes sign and becomes negative. At late times in the linear
regime an overdensity of matter corresponds to a DDE void, and
vice versa. We saw in this paper that this correspondence continues to
hold in the quasi-linear regime, $-1 \lesssim \delta_{m} \lesssim 10$,
that is appropriate for superclusters and voids of matter.  When
$\delta_{m} \sim \Oo(1)$, Dutta and Maor found that
$\vert\delta_{\phi}\vert$ could grow to be relatively large (of the
order of $0.01 - 0.1$), and that this could lead to corrections to the
equation of state parameter, $w$, of the order of a few percent. However, as we noted in the introduction, the $\Oo(1)$ values of $\delta_{m}$ for which this effect was found fall outside the regime
where the linearized field equations which they used are actually
valid.  
One should point out however, that 
in a matter dominated universe ($\Omega_m\sim  1$), there is 
a clear mapping between the the linear regime of 
an overdensity and the exact fully developed non- 
linear overdensity. Presumably, such  mapping 
should occur also in the presence of a scalar field, 
though the multiple change of sign of the DDE 
perturbation complicate things considerably and 
this relation is not straight forward. 
We also saw that at late times in the linear and quasi-linear regimes $\vert\delta_{\phi}/(1+w)\vert$ grows faster than $\vert \delta_{m} \vert$. If it were not for the onset of truly non-linear behaviour in the evolution of $\delta_{m}$, this would indeed lead to large values of $\vert \delta_{\phi}\vert$ seen at late times by \citet{dutta}.

In our study we found that $\delta_{\phi}/(1+w)$ for realistic
astrophysical objects is very small, indeed it is generally $\lesssim
\Oo(10^{-4})$.  The fast growth in $\delta_{\phi}$ seen by \citet{dutta} for $\delta_{m} \sim \Oo(1)$ should therefore been
interpreted as resulting from applying linearized field equations
outside their realm of validity, rather than a true physical effect.
Furthermore, at very late times when the matter perturbation has
virialised and accretion onto the core has all but ceased, we found
that $\delta_{\phi}/(1+w) \rightarrow \mathrm{const}$, and that
$\delta_{\phi}/(1+w) \sim \Oo(GM_{-2}/R_{-2})$, where $R_{-2}$ is the
smallest radius for which $\bar{\delta}$ decreases as quickly as
$1/R^{2}$, $M_{-2}$ is the mass inside $R_{-2}$.

Whilst it is possible with
highly non-linear potentials, in general it seems improbably that
$O(10^{-5}(1+w))$ values of $\delta_{\phi}$, which in turn  result in order
$O(10^{-5}(1+w))$ values for $\delta w$ will result in any pronounced
non-linear effects especially over sub horizon scales. It is fair to say though that the conditions might
be on the brink of breaking down for a $300h^{-1}$ Mpc void, since $HR \sim
0.1$. But for smaller objects provided the conditions we
place on the potential hold (which essentially preclude small changes
in $\phi$ from resulting in huge changes in $V$, $V'$ or $V''$) then  there is no particular
reason to think that perturbations in $w$ will result in $O(1)$ or greater
corrections to $\delta_{\phi}$.
One should also point out that in the case where the scalar field is
strongly coupled to matter  ( e.g. as in \citep{we1,brax,we2}), then the dark energy perturbations might
not be small.  Since it is natural to expect that when matter starts
to infall into the nonlinear regime it will drag along the dark energy
field.

We also used our analytical expression for $\delta_{\phi}/(1+w)$
to calculate the Integrated Sachs Wolfe (ISW) induced by the DDE
density perturbations. We found
that the ISW due to the DDE perturbations is always much smaller than
the ISW effect due to the matter perturbation in these DDE
models by a factor of roughly $H^2R_{-2}^2$. Indeed, even the ISW effect due to the clustering of DDE near
an extremely large $300\,\mathrm{Mpc}$ is found to result in a
temperature anisotropy no larger than $\Oo(10^{-7}(1+w) \lesssim
10^{-8})$.  For smaller objects the ISW effect due to DDE clustering
scales as $R_c^4$ (where $R_c$ is the radius of the cluster).  It is
therefore unlikely that, if dark energy is described by a simple
uncoupled quintessence theory, DDE perturbations will be detected through
the ISW effect.

In summary: we have derived an analytical expression for the DDE
density contrast in uncoupled quintessence models and used it to make
quantitative predictions for DDE clustering.  Our results should also apply to fields with very weak matter coupling. This analysis could also be straightforwardly extended to models that are coupled to matter but we
leave this to a later work.  If the mass of the scalar in these models
is $\Oo(H)$, then $\delta_{\phi}/(1+w)$ is, to leading order,
independent of the specifics of the underlying theory.  As opposed to
what has been suggested elsewhere, we found that
$\delta_{\phi}/(1+w)$ is always small compared to the matter density
contrast, and that even when the matter perturbation goes non-linear,
$\delta_{\phi}/(1+w)$ remains $\lesssim 10^{-3}$ for typical astrophysical objects. If we define the radius, $R_c$,
of a matter over/under - density by the condition that at $R=R_c$, the
mean matter density contrast drops off like $1/R^2$, then we found that $\vert \delta_{\phi} / \bar{\delta} \vert \sim
\Oo((1+w)H^2 R_{c}^2)$. DDE clustering could potentially be detected
through the ISW effect or measurements of the peculiar velocity field
of matter.  Roughly, the magnitude of the DDE
clustering contribution to these effects is a factor $\vert
(1-\Omega_m) \delta_{\phi} / \Omega_m \delta_m \vert \sim
\Oo((1-\Omega_m)(1+w) H^2 R_{c}^2 / \Omega_m) \ll 1$ times the
contribution from the clustering of ordinary matter.  For realistic
astrophysical objects, we found that voids and
super-voids correspond to local overdensities in dark energy, with
$\delta_{\phi}/(1+w) \sim \Oo(10^{-5})$ for clusters and average
voids, and $\delta_{\phi}/(1+w) \sim \Oo(10^{-4})$ for super-voids. If
voids with radii of $100-300\,{\rm Mpc}$ exist within the visible
Universe then $\delta_{\phi}$ may be as large as $10^{-3}(1+w)$.
Linear overdensities of matter and super-clusters generally
correspond to local voids in dark energy; for a typical super-cluster:
$\delta_{\phi}/(1+w) \sim \Oo(-10^{-5})$.  

Astronomical observations
indicate $w = -1 \pm 0.1$ for $z < 1$, and so we conclude that if dark
energy is described by an uncoupled quintessence model, then today
dark energy is almost homogeneous on sub-horizon scales with
perturbations generally of order of $10^{-6}$; the largest
perturbations will be $\Oo(10^{-5}-10^{-3})$ and associated with very
large voids of matter.

%%%%%%%%%%%%%%%%%%%%%%%%%%%%%%%%%%%%%%%%%%%%%%%%%%%%%%%%%%%%%%%%%%%%%%
\section*{Acknowledgements} %%%%%%%%%%%%%%%%%%%%%%%%%%%%%%%%%%%%%%%%%%%
%%%%%%%%%%%%%%%%%%%%%%%%%%%%%%%%%%%%%%%%%%%%%%%%%%%%%%%%%%%%%%%%%%%%%%

We thank S. Dutta and I. Maor for the many 
useful comments and suggestions on this work. DFM is supported by the Alexander von Humboldt Foundation. DJS is supported by PPARC

%\section*{References}


\begin{thebibliography}{99}

\bibitem[\protect\citeauthoryear{Abramo et al.}{2007}]{liberato}Abramo~L.~R., Batista~R.~C., Liberato~L. and Rosenfeld~R. (2007),  \texttt{arXiv:0707.2882 [astro-ph]}.

\bibitem[\protect\citeauthoryear{Adelman-McCarthy et al.}{2006}]{lss}Adelman-McCarthy~J.~K. \emph{et al.} (2006),
  Astrophys.\ J.\ Suppl.\  {\bf 162} 38.
\bibitem[\protect\citeauthoryear{Amendola}{2000}]{amendola}Amendola~L. (2000), Phys.\ Rev.\ D \textbf{62}, 043511.
\bibitem[\protect\citeauthoryear{Amarzguioui et al.}{2006}]{morad}Amarzguioui~M., Elgaroy~O., Mota~D.~F. and Multamaki~T. (2006), Astron. Astrophys. \textbf{454} 707.

\bibitem[\protect\citeauthoryear{Balaguera-Antol\'{\i}nez, Mota \& Nowakowski}{2006}]{bala}Balaguera-Antol{\'{\i}}nez~A., Mota~D.~F. and Nowakowski~M. (2006),
Class. Quant. Grav \textbf{23}, 4497-4510.
\bibitem[\protect\citeauthoryear{Balaguera-Antol\'{\i}nez, Mota \& Nowakowski}{2007}]{bala1}Balaguera-Antol{\'{\i}}nez~A., Mota~D.~F. and Nowakowski~M. (2007), to appear in Mon. Not. Roy. Astro. Soc..
\bibitem[\protect\citeauthoryear{Bardelli et al.}{2000}]{bardelli}Bardelli~S., Zucca~E., Zamorani~G., Moscardini~L. and Scaramella~R. (2000), Mon. Not. Roy. Astro. Soc. 312, 540.
\bibitem[\protect\citeauthoryear{Barrow \& Shaw}{2007}]{shawbarrow1}Barrow~J.~D. and Shaw~D.~J. (2007), Gen. Rel. Grav. to appear in Obregon's Festschrift.
\bibitem[\protect\citeauthoryear{Bartelmann, Doran \& Wetterich}{2005}]{doran}Bartelmann~M., Doran~M. and Wetterich~C. (2005),  \texttt{[arXiv:astro-ph/0507257]}.

\bibitem[\protect\citeauthoryear{Bondi}{1947}]{tolbondi1}Bondi~H., Mon. Not. R. astron. Soc. \textbf{107}, 410 (1947)
\bibitem[\protect\citeauthoryear{Bolejko, Krasi\'nski \& Hellaby}{2005}]{bolejko} Bolejko~K., Krasi\'nski~A. and Hellaby~C. (2005), Mon. Not. Roy. Astron. Soc. 362, 213.

\bibitem[\protect\citeauthoryear{Brax et al.}{2004}]{brax}  Brax et al. 2004 Phys.Rev.D70, 123518. 

\bibitem[\protect\citeauthoryear{Brookfield et al.}{2006a}]{brook} Brookfield~A.~W.  \emph{et al.} (2006),  Phys.\ Rev.\ Lett.\  {\bf 96}, 061301.
\bibitem[\protect\citeauthoryear{Brookfield et al.}{2006b}]{lsss2}Brookfield~A.~W., van~de~Bruck~C., Mota~D.~F. and Tocchini-Valentini~D. (2006b),  Phys.\ Rev.\  D {\bf 73}, 083515.

\bibitem[\protect\citeauthoryear{Caimmi}{2007}]{roberto} Caimmi R., 2007, New Astron.12:327-345. 
 


\bibitem[\protect\citeauthoryear{Capozziello,  Cardone \& Troisi}{2005}]{mod1}Capozziello~S.,  Cardone~V.~F.  and Troisi~A. (2005), 
Phys.\ Rev.\ D {\bf 71}, 043503.
\bibitem[\protect\citeauthoryear{Conley  et al.}{2007}]{HubbleBubble1}Conley~A. \emph{et al.} (2007), \texttt{arXiv:0705.0367 [astro-ph]}; 

\bibitem[\protect\citeauthoryear{Corasaniti, Giannantonio \& Melchiorri}{2005}]{lsss}Corasaniti~P., Giannantonio~T and Melchiorri~A. (2005),  Phys.\ Rev.\  D {\bf 71}, 123521.


\bibitem[\protect\citeauthoryear{Dey, Strauss \& Huchra}{1990}]{dey}Dey~A., Strauss~M.~A. and Huchra~J. (1990), Astron. J. 99, 463.

\bibitem[\protect\citeauthoryear{Doran, Robbers \& Wetterich}{2007}]{georg}Doran~M., Robbers~G. and Wetterich~C. (2007), Phys.\ Rev.\  D {\bf 75}, 023003.

\bibitem[\protect\citeauthoryear{Dutta \& Maor}{2007}]{dutta}Dutta~S. and Maor~I. 2007, Phys.\ Rev.\  D {\bf 75}, 063507.

\bibitem[\protect\citeauthoryear{Dvali,  Gabadadze \& Porrati}{2000}]{dgp}Dvali~G.~R.,  Gabadadze~G. and Porrati~M. (2000),
Phys.\ Lett.\ B {\bf 485}, 208.

\bibitem[\protect\citeauthoryear{Geller, Diaferio \& Kurtz}{1999}]{geller}Geller~M.~J., Diaferio~A. and Kurtz~M.~J. (1999), Astrophys. J. 517, L23.

\bibitem[\protect\citeauthoryear{Giovanelli et al.}{1999}]{HubbleBubble3}Giovanelli~R. \emph{et al.} (1991), Astrophys. J. 525 25.

\bibitem[\protect\citeauthoryear{Hamilton et al.}{1991}]{hamilton}Hamilton~A.~J.~S., Kumar~P., Lu~E., Mathews~A. (1991), Astrophys J., 374, L1.
\bibitem[\protect\citeauthoryear{Hannestad \& Mortsell}{2002}]{lsss1}Hannestad~S. and Mortsell~E. (2002),  Phys.\ Rev.\  D {\bf 66}, 063508.
\bibitem[\protect\citeauthoryear{Hinch}{1991}]{hinch}Hinch~E.~J. (1991), \emph{Perturbation methods}, (CUP, Cambridge).
\bibitem[\protect\citeauthoryear{Hoyle \& Vogeley}{2004}]{hoyle}Hoyle~F. and Vogeley~M.~S. (2004), Astrophys. J. 607, 751.

\bibitem[\protect\citeauthoryear{Hoffman}{1986}]{hoffman}Hoffman~Y. (1986), Astrophys. J. 308, 493.

\bibitem[\protect\citeauthoryear{Inoue \& Silk}{2006}]{SilkVoid}Inoue~K.~T. and Silk~J. (2006), \texttt{[arXiv:astro-ph/0612347]}

\bibitem[\protect\citeauthoryear{Kirschner et al.}{1981}]{kirscher}Kirschner~R.~P., Oemler~A., Schechter~P. L. and Shectman~S.~A. (1981), Astrophys. J. 248, L57.
\bibitem[\protect\citeauthoryear{Kirschner et al.}{1987}]{kirscher2}Kirschner~R.~P., Oemler~A., Schechter~P. L. and Shectman~S.~A. (1987), Astrophys. J. 314, 493.
\bibitem[\protect\citeauthoryear{Koivisto \&
    Mota}{2006}]{koivisto}Koivisto~T. and  Mota~D.~F. (2006), Phys.\ Rev.\  D
  {\bf 73}, 083502  [arXiv:astro-ph/0512135].
\bibitem[\protect\citeauthoryear{Koivisto \&
    Mota}{2007a}]{tomi1}Koivisto~T. and  Mota~D.~F. (2007a), Phys.\ Lett.\  B {\bf 644}, 104 (2007).
\bibitem[\protect\citeauthoryear{Koivisto \&
    Mota}{2007b}]{tomi2}Koivisto~T. and  Mota~D.~F. (2007b), Phys.\ Rev.\  D {\bf 75}, 023518 (2007).

\bibitem[\protect\citeauthoryear{Kolb Matarrese \& Riotto}{2006}]{nodark}Kolb~E., Matarrese~S. and Riotto~A. (2006), New J.Phys. 8 322.

\bibitem[\protect\citeauthoryear{Lahav et al.}{1991}]{Lahav1991}Lahav~O. \emph{et al.} (1991), Mon. Not. Roy. Astro. Soc. 251 128.
\bibitem[\protect\citeauthoryear{Linder \& White}{2005}]{linder}Linder~E. and White~M. (2005), Phys.\ Rev.\  D {\bf 72}, 061304.

\bibitem[\protect\citeauthoryear{Mainini}{2005}]{Mai}Mainini~R. (2005), Phys.\ Rev.\  D {\bf 72}, 083514.
\bibitem[\protect\citeauthoryear{Manera \& Mota}{2006}]{manera}Manera~M. and Mota~D.~F. (2006), Mon.~Not.~Roy.~Astron.~Soc. \textbf{371} 1373 \texttt{[arXiv:astro-ph/0504519]}.
\bibitem[\protect\citeauthoryear{Moffat}{2006}]{nodark1}Moffat~J.~W. (2006),  J. Cosmol. Astropart. Phys. 05.
\bibitem[\protect\citeauthoryear{Maor \& Lahav}{2005}]{maor} Maor~I. and Lahav~O. (2005), \texttt{[arXiv:astro-ph/0505308]}.
\bibitem[\protect\citeauthoryear{Mota \& van de Bruck}{2004}]{carsten} Mota~D.~F. and van~de~Bruck~C.(2004), Astron. Astrophys. {\bf 421}, 71.


\bibitem[\protect\citeauthoryear{Mota \& Shaw}{2006}]{we1} Mota D. F., Shaw D. J., 2006, Phys.Rev.Lett. 97, 151102 
\bibitem[\protect\citeauthoryear{Mota \& Shaw}{2006}]{we2}Mota D. F., Shaw D. J. ,2007, Phys. Rev. D 75, 063501 


\bibitem[\protect\citeauthoryear{Navarro, Frenk \& White}{1997}]{nfw}Navarro~J.~F., Frenk~C.~S. and White~S.~D.~M. (1997), Astrophys. J. 490, 493.
\bibitem[\protect\citeauthoryear{Nojiri  \& Odintsov}{2003}]{mod}Nojiri~S. and Odintsov~S.~ D. (2003), 
Phys.\ Rev.\ D {\bf 68}, 123512. 
\bibitem[\protect\citeauthoryear{Nunes \& Mota}{2006}]{nelson}Nunes~N.~J. and Mota~D.~F. (2006), Mon. Not. Roy. Astron. Soc. 368:2 751 \texttt{[arXiv: astro-ph/0409481]}.
\bibitem[\protect\citeauthoryear{Nunes, da~Silva \& Aghanim}{2005}]{nunes}Nunes~N.~J., da~Silva~A.~C. and Aghanim~N. (2005), \texttt{[arXiv:astro-ph/0506043]}.

\bibitem[\protect\citeauthoryear{Peebles \& Ratra}{1988}]{quint1}Peebles~P.~J.~E. and Ratra~B. (1998), ApJ {\bf 325}, L17. 
\bibitem[\protect\citeauthoryear{Peebles}{1980}]{Peebles1980}Peebles~P.~J.~E. (1980), The large scale structure of the Universe, Princeton University Press.
\bibitem[\protect\citeauthoryear{Pettorino, Baccigalupi \& Mangano}{2005}]{valeria}Pettorino~V., Baccigalupi~C. and Mangano~G. (2005), JCAP {\bf 0501}, 014.
\bibitem[\protect\citeauthoryear{Percival}{2005}]{perc}Percival~W.~J. (2005), \texttt{[arXiv:astro-ph/0508156]}.
\bibitem[\protect\citeauthoryear{Perlmutter et al.}{1999}]{sn1aa}Perlmutter~S. \emph{et al.} (1999), Astrophys. J. {\bf 517}, 565.

\bibitem[\protect\citeauthoryear{Riess et al.}{1998}]{sn1a} Riess~A.~G. \emph{et al.} (1998), Astron. J. {\bf 116}, 1009.
\bibitem[\protect\citeauthoryear{Riess et al.}{2004}]{sn1anew} Riess~A.~G. \emph{et al.} (2004), Astron. J. {\bf 607}, 665.
\bibitem[\protect\citeauthoryear{Riess et al.}{2006a}]{wbounds} Riess~A.~G. \emph{et al.} (2006a), to appear in Astrophys. J. 
\bibitem[\protect\citeauthoryear{Riess et al.}{2006b}]{sn1anew1}Riess~A.~G. \emph{et al.} (2006b), \texttt{[arXiv:astro-ph/0611572]}. 
\bibitem[\protect\citeauthoryear{Rudnick, Brown \& Williams}{2007}]{WMAPcold} Rudnick~L., Brown,~S. and Williams~L.~R. (2007), \texttt{arXiv:0704.0908 [astro-ph]}


\bibitem[\protect\citeauthoryear{Schwarz \& Weinhorst}{2007}]{HubbleBubble}Schwarz~D.~J., Weinhorst~B. (2007), \texttt{arXiv:0706.0165 [astro-ph]}; 
\bibitem[\protect\citeauthoryear{Shaw \& Barrow}{2006a}]{shawbarrow}Shaw~D.~J. and Barrow~J.~D. (2006a), Phys. Rev. D 73, 123505.
\bibitem[\protect\citeauthoryear{Shaw \& Barrow}{2006b}]{shawbarrowb}Shaw~D.~J. and Barrow~J.~D. (2006b), Phys. Rev. D 73, 123506.
\bibitem[\protect\citeauthoryear{Shaw \& Barrow}{2006c}]{shawbarrowc}Shaw~D.~J. and Barrow~J.~D. (2006c), Phys. Lett. B 639, 596.
\bibitem[\protect\citeauthoryear{Shaw \& Mota}{2007}]{improvesc}Shaw~D.~J. and Mota~D.~F. (2007), to appear in Astrophys. J. Supp. 
\bibitem[\protect\citeauthoryear{Spergel et al.}{2003}]{wmap}Spergel~D. \emph{et al.} (2003), Astrophys. J. Suppl. {\bf 148}, 175.
\bibitem[\protect\citeauthoryear{Spergel et al.}{2007}]{wmap3}Spergel~D. \emph{et al.} (2007), Astrophys. J. Suppl. {\bf 170}, 377.
\bibitem[\protect\citeauthoryear{Tegmark et al.}{2006}]{lss1}Tegmark~M. \emph{et al.} (2006), Phys. Rev. D {\bf 74} 123507.
\bibitem[\protect\citeauthoryear{Tolman}{1934}]{tolbondi}Tolman~R.~C. (1934), Proc. Nat. Acad. Sci. USA 20, 169. 
\bibitem[\protect\citeauthoryear{Tully}{1982}]{tully}Tully~R.~B. (1982), Astrophys. J. 257, 389.

\bibitem[\protect\citeauthoryear{Wang}{2006}]{peng}Wang~P. (2006), Astrophys.\ J.\  {\bf 640}, 18.
\bibitem[\protect\citeauthoryear{Wang \& Steinhardt}{1998}]{wangstei}Wang~L. and Steinhardt~P. (1998), Astrophys.\ J.\  {\bf 508}, 483.

\bibitem[\protect\citeauthoryear{Weller \&  Lewis}{2003}]{weller}Weller~J. and Lewis~A.~M. (2003),  Mon.\ Not.\ Roy.\ Astron.\ Soc.\  {\bf 346}, 987.
\bibitem[\protect\citeauthoryear{Wetterich}{1988}]{quint}Wetterich~C. (1988), Nucl. Phys. {\bf B302}, 668.
\bibitem[\protect\citeauthoryear{Wood-Vasey et al.}{2007}]{sn1anew2}Wood-Vasey~W.~M. \emph{et al.} (2007), \texttt{astro-ph/0701041}.



\bibitem[\protect\citeauthoryear{Zehavi et al.}{1998}]{HubbleBubble2}Zehavi~I. \emph{et al.} (1998), Astrophys. J. 503 483.


\end{thebibliography}
\end{document}